\documentclass[journal=mamobx,manuscript=article,layout=twocolumn]{achemso}

\usepackage{amsthm}
\usepackage{mathtools}
\usepackage{physics}
\usepackage{xcolor}
\usepackage{graphicx}
\usepackage{csquotes}
\usepackage{amsmath}
\usepackage{amssymb}
\usepackage{bm}
\usepackage{caption}
\usepackage{subcaption}
\usepackage[english]{babel}
\usepackage{algorithm}
\usepackage{algpseudocode}
\usepackage{tikz}
\usepackage{cuted}
\usepackage[normalem]{ulem}

\usepackage[version=3]{mhchem} 

\newcommand{\Tensor}[1]{\bm{\ensuremath{\mathsf{#1}}}}
\newcommand{\Vector}[1]{\ensuremath{\mathbf{#1}}}
\def\cg{$c_g$}
\def\cbyc{$c/c^*$}
\def\est{$\epsilon_{st}$}
\usepackage[colorlinks,allcolors=cyan!70!black]{hyperref} 

\author{Dominic Robe}
    \affiliation{Department of Chemical and Biological Engineering, Monash University, Melbourne, VIC, 3800, Australia}
    \affiliation{Current address: Department of Mechanical Engineering, University of Melbourne, Melbourne, VIC, 3052, Australia}
\author{Aritra Santra}
 \affiliation{Department of Chemical and Biological Engineering, Monash University, Melbourne, VIC, 3800, Australia}
 \affiliation{Current address: Department of Chemical Engineering, Indian Institute of Technology (Indian School of Mines), Dhanbad, Jharkhand,  826004, India}
\author{Gareth H. McKinley}
 \affiliation{Department of Mechanical Engineering, Massachusetts Institute of Technology, Cambridge, MA, 02139, United States}
\author{J. Ravi Prakash}
    \email{ravi.jagadeeshan@monash.edu}
    \affiliation{Department of Chemical and Biological Engineering, Monash University, Melbourne, VIC, 3800, Australia}

\title{Evanescent Gels: Competition Between Sticker Dynamics and Single Chain Relaxation}


\begin{document}

\begin{abstract}
Solutions of polymer chains are modelled using non-equilibrium Brownian dynamics simulations, with physically associative beads which form reversible crosslinks to establish a system-spanning physical gel network. Rheological properties such as the zero-shear-rate viscosity and relaxation modulus are investigated systematically as functions of polymer concentration and the binding energy between associative sites. It is shown that a system-spanning network can form regardless of binding energy at sufficiently high concentration. However, the contribution to the stress sustained by this physical network can decay faster than other relaxation processes, even single chain relaxations. If the polymer relaxation time scales overlap with short-lived associations, the mechanical response of a gel becomes ``evanescent'', decaying before it can be rheologically observed, even though the network is instantaneously mechanically rigid. In our simulations, the concentration of elastically active chains and the dynamic modulii are computed independently. This makes it possible to combine structural and rheological information to identify the concentration at which the sol-gel transition occurs as a function of binding energy. Further, it is shown that the competition of scales between the sticker dissociation time and the single-polymer relaxation time determines if the gel is in the evanescent regime.
\end{abstract}

\section{Introduction}
\label{sec:introduction}

\subsection{Associative Polymers}
\label{sec:Associative Polymers}

\begin{figure*}[t]
    \includegraphics[width=\textwidth]{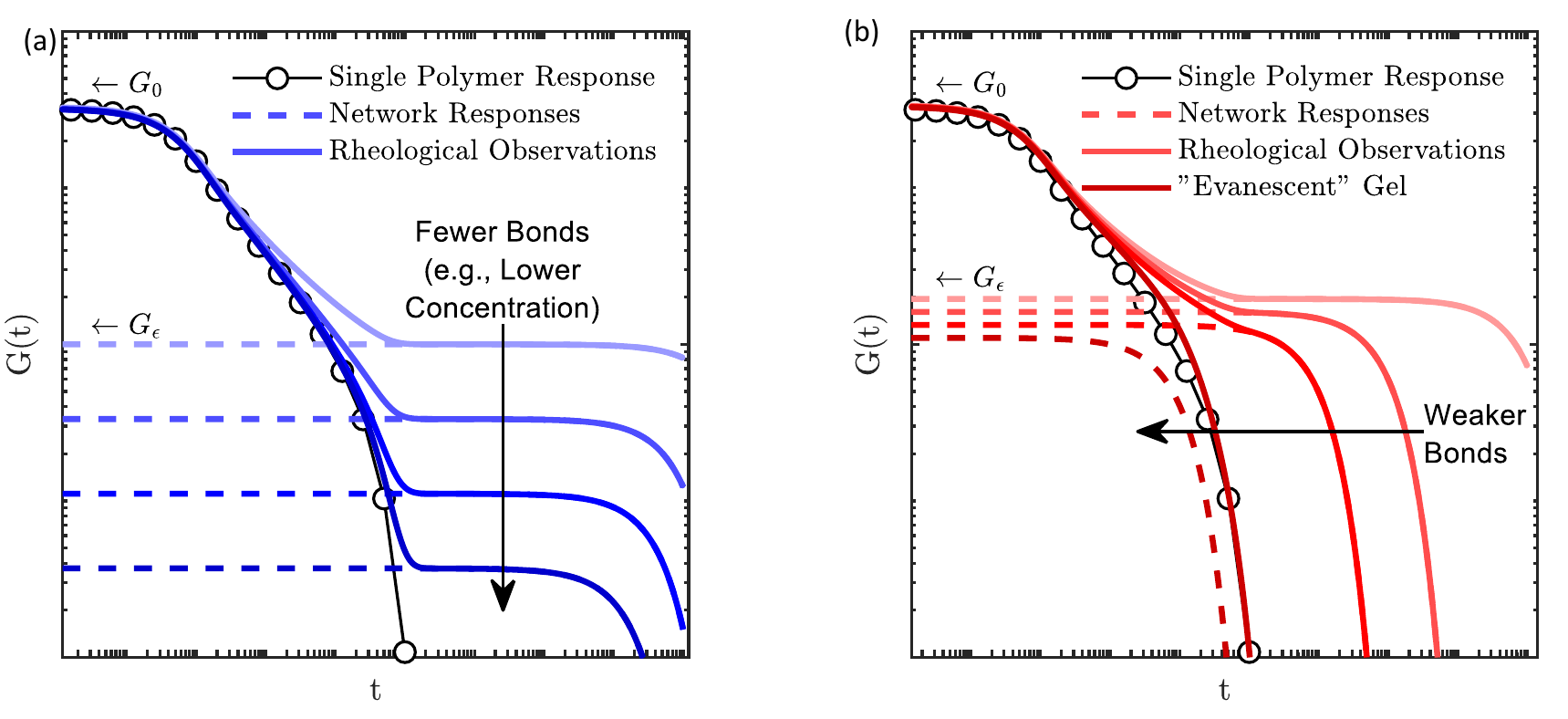}
     \caption{\small{Schematic representations of the hypothesized linear viscoelastic responses of physical gels and their dependence on the number and strength of reversible bonds. (a) Given a fixed sticker strength, the blue curves represent how the relaxation modulus $G(t)$ changes as we decrease the density of bonds (for example, by changing the polymer concentration). Darker shades correspond to a “weaker” gel with a lower elastic plateau. (b) Given a fixed concentration, red curves represent how changing the bond lifetime (for example, by varying the sticker interaction energy) modifies the relaxation modulus. Darker shades approach the “evanescent” gel behaviour, in which the elastic plateau is subsumed by the single chain response. In both figures, the broken lines represent the hypothesized viscoelastic response of the network, which is not readily separable from the single chain response, such that only their combined response (solid coloured lines) is observable rheometrically.}}
    \label{fig:Goft Schematics}
        \vspace{-10pt}
\end{figure*}

Associative polymers are polymer chains that are decorated with functional groups, or stickers, which can form thermo-reversible bonds with each other~\cite{RubColby2003,RubDob97}. This reversibility distinguishes them from polymers which have been chemically cross-linked, essentially permanently \cite{Martin1991-od}. With a sufficiently high number of bonds or cross-links, individual polymer chains become united in a system-spanning mesh capable of bearing mechanical load, and the system has become a gel \cite{Flory1974-tm}. In a permanent chemical gel, this load is supported indefinitely, but in a reversible ``physical'' gel, thermally activated dissociations release stress over time and the gel will eventually flow \cite{Douglas2018-xh}. In chemical gels, the onset of gelation is generally triggered by an increase in concentration or the extent of the cross-linking reaction \cite{Winter:2000gw,Nicolai1997-su}. In physical gels, concentration, temperature, and sticker strength are typical control variables \cite{Parada2018-fx}. In this work these controls are collectively referred to as triggering variables. In both classes of gel, the molecular architecture of the unbound precursor chain (branches and sticker locations) can also play a role in the ultimate material properties of the gel\cite{Chassenieux2011-fv}.

Physical gels have attracted attention due their combination of practically elastic behavior, ultimate fluidity, and tunable properties\cite{Parada2018-fx}. Since the relaxation is driven by thermal activation, physical gels have strongly temperature-dependent rheological properties. Further, solution variables such as pH and concentration \cite{Yan2004-zl,Furikado2014-vr} which can be modified after formulation can also affect the gel properties without depending on the preparation state. These qualities provide physical gels with an adaptability or tunability that is hard to emulate in chemical gels. There are also a range of different associative groups that can be used to decorate the individual chains with \cite{Webber2016-hj,Kloxin2013-ci,Wang2015-kv,Voorhaar:2016jt,Zhang2018-qb,Zhang2017-gs}, which make the sticker strength and functionality design variables as well. The proliferation of control parameters for such physical gels makes them ideal materials for applications that call for rapid prototyping or precisely tuned rheological properties \cite{Yan2004-zl,Furikado2014-vr,Parada2018-fx}.

They have two other valuable qualities as well. Firstly, the associative mechanism for gelation means that if a gel is torn apart by excessive load, the polymer chains will dissociate rather than break the polymer chain, and the stickers can then form new associations to rebuild the mesh in a self-healing process \cite{Stukalin2013-hu,Herbst2013-gi,Chaudhuri2016-lf,Holten-Andersen2011-ke}. In addition, the rheology of associative polymers is driven by the statistical physics of reversible bonds instead of a particular chemistry. This means that applications with limited chemical options (such as biomedical applications \cite{Graham1987-mp,Fullenkamp2013-da,Tang2018-vm,Lee2010-yl}) are readily developed.

The focus of this work is to investigate the effects of polymer concentration and ``sticker strength'' (the activation energy needed to break a reversible bond) on the linear rheological response of a physical gel. Schematic representations of the effects of these parameters are shown in Fig.~\ref{fig:Goft Schematics}.  Broadly, there are obvious limiting behaviors for these two parameters, but a clear understanding of the crossover between these limits is not yet established. In the limit of very high sticker strength, the behavior of a physical gel must approach that of (thoroughly studied) chemical gels. That is, at low concentration, stickers can only bond with other stickers on the same chain, and no network forms. At very high concentration, all stickers are bound, most to different chains, and a robust elastic network is formed, with some defects or dangling chains. In the limit of low sticker strength, associative polymers must behave as undecorated ``homopolymers'' which do not form a gel at any concentration. Somewhere in between these sticker strength limits, as we decrease the sticker strength there must be a point at which the signatures of gelation disappear. In the present work, molecular simulations of a model physical gel system are used to resolve this point and describe the differences in behavior on either side of it. A curious discovery from this investigation has been that a system-spanning network can exist, and yet not exhibit the distinct rheological signatures that are conventionally used to identify a gel.

It is necessary to clarify some terminology about gels and associations. The long-time (or low-frequency) plateau in the elastic modulus of a physical gel (labelled $G_\epsilon$ in Fig.~\ref{fig:Goft Schematics}) is sometimes referred to as $G_0$, but this notation  sets aside the relaxation processes within the polymer chains. Since our work focuses on a regime where both phenomena are relevant, we denote the (persistent but ultimately finite-life time) elastic modulus of a physical gel as $G_\epsilon$. The instantaneous shear modulus, including the entire polymeric contribution at time $t=0$ following imposition of a step strain is then denoted as $G_0$, as indicated in Fig.~\ref{fig:Goft Schematics}. Secondly, we note that, the term ``weak gel'' has been used to describe physical gels quite generically~\cite{Douglas2018-xh}. This term is applied because a physical gel can exhibit a lower apparent elastic modulus than a chemical gel with similar molecular weight, concentration, and bond density. In some cases, the term ``weak gel'' has been used to specifically refer to an associative polymer solution which exhibits no observable elastic plateau, but this manuscript explicitly avoids that usage. There is a distinct difference between a weak sticker and a weak gel. The ``strength'' of a gel is a reference to the magnitude of its elastic modulus, as represented schematically in Fig.~\ref{fig:Goft Schematics}(a). The ``strength'' of a sticker is effectively a reference to the lifetime of a bond, as in Fig.~\ref{fig:Goft Schematics}(b). If one assumes a particular polymer concentration and sticker strength produce a robust gel, then a reduction of the concentration reduces the number of bonds, weakening the gel. If instead the concentration is fixed but the sticker strength is reduced, then the depth of each energy well representing the strength of the stickers binding is reduced, but the high concentration of available stickers will ensure that the number of bonds remains high. The height of the elastic modulus is therefore only marginally affected by the strength of the stickers, while the dissociation time scale is much more sensitive to it. A system with stickers so weak that the gel-like behavior is disrupted by thermal fluctuations is therefore more suitably referred to with a temporal qualifier, so the term ``evanescent'' is used

\subsection{Gel Transition}
\label{sec:Gel Transition}

Experimentally, gelation in associative polymeric solutions is typically characterised by applying small amplitude oscillatory shear and calculating the viscoelastic response in terms of the storage modulus ($G'$), loss modulus ($G''$), and loss tangent ($\tan\,\delta=G''/G'$)~\cite{WinterGel,Bromberg,Dennis09,polym9110556,Ruyumbeke2017,SumanJOR,Indei17}. The dynamic moduli as functions of the oscillation frequency, $\omega$, for an unentangled homopolymer solution are well described by the spectrum of relaxation times predicted by the Zimm model (or, in simulations that neglect hydrodynamic interactions, by the Rouse model), so that there always exists a terminal flow regime in the limit of sufficiently low frequency. However, multi-sticker associative polymer solutions may exhibit broad power law relaxation spectra~\cite{Mewis2001,Andersen2014,Wagner2017} with no readily discernible characteristic relaxation time scale, such that $G'\sim G''\sim \omega^{n}$ over a wide range of oscillatory forcing frequency~\cite{WinterGel,SumanJOR}. The onset of gelation for such systems is typically identified by plotting the loss tangent ($\tan\,\delta$) as a function of the triggering variable for a range of frequency, and then identifying the threshold value of the triggering variable for which $\tan\,\delta$ becomes independent of frequency~\cite{WinterGel,SumanJOR}. 

Apart from the dynamic moduli, gelation in associative polymer solutions is also characterized based on the scaling of zero-shear-rate viscosity ($\eta_{p0}$) and terminal relaxation time ($\tau$) with variations in the triggering variable such as concentration or temperature~\cite{Bromberg,SumanJOR,RnSdynamics}. For instance, the divergence of the zero-shear-rate viscosity and terminal relaxtion time at the onset of the sol-gel transition are prominent dynamic signatures of gelation for strong elastically active gels~\cite{SumanJOR}. However, for weak stickers, instead of divergence, both the zero-shear-rate viscosity and the terminal relaxation time can exhibit non-divergent power-law scaling in the gel regime~\cite{RnSdynamics,RnS2001,Bromberg}. Another characteristic of the gel transition is the appearance of an equilibrium shear modulus at long times (or low frequency) as the system becomes mechanically rigid. The magnitude of this modulus (visually, the height of a plateau in $G(t)$ or $G'(\omega)$, denoted as $G_\epsilon$ in Fig.~\ref{fig:Goft Schematics}) increases with the triggering variable in accordance with percolation theory \cite{10.1007/3-540-11471-8_4,Stauffer,SumanJOR,Winter1997}.

The onset value of a triggering variable at which the transition from the ``sol'' (phase with no system-spanning network) to the ``gel'' phase (where there is such a network) occurs is called the gel point and is denoted by $c_g$ when concentration is the control variable. At this point the solution forms a ``critical gel'', meaning that a fractal network has formed \cite{SumanJOR,Muthukumar1989-kk} which spans the sample domain, but contains a vanishing number of loops. Such a network exhibits a power law distribution of relaxation times, and therefore a divergent viscosity. Simultaneously, the critical gel lacks interconnected structures (discussed in the next section), necessary to exhibit an elastic plateau. The development of this network across the gel point represents a percolation transition, which exhibits several scaling laws near the transition on either the gel or the sol side. In particular, we note that when $c$ is near $c_g$, we expect
\begin{eqnarray}
    G(t) & =G_0\left(\dfrac{t}{{\hat \tau}_0}\right)^{-n}, \label{eqn:Gtscaling} \\[10pt]
    \eta_{p0} & \sim\left(\dfrac{c_g-c}{c_g}\right)^{-s},  \label{eqn:etascaling} \\[10pt]
     G_\epsilon & \sim\left(\dfrac{c-c_g}{c_g}\right)^{z}. \label{eqn:Gcscaling}
\end{eqnarray}
Here ${\hat \tau}_0$ refers to the relaxation time of the smallest possible connection between branch points in a network (in this work, the number of springs between adjacent stickers) and $\eta_{p0}$ is the polymer contribution to the zero-shear-rate viscosity. The scaling exponents $n,s,$ and $z$ depend on the topology of the individual associative polymers, but the relationships between these exponents is universal, and takes the form $n = z/(z+s)$~\cite{SumanJOR}.

\subsection{Sticky Theories}
\label{sec:Sticky Theories}

The mechanical rigidity of a gel originates from the concentration of elastically active chains within the network. Elastically active chains are distinguished from dangling or looping chains or free clusters. A dangling chain is only connected to the network at one end. Looping chains connect back to the same point in the network. Free clusters (and free chains) aren't associated to the system-spanning network. All of these other types of chains can fully relax any stress on a finite time scale, but each elastically active chain contributes to the overall elastic modulus of the network. The relationship between the concentration of elastically active chains $\nu_e$ and the shear modulus $G_\epsilon$ is often expressed as $G_\epsilon=\nu_ek_BT$. However, $\nu_e$ is not directly measurable in a rheological experiment. To the authors' knowledge, neither has it been evaluated explicitly by previous simulation studies, likely due to the subtlety of distinguishing elastically active chains from fully relaxable ones. The number of elastically active chains is related to the circuit rank of the elastic network \cite{KROLL201582}. The circuit rank is a measure of the number of independent closed loops within a network, which can be extracted from the bond configuration in a simulation. The number density $\nu_e$ is usefully related to the rheological signatures of gelation. Below the gel point, $\nu_e$ is zero by definition, since there is no system-spanning network. At the gel point, $\nu_e$ is still technically zero, since the circuit rank at the percolation transition is zero. At this point there is a scale free distribution of relaxation times, since the fractal nature of the network amounts to a scale free distribution of dangling chains lengths. As more bonds form beyond the gel point, new circuits in the network introduce mechanical rigidity, so $\nu_e$ (and therefore $G_\epsilon$) increases. The appearance of this bond-concentration-dependent elastic modulus is what is represented in Fig.~\ref{fig:Goft Schematics}(a).

In associative polymer solutions, formation of a percolating network does not always guarantee the existence of a gel with a measurable elastic modulus. Polymer chain length, monomer concentration, density of sticky groups along the polymer backbone are all parameters that modify the viscoelastic response~\cite{Ruyumbeke2017,Indei17}. It is interesting to note that depending on the elasticity of the network formed after gelation, the dynamic modulii ($G'$ and $G''$) may show power law scaling with no characteristic relaxation, terminal flow, or a distinct plateau (eventually also reaching a terminal regime)~\cite{Ruyumbeke2017,Indei17,Bromberg}. For instance, $G'$ and $G''$ can exhibit a strong elastic response in a system with high molecular weight, yet show a terminal flow behaviour at low molecular weight with the same spacing and strength of stickers\cite{Ruyumbeke2017}. Such work demonstrates that the crossover between gel-like and fluid-like behavior is multi-faceted. There are not always well-established guiding principles for determining under which conditions a gel forms.

Since the unique rheological properties in solutions of associative polymers arise from the statistics of sticker association, \citet{AritraStatJoR} have thoroughly investigated the static properties related to the scaling of intra-chain and inter-chain associations and different static signatures of gelation based on percolation, maxima in the free-chain concentration, and onset of bimodality in the cluster size distribution. However, the relationship of these static or instantaneous signatures with the viscoelasticity and dynamic signatures of gelation is currently unknown. The key findings from that earlier study on the static signatures of gelation are that each of these different signatures occur at different values of scaled concentration, $c/c^*$, and these concentrations are independent of the chain length ($N_b$) for a given distance between stickers on a chain ($\ell$) and sticker strength ($\epsilon_{st}$). In the present work, this ambiguity is dispelled by combining microscopic structural measurements with conventional rheological bulk measurements of the shear relaxation modulus to reveal the effect that changing sticker strength has on physical gelation.

The paper is organised as follows. Several aspects of our simulation methods are discussed in the \nameref{sec:Model} section. The subsection \nameref{sec:sticker algorithm} is noted as a pivotal development which is instrumental to this work, in which a method of calculating the concentration of elastically active chains explicitly from simulation configurations is presented. The \nameref{sec:Results} section presents the data obtained from these simulations. In the \nameref{sec:Conclusion} we discuss several insights and conjectures drawn from the simulation data.

\section{Model for Associative Polymer Solutions} \label{sec:Model}

\subsection{Governing Equation}
\label{sec:governing equation}

A bead-spring chain model was used to simulate solutions of associative polymers using Brownian dynamics (BD) within the HOOMD-Blue simulation toolkit \cite{ANDERSON2020109363,HOWARD2019139}. The governing equation in BD simulations is a stochastic differential equation describing the evolution of the position vector $\Vector{r}_\mu(t)$ of a bead $\mu$ with time $t$. The Euler integration algorithm for the non-dimensional version of this It\^{o} stochastic differential equation is given in its most general form as~\cite{Stoltz2006},
\begin{equation}
\label{eqn:governing}
    \begin{split}
        \Vector{r}_\mu(t+\Delta t)=&\Vector{r}_\mu(t)+(\pmb{\kappa}\cdot\Vector{r}_\mu(t))\Delta t\\
        &+\frac{\Delta t}{4}\sum_{\nu=1}^N\Tensor{D}_{\mu\nu}\cdot\left(\Vector{F}^\Vector{s}_\nu+\Vector{F}^\Vector{SDK}_\nu\right)\\
        &+\frac{1}{\sqrt{2}}\sum_{\nu=1}^N\Tensor{B}_{\mu\nu}\cdot\Vector{\Delta W}_\nu
    \end{split}
\end{equation}
This equation is nondimensionalized with the length scale $l_H=\sqrt{k_BT/H}$ and time scale $\lambda_H=\zeta/H$, where $H$ is the spring constant and $\zeta$ is hydrodynamic friction coefficient of a bead. There are three qualitatively different terms in this equation. The term containing $\pmb{\kappa}$ accounts for the unperturbed solvent flow field \Vector{v} through $\pmb{\kappa}=(\nabla\Vector{v})^\top$.

The second term accounts for the forces acting on each bead, where $\Vector{F}^\Vector{s}$ and $\Vector{F}^\Vector{SDK}$ are the net forces on each particle due to the connecting springs and pairwise interactions, which will be detailed later. The spring forces $\Vector{F}^\Vector{s}$ are drawn from a finitely extensible nonlinear elastic (FENE) spring force law with extensibility parameter $b=50.0$. The force $\Vector{F}^\Vector{SDK}$ includes contributions from the excluded volume (EV) interaction between all overlapping pairs of beads, as well as the associative interaction between reversibly bound pairs of stickers. Both of these types of interactions are modeled by the piecewise potential proposed by Soddemann, Dünweg, and Kremer (SDK)\cite{soddemann2001generic}.
\begin{strip}
    \begin{align}
        U_{\textrm{SDK}}=\left\{
        \begin{array}{l l l}
            &4\left[ \left( \dfrac{\sigma}{r} \right)^{12} - \left( \dfrac{\sigma}{r} \right)^6 + \dfrac{1}{4} \right] - \epsilon;  & r\leq 2^{1/6}\sigma \vspace{0.5cm} \\
            & \dfrac{1}{2} \epsilon \left[ \cos \,(\alpha \left(\dfrac{r}{\sigma}\right)^2+ \beta) - 1 \right] ;& 2^{1/6}\sigma \leq r \leq r_c \vspace{0.5cm} \\
            & 0; &  r \geq r_c
        \end{array}\right.
    \end{align}
\end{strip}
This potential can represent interactions from an athermal excluded volume interaction if $\epsilon=0$, to a strong associative interaction if $\epsilon\gg1$. We denote strength of the excluded volume interaction between non-associative backbone beads as $\epsilon_{bb}$, This same interaction applies to non-associative beads interacting with a sticker. Meanwhile, the sticker strength is $\epsilon_{st}$. In principle, the EV interaction can interpolate between good solvent and poor solvent limits by varying $\epsilon_{bb}$ \cite{Aritra2019}, but in this work, it was fixed in the good solvent limit $\epsilon_{bb}=0$.

The diffusion tensor $\Tensor{D}_{\nu \mu} = \delta_{\nu \mu} \Tensor{\delta} + (1-\delta_{\nu \mu})\pmb{\varOmega}(\Vector{r}_{\nu}-\Vector{r}_{\mu})$,  where $\Tensor{\delta}$ and $\delta_{\mu \nu}$ represent a unit tensor and Kronecker delta respectively, and $\pmb{\varOmega}$ is the Rotne-Prager-Yamakawa (RPY) hydrodynamic interaction tensor \cite{Yamakawa1971}. The vector $\Vector{W}_\nu$ represents a collection of independent standard Wiener process, so each element is drawn from a real-valued Gaussian distribution with zero mean and variance $\Delta t$. The non-dimensional $\Tensor{B}_{\nu \mu}$ acts on $\Vector{W}_\nu$ to produce multiplicative noise\cite{Ottinger1996}. It is sometimes noted as the ``square root'' of the diffusion tensor by constructing block matrices $\bm{\mathcal{D}}$ and $\bm{\mathcal{B}}$ where the $(\mu,\nu)$ block contains $\Tensor{D}_{\mu\nu}$ or $\Tensor{B}_{\mu\nu}$, respectively, so $\bm{\mathcal{B}}\cdot\bm{\mathcal{B}}^T=\bm{\mathcal{D}}$. Efficiently computing this decomposition is challenging, but has recently been implemented as a plugin to HOOMD-Blue using an efficient "positively split" Ewald sum\cite{Fiore2017}. That work brought the cost of large many-particle simulations with hydrodynamic interactions down to $\mathcal{O}(N\sqrt{\log{N}})$. This was a key development for the present work, as large systems are necessary to capture the dynamics of the fractal-like structure of the super-molecular network near the gel transition. We studied systems here with up to thousands of chains, or tens of thousands of particles, but this algorithm could support systems with millions of particles. The super-linear computational cost due to the logarithm would increase from a factor of 2 at $N=10^4$ to 2.45 at $N=10^6$. RPY hydrodynamics are characterized by the hydrodynamic radius of a bead $a$, which was fixed at $a=1$ in the original positively split Ewald implementation. We have modified it to allow for the typical hydrodynamic interaction parameter $h^*=a/\sqrt{\pi}$ in polymer theory, and whose value has been set in the current simulations to be $h^* =0.2$.

The simulation workflow is as follows. Polymer chains with $N_b$ total beads are initialized in random walk configurations and equilibrated in the dilute limit with no associations to measure the radius of gyration in order to calculate the overlap concentration $c^*$. The homopolymer relaxation time $\tau_0$ is estimated using the Thurston approximation to the Zimm model \cite{THURSTON1974569}. Systems are then equilibrated at a chosen \cbyc{} with \est{} = 0 for 3$\tau_0$. Associations are then equilibrated by increasing \est{} in increments of 3 and running for 3$\tau_0$ at each \est. The final simulation snapshot at each \est{} is used as the initial configuration for a step strain measurement. Planar shear flow is applied via the velocity gradient
\begin{equation}
    \nabla\Vector{v}=\begin{pmatrix}
0 & 0 & 0\\
\dot{\gamma} & 0 & 0\\
0 & 0 & 0
\end{pmatrix}
\end{equation}
with Lees-Edwards periodic boundary conditions. The strain rate $\dot{\gamma}=10^6$ is applied for $t=2\cdot 10^{-7} \lambda_H$, to produce a final strain of $\gamma=0.2$. After the step, systems were held at this strain for $10^3\lambda_H$ to measure $G(t)$. Larger strains were found to separate a small fraction of associated stickers. The strain used here produced a response consistent with smaller strains, but with a clearer signal to noise ratio. 

\subsection{Dynamic Functional Associations}
\label{sec:sticker algorithm}

By utilizing various polymer chemistries, associative groups can be incorporated into physical gels with a wide variety of properties \cite{Douglas2018-xh,djabourov_nishinari_ross-murphy_2013,doi:10.1021/cr990125q}. Two key parameters to describe this interaction are the energy barrier against dissociation (sticker strength \est) and the number of associations that a functional group can form (functionality $f$). If the association is based on hydrophobic interactions, the bond strength varies smoothly with the length of a hydrophobic block, and the functionality can be very high. For other chemistries, the functionality is generally 1-3, and the strength depends on chemistry and pH. Another feature of associative systems is the notion of selective ``species'' of stickers which either only associate with themselves, or only with a particular other species. An assortment of such qualities can be leveraged to enhance material performance\cite{Wu2023-sg}.

\begin{algorithm}[t]
 \caption{\small{Association Update}}
\label{alg:associations}
\begin{algorithmic}[1]
\State possible pair list $\gets$ conventional neighbour list
\State new pair list $\gets \varnothing$
\ForAll{stickers $i$}
    \State $N_{bonds}[i]\gets0$
\EndFor
\State shuffle possible pair list
\ForAll{pairs $(i,j) \in$ possible pair list}
    \State $\Delta E_{ij} \gets U_{bound}(r_{ij})-U_{unbound}(r_{ij})$
    \If {$(i,j) \in$ previous pair list}
        \If {$X\sim U(0,1)>\exp(-\Delta E_{ij})$}
            \State append $(i,j)$ to new pair list
            \State $N_{bonds}[i]\gets N_{bonds}[i]+1$
            \State $N_{bonds}[j]\gets N_{bonds}[j]+1$
        \EndIf
    \ElsIf{$N_{bonds}[i]<functionality$ \textbf{ and } $N_{bonds}[j]<functionality$}
        \If {$X\sim U(0,1)<\exp(-\Delta E_{ij})$}
            \State append $(i,j)$ to new pair list
            \State $N_{bonds}[i]\gets N_{bonds}[i]+1$
            \State $N_{bonds}[j]\gets N_{bonds}[j]+1$
        \EndIf
    \EndIf
\EndFor
\State previous pair list $\gets$ new pair list
\end{algorithmic}
\end{algorithm}

\begin{table*}[t]
    \centering
    \begin{tabular}{l | l | l | l}
        Parameter Name                     & Symbol          & Values                \\
        \hline
        Backbone interaction strength      & $\epsilon_{bb}$  & 0                    \\
        Sticky beads per chain             & $f$              & 4                    \\
        Spacer beads between stickers      & $\ell$           & 4,0                  \\
        Sticker strength                   & $\epsilon_{st}$  & 3-12                 \\
        Hydrodynamic interaction parameter & $h^*$            & 0,0.2                \\
        Beads per chain ($(\ell+1)f+\ell$) & $N_b$            & 4,24                 \\
        Concentration                      & $c/c^*$          & 0,0.1-10             \\
        Integration time step              & $\Delta t$       & 0.0001, 0.001, 0.005 \\
        Simulation duration                & $t$              & 1000                 \\
        Polymer chains per simulation      & $N_c$            & 8-30, 300, 3000      \\
        Independent simulation instances   & $N_\mathrm{run}$ & 32-1000              \\
        Step Strain                        & $\gamma$         & 0.2                  \\
        Step Strain rate                   & $\dot{\gamma}$   & $10^6$               
    \end{tabular}
     \caption{\small{Parameter values used in this work.}}
    \label{tab:parameters}
      \vspace{-5pt}
\end{table*}

In this work, if a pair of beads are associated, they simply interact with an SDK potential with a larger value of the well depth than the “backbone” excluded volume interaction that most bead pairs feel. The decision to bind or unbind a pair is made using a typical Monte Carlo (MC) process. If two stickers are within the cutoff distance of each other, the change in energy $\Delta E$ if the bond state were changed is calculated. A pseudo-random number is drawn from a uniform distribution between 0 and 1. If the random number is less than $\exp(-\Delta E/k_bT)$, then the change of state is carried out. In an update sweep, each existing bond attempts to break in this manner, then bond formations are attempted. If two stickers are within the cutoff distance, but at least one of them already has as many bonds as the functionality setting permits, then the bond formation for the new pair is not attempted.

The bond update algorithm is detailed in Algorithm 1. A simulation contains $N_{stick}$ stickers each with functionality $f$. The algorithm maintains an $N_{stick}\cross f$ array in which the $i$th row contains the list of indices of stickers associated with sticker $i$. There is a persistent copy of this array called the previous-partners list, which is accessible to the machinery within HOOMD for calculating pairwise forces based on a neighbour list. This persistent array is initialized to the no-bonds state, then in each update step, Algorithm 1 is applied. 

This association update sweep is applied once before each Brownian dynamics time step. If the BD simulation were frozen and the MC bond update applied iteratively, it would sample the Boltzmann distribution for the possible bond configurations of the system, given that instantaneous particle position configuration. This is clear because all possible bond state changes are directly reversible, and have the complementary probabilities required for detailed balance \cite{manousiouthakis1999strict,norris1998markov}. That is, the probability of moving into a state decays exponentially  with the energy difference between the states. As BD and MC update steps are interleaved, it is expected that the infinitesimal changes in particle positions during a BD step will only infinitesimally change the possible bond configuration energies. Therefore, the Boltzmann distribution from one update step to the next will be similar enough for the MC bond update steps to approximate it.

\begin{figure*}[t]
    \centering
    \includegraphics[width=\textwidth]{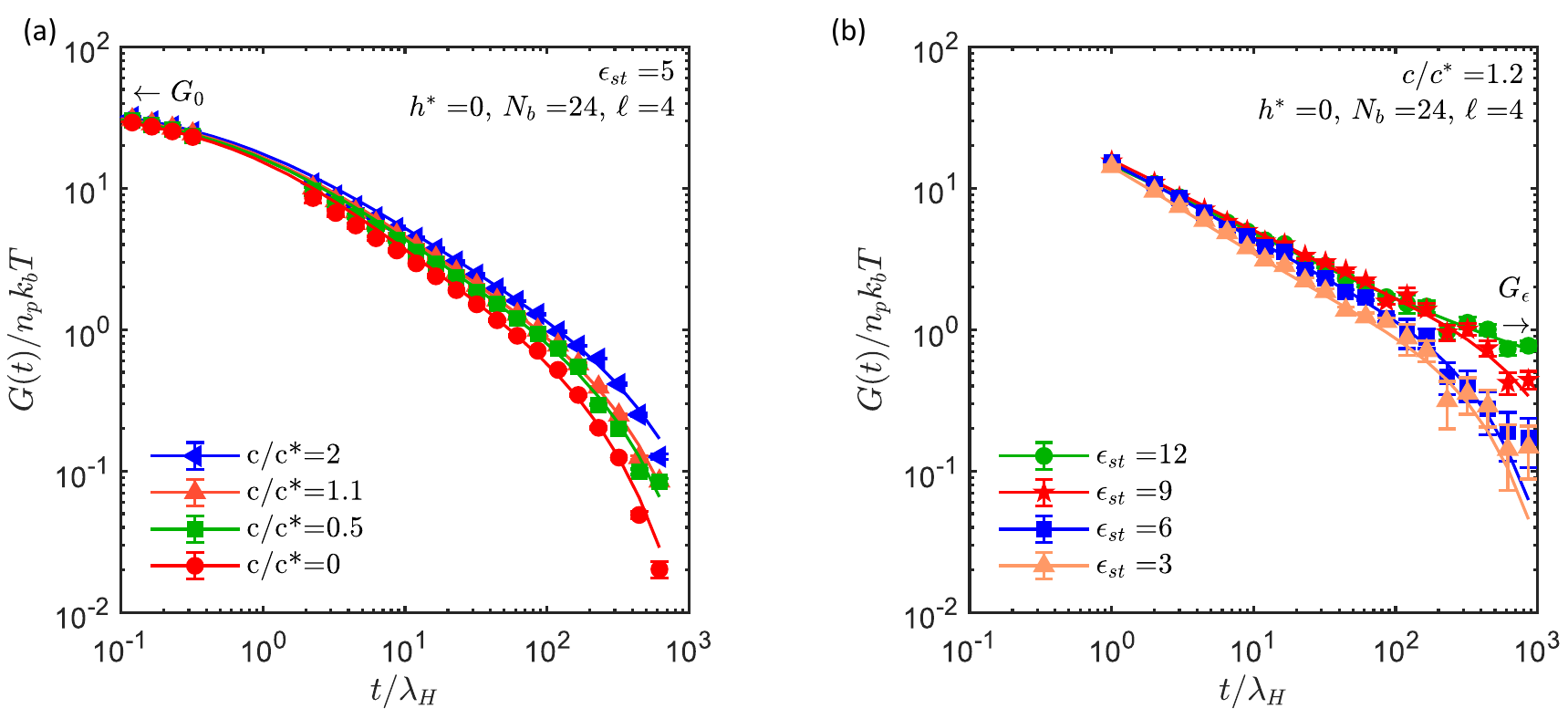}
     \caption{\small{Simulated linear viscoelastic stress response $G(t)$ (a) for sticker strength \est$=5$ while varying \cbyc and (b) for \cbyc$=1.2$ while varying \est{} in the absence of HI.}}
    \label{fig:Goft Weak and Goft Dense}
      \vspace{-10pt}
\end{figure*}

\subsection{Parameter Values}
\label{sec:parameter values}

The simulations carried out to study the dynamic properties consist of multi-sticker associative polymer solutions at finite concentrations with $f=4$ stickers per chain and spacer length, $\ell=4$ or 0, implying the total number of beads per chain $N_b=(\ell+1)f+\ell=24$ or 4. All simulations used an athermal solvent quality for the backbone, $\epsilon_{bb}=0$, and a range of sticker strengths $\epsilon_{st}$. All relevant parameters are compiled in Table~\ref{tab:parameters}. For the specific system $f=4, \ell=4, \epsilon_{st}=5$, the three static signatures of gelation are observed at $c_{g_1}/c^* \approx 0.3$, $c_{g_2}/c^* \approx 1.0$ and $c_{g_3}/c^* \approx 0.5$, where the subscripts $g_1$, $g_2$ and $g_3$ denote three distinct gelation signatures at the percolation transition, onset of cluster size bimodality, and free chain concentration maximum, respectively~\cite{AritraStatJoR}. During the production run, dynamic properties are calculated as a function of time from each independent trajectory. Ensemble averages and error of mean estimates of different dynamic properties are then computed over a collection of 500 to 1000 such independent trajectories. All simulations with HI (with $h^*=0.2$) have been carried out with a non-dimensional time-step $\Delta t=0.005$. Simulations are carried out using the non-dimensional time scale $\lambda_H$ and length scale $l_H$.

\section{Results}
\label{sec:Results}

The bulk of the data examined in this work is extracted from the linear viscoelastic stress response $G(t)$. In BD simulations, this response can be measured either using the stress auto correlation function in equilibrium, or by deforming the simulation volume to perform a step strain protocol and measuring the resulting stress. Examples of the resulting relaxation curves are presented for a few conditions in Fig.~\ref{fig:Goft Weak and Goft Dense}. Detailed expressions for the evaluation of the stress auto-correlation and relaxation modulus, as well as the resulting numerical measurements are provided in the Supplementary Information for all combinations of sticker strength, concentration, chain length, and hydrodynamic interactions which were evaluated for this work. A further validation of our $G(t)$ measurements via selected small amplitude oscillatory shear flow simulation is also presented there. Fig.~\ref{fig:Goft Weak and Goft Dense}(a) shows data from simulations with relatively weak stickers, \est=5, ranging from the dilute limit to twice the overlap concentration. These responses exhibit a Rouse-like distribution of relaxation modes, as well as a modest push toward longer relaxation times with increased concentration. In contrast, Fig.~\ref{fig:Goft Weak and Goft Dense}(b) shows data from simulations just above $c^*$ with a range of \est{} from 3 to 12. The curves for higher sticker strengths exhibit a more notable response at later times. It is expected that for sufficiently high concentration and sticker strength, $G(t)$ should include a plateau due to the elastic response of the network. This interpretation is plausible for the \est=12 case in Fig.~\ref{fig:Goft Weak and Goft Dense}(b), but more data are needed for a clearer picture.

\subsection{Strong Stickers}
\label{sec:Strong Stickers}

\begin{figure}[t]
    \centering
    \includegraphics[width=\columnwidth]{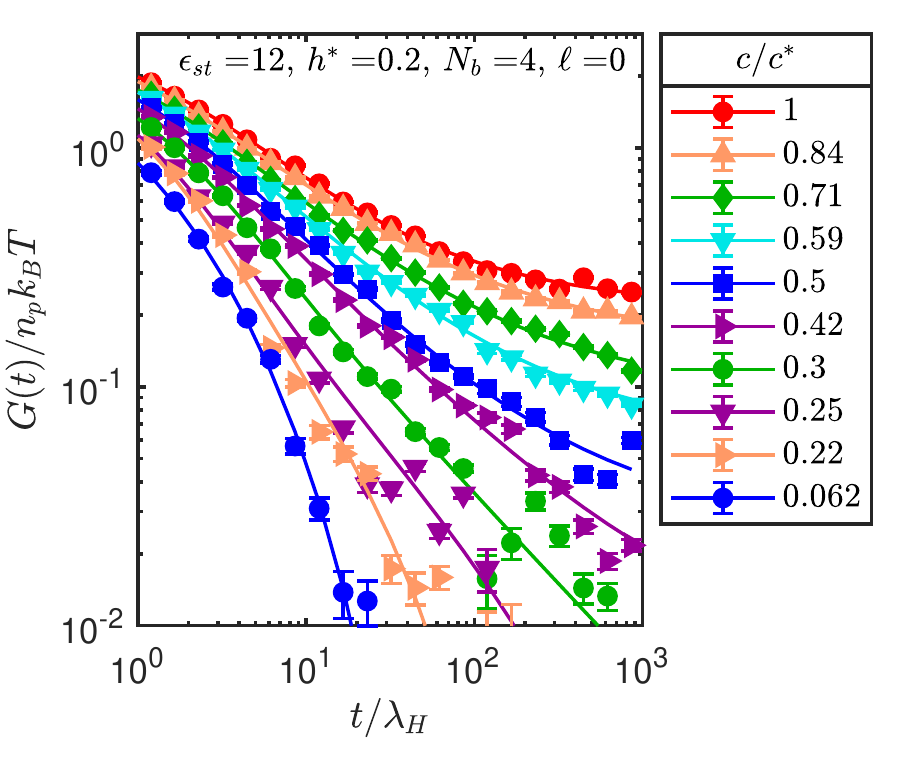}
    \caption{\small{Conventional rheological signature of gelation with concentration as a triggering variable. $G(t)$ exhibits a power law relaxation near $c_g/c^*\approx0.30$, and a growing plateau as $c$ is increased further.}}
        \label{fig:Goft and hyperscaling}
        \vspace{-10pt}
\end{figure}

In order to reduce the computational cost of data collection, a system of very short chains is modelled so that the single-chain relaxation modes are limited to short times, rendering the response of the network more distinguishable at later times. Fig.~\ref{fig:Goft and hyperscaling} shows $G(t)$ for a system of 4-bead chains, in which each bead is a sticker. Fig.~\ref{fig:Goft and hyperscaling}(a) clearly demonstrates the transition from a sol phase with a very short terminal relaxation time, to a critical gel with a power-law relaxation process at \cg{}$/c^*\approx0.30$ causing the viscosity to diverge. Finally,  a gel with a long-lasting elastic plateau is formed. Without running longer simulations, it is impossible to measure directly how long this plateau will persist before dissociations release stress, but the computational expense of BD limits the duration of routine simulations. In the present analysis, $G(t)$ curves are fit with the functional form
\begin{equation}
    G(t)=\left[\mathbb{G}E_\alpha\left(-\frac{\mathbb{G}}{\mathbb{V}}t^\alpha\right)+G_\epsilon\right]\exp\left(-\frac{t}{\tau_\epsilon}\right),
    \label{eqn:MLphys}
\end{equation}
where $E_\alpha(z)$ is the Mittag-Leffler function
\begin{equation}
    E_\alpha(z)=\sum_{k=0}^\infty\frac{z^k}{\Gamma({\alpha}k+1)}.
\end{equation}
This form captures the phenomenology of a broad distribution of relaxation modes due to chain segment relaxation processes, in addition to a single extra mode due to the load supported by a rigid network. The Mittag-Leffler function approaches a stretched exponential at short times and a power law at long times \cite{Podlubny1999-ba,Schiessel1995-ss,Metzler2002-pn}. This function has been derived using fractional calculus as a model for viscoelastic materials \cite{Jaishankar2013-uo,Jaishankar2014-hg}. Here it is simply employed as a phenomenological model to extract properties from $G(t)$ data. We multiply the Mittag-Leffler function by an exponential to account for the terminal flow behavior inherent in physical gels, and to extract the longest relaxation time. For a system in the sol state, the plateau modulus $G_\epsilon$ is 0. In the limit of a chemical gel, the terminal relaxation time $\tau_\epsilon$ is infinite. At the gel transition, the terminal relaxation time has diverged, but the plateau modulus is still 0. In each of these cases, a simplified model can be used to fit $G(t)$ with more robust values of the remaining parameters. A discussion of this fitting and the use of information criteria to determine the appropriate model is available in the Supplementary Information. The zero-shear-rate viscosity $\eta_{p0}$ in the sol phase is measured using the integral of the $G(t)$ fit as $\eta_{p0}=\int_{t=0}^\infty G(t)\dd t$. A validation study comparing this integration to a steady shear protocol is available in the Supplementary Information. The elastic modulus in the gel phase is taken from the fit parameter $G_\epsilon$.

\begin{figure*}[t]
    \includegraphics[width=\textwidth]{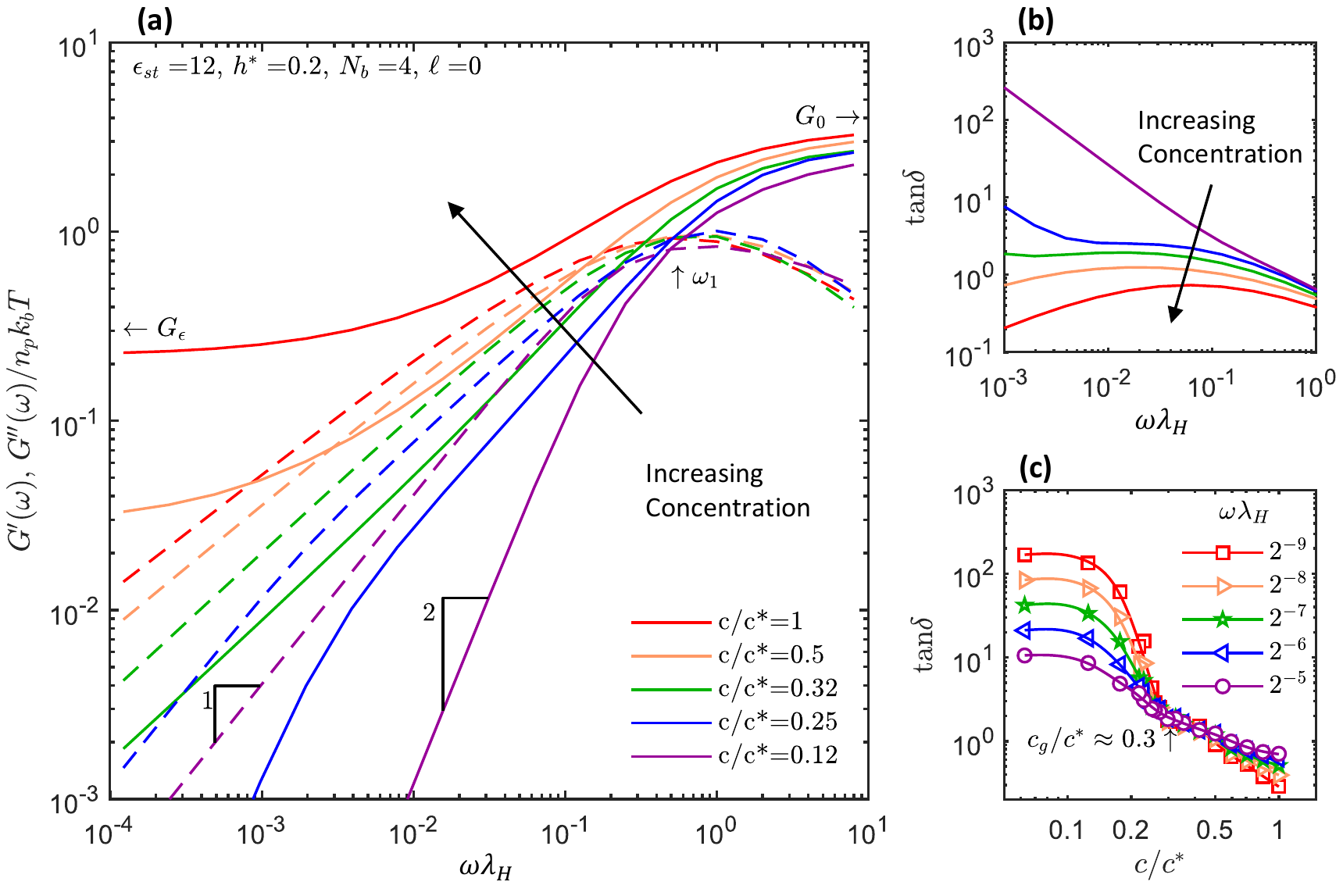}
     \caption{\small{ (a) Storage $G'(\omega)$ (solid) and loss $G''(\omega)$ (dashed) modulii as functions of dimensionless frequency, $\omega \lambda_H$ at fixed \est{} and varied \cbyc. The characteristic frequency $\omega_1$ is indicated at the intersection of $G'$ and $G''$. $G_\epsilon$ and $G_0$ are indicated as the low- and high-frequency limits, respectively, of $G'$. The low-frequency, low-concentration slopes of $G'$ and $G''$ are labelled to indicate terminal flow. (b) The loss tangent $\tan\delta=G''/G'$, for the same values of \cbyc. (c) $\tan\delta$ as a function of concentration. The intersection of curves for various frequencies indicates the gel point.}}
    \label{fig:modulii and tandelta}
      \vspace{-10pt}
\end{figure*}

Small amplitude oscillatory shear experiments are typically used to measure the frequency dependent storage and loss modulii $G'(\omega)$ and $G''(\omega)$. Fig.~\ref{fig:modulii and tandelta}(a) shows a construction of these modulii from simulation data by taking the real and imaginary parts of the Fourier transform of $G(t)$ (further details of these calculations and a validation of the Fourier transform method using a selection of oscillatory shear simulations are given in the Supplementary Information). For numerical stability, the fit to Eqn.~\ref{eqn:MLphys} is used instead of the raw data. These curves exhibit all of the conventional features of gel-forming systems. At low concentration, the terminal flow regime is observed at low frequency. The characteristic frequency $\omega_1$ is also identified at the intersection of $G'$ and $G''$. Near the critical point (the green line for \cbyc = 0.32), the modulii are parallel power laws over a broad range of frequencies. Above \cg, the elastic modulus is seen as a low-frequency plateau in $G'$. The gel transition is sometimes located using the frequency-independence of the loss tangent $\tan \delta=G''/G'$, as plotted in Figs.~\ref{fig:modulii and tandelta}(b) (vs frequency) and (c) (vs concentration). The flatness of $\tan \delta(\omega)$ near \cbyc~$\approx 0.32$ or equivalently the intersection of $\tan \delta(c)$ curves for various frequences at \cbyc~$\approx0.3$ are consistent with the indications drawn from $\eta_{p0}$ and $G_\epsilon$. 

\subsection{Weak Stickers}
\label{sec:Weak Stickers}

\begin{figure}[t]
    \centering
    \includegraphics[width=\columnwidth]{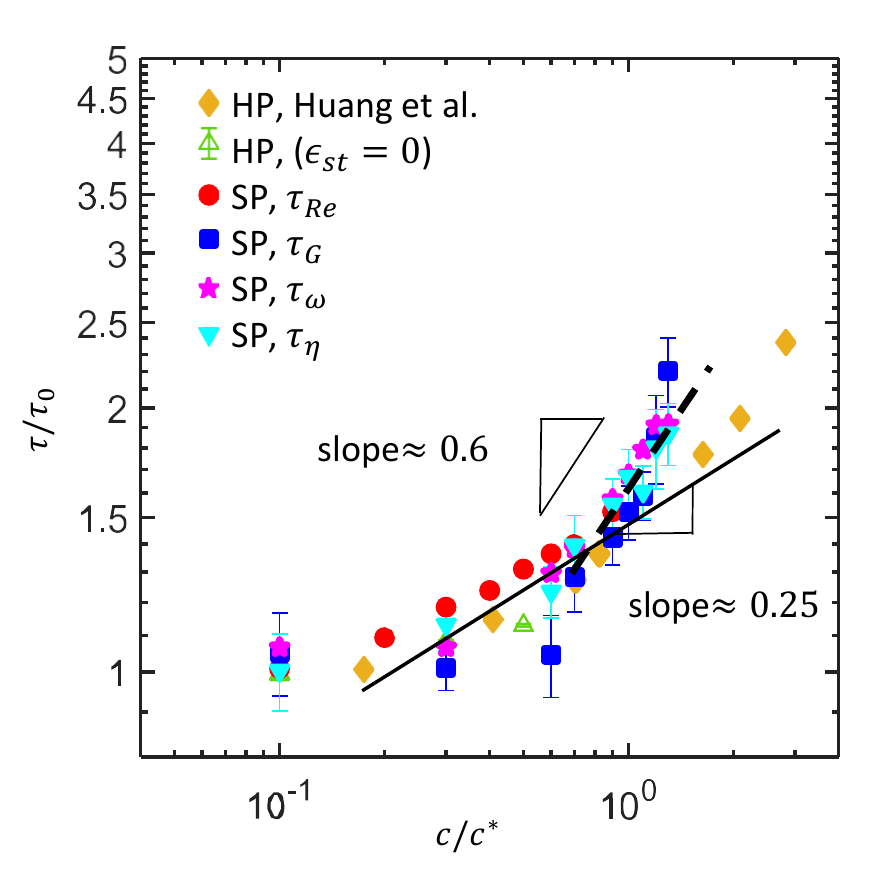}
     \caption{\small{Multiple independent measurements of the longest relaxation time in a solution of associative polymers with \est=5 show self-consistent scaling with \cbyc, but no indication of the diverging time scale expected on approach to the gel transition. Yellow diamonds represent simulation data from \citet{Huang2010} for homopolymer solutions.}}
    \label{fig:tausVc}
      \vspace{-10pt}
\end{figure}

Returning now to the weaker stickers represented in Fig.~\ref{fig:Goft Weak and Goft Dense}(a), the properties of associative polymers are again considered as a function of concentration, despite the absence of a clear plateau due to an elastic network. The scaling of the terminal relaxation time ($\tau$) with concentration is often used to characterize the dynamic signatures of gelation in associative polymer solutions~\cite{SumanJOR,RnSdynamics}. Fig.~\ref{fig:tausVc} presents several dynamical measurements of this system, in an effort to discover any macroscopically observable evidence of the divergence seen in Fig.~\ref{fig:Goft and hyperscaling}. Measurements are reported of time scales extracted from the end-to-end unit vector auto-correlation ($\tau_{Re}$), the tail of the stress relaxation ($\tau_G$), the inverse of the characteristic frequency ($\tau_\omega$), and the zero viscosity ($\tau_\eta$). Plots of the raw numerical measurements of these observables, procedures for extracting relaxation times from each of them, and measurements of the instantaneous shear modulus are presented in the Supplementary Information. All of these time scale measurements $\tau_i$ are normalized by their respective dilute limits $\tau_i^0$. The dependence of all these measurements on concentration collapse to a power law with an exponent near 0.6 above \cbyc$\approx0.5$, as opposed to the divergence leading up to \cg{} observed with higher \est. This raises several notable differences between the rheology of solutions with weak or strong stickers. The system with weak stickers exhibits finite relaxation times at all concentrations. In addition, there is no plateau in $G(t)$ in Fig.~\ref{fig:Goft Weak and Goft Dense}(a). Similarly, presented in the Supplementary Information, there is no frequency-independent region of the loss tangent. These results suggest that a gel has not formed in the weak system. Yet there is a clear quantitative cross-over in the scaling of relaxation times with concentration. Interestingly, the maxima in the free chain concentration for this system when studied by \citet{AritraStatJoR}, which occurs at $c/c^*=0.5$, coincides with the cross-over concentration of the $\tau\sim c^{0.6}$ regime.

The scaling of the ratio $\tau/\tau_0$ with $c/c^*$ for associative polymer solutions is compared with that of homopolymer solutions, where the terminal relaxation time for homopolymers are estimated from end-to-end unit vector auto-correlation function. Note that for homopolymers we have performed only few simulations in the dilute regime, and data has been acquired from the work by Huang et al.~\cite{Huang2010} (yellow diamonds in Fig.~\ref{fig:tausVc}), for the purpose of comparing with associative polymers. As compared to associative polymer solutions, the normalised relaxation time for the homopolymers goes through a broad cross-over with a scaling exponent of $(2-3\nu)/(3\nu-1) = 0.25$, where the Flory exponent $\nu=0.6$, which is a well known scaling law for relaxation time in the semi-dilute unentangled regime~\cite{deGennes,Huang2010}. It is interesting to note that the concentration dependence of homopolymer and associative polymer relaxation times is roughly similar until the associative solution crosses over to the gel scaling regime. It should also be noted that if the strong-sticker simulations could be run for a few more orders of magnitude of time, the dissociation of the network would be captured and the scaling of the terminal time scale could be analyzed.

\begin{figure}[t]
    \centering
    \includegraphics[width=\columnwidth]{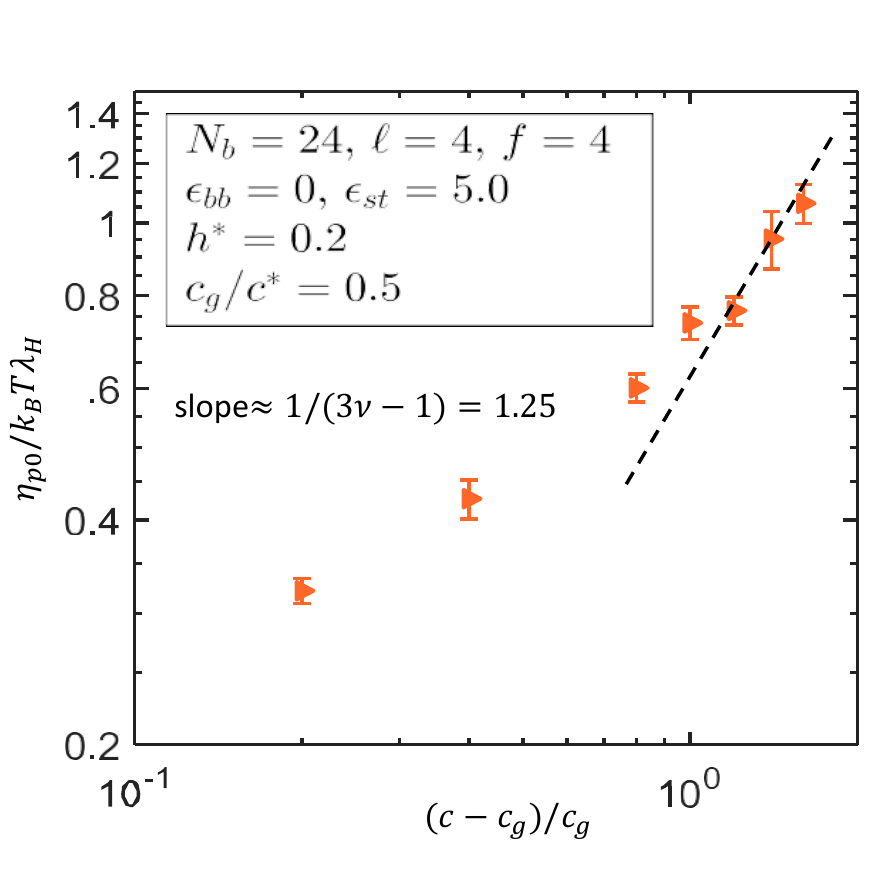}
     \caption{\small{Concentration scaling of the polymer contribution to the zero shear-rate viscosity for a solution of associative polymers with \est=5. The value $c_g/c^*$=0.5 has been estimated based on the scaling regimes in Fig.~\ref{fig:tausVc} to enable this comparison with theory~\citep{RnSdynamics} and experiment~\citep{Bromberg}.}}
    \label{fig:viscosityScaling}
      \vspace{-10pt}
\end{figure}

In the mean-field theory developed by Rubinstein and Semenov~\citep{RnSdynamics} for associative polymer solutions, the zero-shear-rate viscosity ($\eta_{p0}$) in the post gel regime, close to the gel-point, is shown to scale with the relative distance, $\Delta$, from the gelation concentration ($c_g$), where $\Delta = (c_g-c)/c_g$, as 
\begin{equation}
    \label{DyEq:eta0_gel}
    \eta_{p0} \sim \Delta^{1/(3\nu-1)},
\end{equation} 

\noindent{}where $\nu=0.6$ is the Flory exponent, which implies that the exponent $1/(3\nu-1)=1.25$. Considering $c_g/c^*=0.5$, we find $\eta_{p0}$ to scale with a slope of $1.25$ in the post-gel regime, as shown in Fig.~\ref{fig:viscosityScaling}. Note that it is not scaled with the number density of polymer chains, $n_p$ (which is related to monomer concentration $c$), in order to bring out the concentration dependence explicitly. This is one of the dynamic signatures of gelation which is in agreement with the prediction by the mean-field theory~\citep{RnSdynamics} and as observed by Bromberg~\citep{Bromberg} in experiments with thermo-reversible hydrogels.

\begin{figure}[t]
    \centering
    \includegraphics[width=\columnwidth]{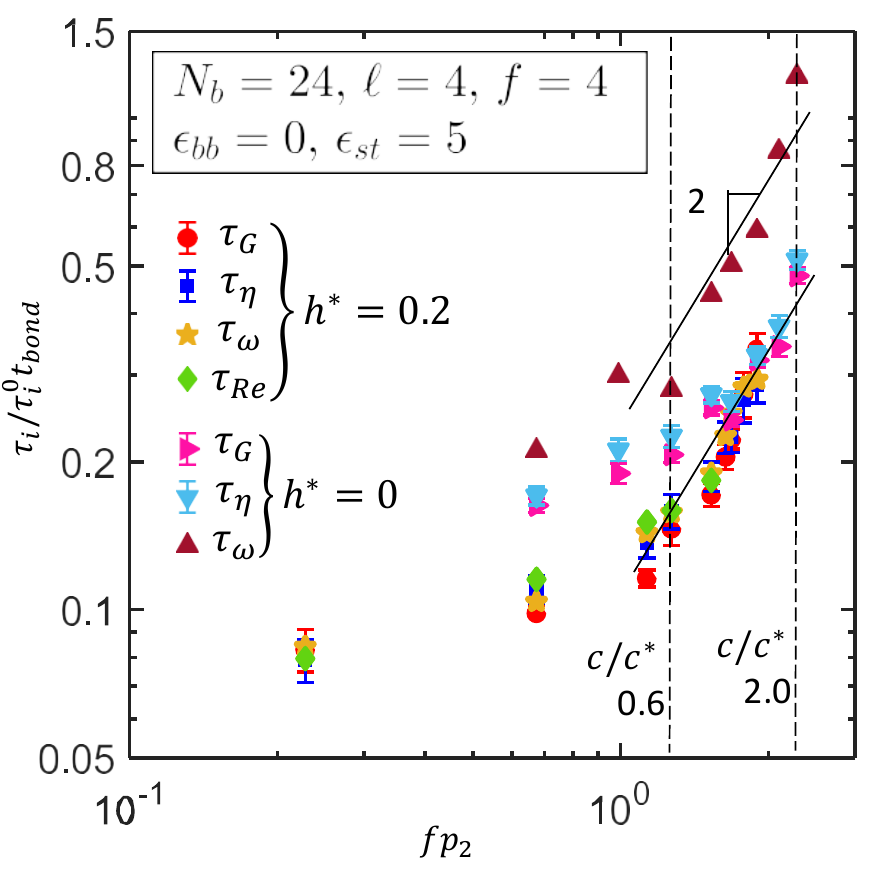}
     \caption{\small{Scaling of terminal relaxation times with the number of inter-chain associations per chain $fp_2$. Vertical lines indicate values of $fp_2$ corresponding to \cbyc=0.6 and 2 for these values of $f,\ell,$ and \est.}}
    \label{fig:fp2Scaling}
      \vspace{-10pt}
\end{figure}

Rubinstein and Semenov~\cite{RnSdynamics,RnS2001} have proposed the sticky Rouse model, which predicts the scaling of the relaxation time, $\tau_{\text{relax}}$, in unentangled solutions of associative polymers well above the gel point. According to the prediction
\begin{equation}
    \label{DyEq:tauRelax}
    \tau_{\text{relax}} \approx t_{\text{bond}}\,(f\,p_2)^2,
\end{equation}

\noindent{}where $t_{\text{bond}}$ is the bond lifetime of associated stickers, $f$ is the number of stickers per chain and $p_2$ is the fraction of inter-chain associated stickers. From its definition, the product $f\,p_2$ is essentially the total number of inter-chain associated stickers in a chain. In order to verify this prediction for our model, we first calculate the average bond lifetime of associated stickers by considering all possible associating pairs for the systems of associative polymer solutions at different concentrations. According to the mean-field theory~\cite{RnSdynamics,RnS2001}, $t_{\text{bond}}$ is effectively a function of only the sticker strength, $\epsilon_{st}$, however, from the Brownian dynamics simulations we find that at a constant value of sticker strength, $\epsilon_{st}=5.0$, the bond lifetime varies weakly with concentration, for both HI and no HI cases (as shown in the Supplementary Information). The average bond lifetime of the stickers is relatively higher for simulations carried out with HI. The reason for such a behaviour may be attributed to the influence of the back flow generated by the long-range hydrodynamic perturbations, which increases the contact time between the stickers. It is noteworthy that the values of bond lifetime are of the same order as that of the large scale relaxation times. The ratio $\tau_i/(\tau_i^0\,t_{\text{bond}})$ is plotted as a function of $f\,p_2$ at different values of monomer concentration in the pre and post-gel regimes, as shown in Fig.~\ref{fig:fp2Scaling}. Here, $\tau_i$'s are the large scale relaxation time estimated from various methods as discussed earlier and $\tau_i^0$'s are the corresponding values of relaxation time in the dilute limit. It is necessary to divide the relaxation times by the concentration-dependent non-dimensional bond lifetime $t_{\text{bond}}$ to expose the $(f\,p_2)^2$ scaling in Eqn.~\ref{DyEq:tauRelax}. In order to investigate the effect of HI on the relaxation dynamics of the associative polymer solutions, results from the simulations with HI are compared with those obtained from simulations without HI. From Fig.~\ref{fig:fp2Scaling}, it is clear that in the HI case the ratio $\tau_i/(\tau_i^0\,t_{\text{bond}})$ is independent of the methods used to evaluate the relaxation time. Moreover, the scaling of the ratio $\tau_i/(\tau_i^0\,t_{\text{bond}})$ with $f\,p_2$ goes through a cross-over and follows an asymptotic exponent of $2$ for a range of concentration well above the gel-point ($c_g/c^*=0.5$), which is in agreement with the prediction of the sticky Rouse model, given in Eq.~(\ref{DyEq:tauRelax}). At low concentration, relaxation seems to be faster with HI, but at high concentration, HI is screened and the two conditions converge. This may indicate that neglecting HI increases the observed \cg.

As shown in Fig.~\ref{fig:fp2Scaling}, in the no HI case the normalised relaxation times computed from $R_e$, $G(t)$, and $\eta_{0p}$ are found to match with each other, however, the relaxation time $\tau_{1/\omega}=1/\omega_1$,  calculated from the intersection of $G'$ and $G''$ diverges from the universal curve. The divergence in the behaviour of $\tau_{1/\omega}$ comes from the difference in the scaling of $G'$ and $G''$ in the intermediate frequencies, based on whether the model includes HI or not. In the Rouse model, where HI is not considered, the dynamic moduli follow the power law scaling $G'\sim G''\sim \omega^{0.5}$ in the intermediate range of frequencies, whereas, in case of Zimm model, which considers HI, $G'\sim G''\sim \omega^{2/3}$ (as shown in the Supplementary Information). Thus the point of intersection of $G'$ and $G''$ is very sensitive to the slope of dynamic moduli in the intermediate frequencies. Given the sensitivity of the point of intersection between $G'$ and $G''$ on the presence or absence of HI, since, $\tau_{1/\omega}$ is estimated from this intersection point, there exist a significant difference in the prefactor of $\tau_{1/\omega}$ calculated with and without HI. However, it is worthy to note that in the limit of high concentration ($c/c^* > 1$) at the post gel regime, the normalised relaxation times in the no HI case also follow the same asymptotic scaling as predicted by the mean-field theory. Furthermore, except the estimate of relaxation times based on intersection of $G'$ and $G''$, the HI and no HI case follow the same universal curve at the limit of high concentration in the post gel regime, indicating screening of HI due of formation of a dense gelation network.

\begin{figure*}[t]
    \includegraphics[width=\textwidth]{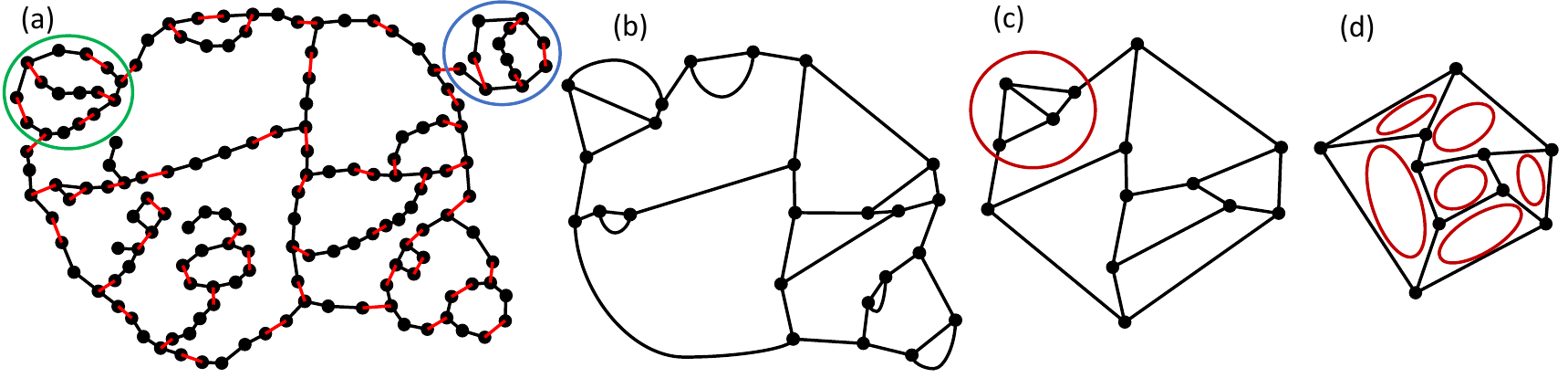}
     \caption{\small{Depiction of the process used to efficiently enumerate the elastically active elements in a super-molecular network. Black edges in (a) correspond to polymer springs, while red edges represent associations. In (b),(c), and (d), edges represent paths between junctions in the original network.}}
    \label{fig:network simplification}
     \vspace{-5pt}
\end{figure*}

\subsection{Identification of Elastically Active Chains}
\label{sec:ElasticallyActive}

The data presented thus far for strong stickers have exhibited the standard rheological signatures of gelation in the permanent-associations limit. Meanwhile the weaker stickers have shown a terminal flow behavior that agrees with theories of thermoreversible associations, but fails the rheological test for the formation of a mechanically rigid network. These results raise two questions. First, how do the weak stickers exhibit the viscosity scaling predicted by assuming the presence of a system-spanning network without exhibiting the elastic response expected from such a network. Second, at what level of sticker strength does the elastic response of the network become apparent? It is noted that a simple consideration of the sticker strength as the triggering variable to find a gel point $\epsilon_g$ at a fixed concentration is uninteresting because the dissociation energy barrier is often determined by chemistry and is not always trivial to adjust. A useful answer to the second question requires a systematic prediction for whether or not a gel is possible at any concentration given a particular \est. To resolve this ambiguity about the existence of a spanning network based on rheological data, the microscopic network structure must be analyzed carefully to aid the interpretation of macroscopic measurements.

The number of elastically active chains in a simulation volume is usually described as the number of independent cycles in a network, or the number of branch points in the network. An important detail to this description is that only those cycles or branches which are attached to the network by at least three independent paths may contribute to the elastic modulus of the network. Here we present an algorithm to identify  the elastically active elements in the simulated molecular network. An illustration of this process is presented in Fig.~\ref{fig:network simplification}. Consider for example a localized, highly inter-connected bundle of polymers which are only connected to the system-spanning network by a single strand, such as the subnetwork circled in blue in the upper right corner of Fig.~\ref{fig:network simplification}(a). This bundle, though it contains many branches and cycles, is ultimately a dangling end which is free to relax after a deformation. Consider further the case that the bundle is connected by two strands, such as the section circled in green if the upper left corner of Fig.~\ref{fig:network simplification}(a). Now the bundle as a whole acts merely as a single bridging chain, and all of the structures internal to the bundle can still relax. Only if the bundle is connected to the network by at least three independent paths may it contribute to the rigidity of the network.

The enumeration of the triply-connected network elements can be accomplished with off-the-shelf graph analysis tools. A graph must be constructed of the super-molecular network, using the $N_{bead}$ beads as nodes and $N_{connect}$ springs and associations as edges. Existing algorithms can identify the tri-connected components of such graphs in $O(N_{bead}+N_{connect})$ time. However, the pre-factor on this computational cost is large enough that systematically applying the algorithm to the full simulation data set is unnecessarily time-consuming. The process is sped up dramatically by first simplifying the network graph to only include branch points as nodes and the bridging chains between them as single edges. This process of simplifying the network graph is depicted in Fig.~\ref{fig:network simplification}(b). The algorithm used is detailed in Algortihm~\ref{alg:Elastic}. This algorithm considers the association matrix $M$ defined by $M_{ij}(t)=1$ if and only if sticker $i$ is bound to sticker $j$ at time t \cite{SingKatz2011}. The algorithm for enumerating elastically active chains then constructs a graph using $M$ as an adjacency matrix, then iteratively replaces branchless paths in this graph with single edges until only the simplest web of elastically active elements remains.

\begin{algorithm}
 \caption{\small{Indentifying Elastically Active chains}}
\label{alg:Elastic}
\begin{algorithmic}[1]
\State $M_{ij}(t_1,t_2) \gets 1$ if $(i,j)$ in pair list at both $t_1$ and $t_2$ or $(i,j)$ are connected by a spring
\State Graph $G$ is defined by adjacency matrix $M$
\State Graph $H$ is the largest connected component of $G$
\While{any nodes in $H$ have degree $<3$}
    \State replace all branchless paths in $H$ with a single edge
    \State remove duplicate edges from $H$
\EndWhile
\State Construct $S$, the SPQR tree of $H$
\State Graph $T$ is the largest node of $S$
\State Return (edges in $T$) - (nodes in $T$) + 1
\end{algorithmic}
\end{algorithm}

\begin{figure*}
    \includegraphics[width=\textwidth]{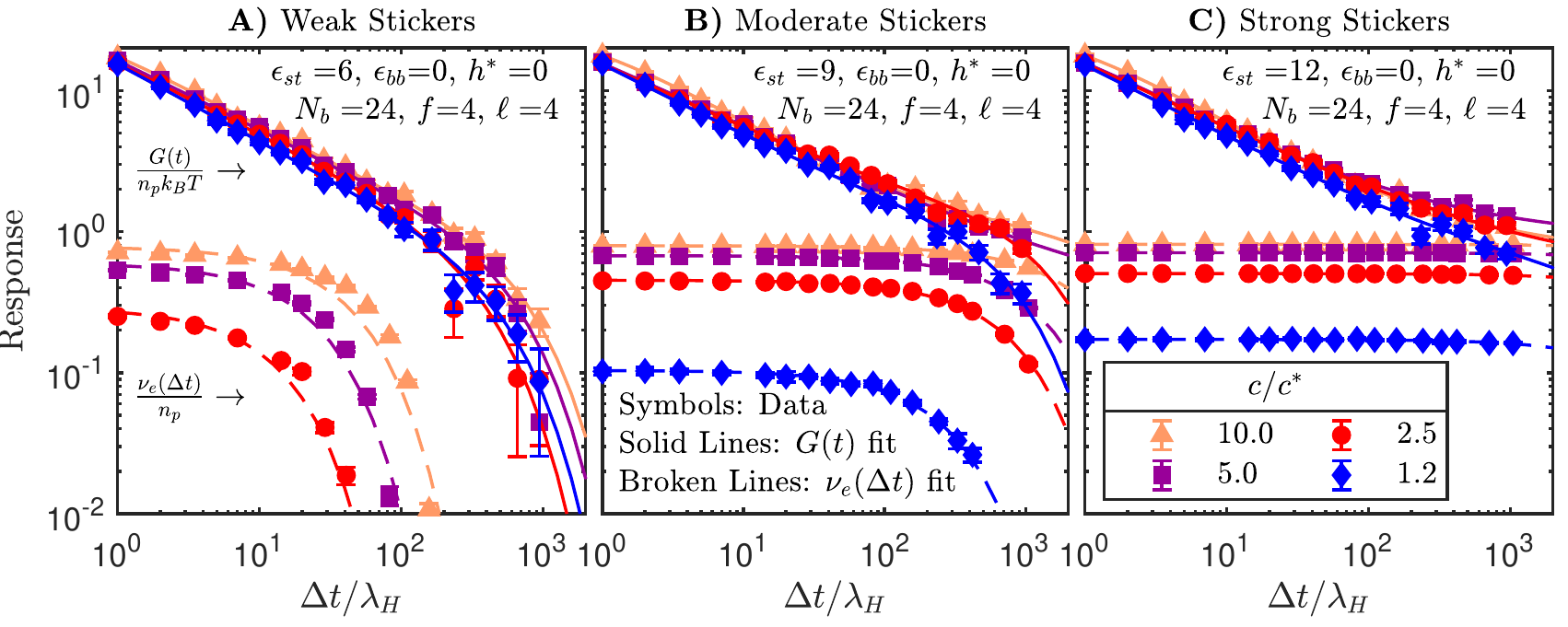}
     \caption{\small{Comparison of the rheologically measured $G(t)$ with the microscopically measured $\nu_e(\Delta t)$ for (a) \est=6 (b) \est=9 (c) \est=12. Weak stickers form an evanescent gel, which contains mechanically rigid structures that dissociate before their elastic response is observable. As \est{} is increased, the elastic plateau emerges from the single-chain relaxations. Solid lines through $G(t)$ data represent fits using Eqn.~\ref{eqn:MLphys}. Broken lines through $\nu_e(\Delta t)$ represent fits with a single exponential.}}
    \label{fig:elastic emergence}
     \vspace{-10pt}
\end{figure*}

Fig.~\ref{fig:network simplification}(a) depicts the single large graph H acquired in step 4. This graph contains several dangling ends and hierarchies of bridging chains. Fig.~\ref{fig:network simplification}(b) depicts the first application of step 5. All nodes and only the nodes which were junctions in the previous step have been kept. Sequences of beads between junctions have been replaced with a single graph edge. Duplicate bonds between junctions have been drawn for visual clarity, but the actual graph constructed in the algorithm only keeps one such edge. Therefore, upon successive applications of step 5, nested bridging chains are successively simplified until only a single edge remains, as show in Fig.~\ref{fig:network simplification}(c). However, there still remains the possibility that a tri-connected subgraph is only connected to the rest of the network by two edges. One such cluster is circled in Fig.~\ref{fig:network simplification}(c). Naively, one would have to test for such clusters by removing every possible pair of edges (an $O(N_{edge}^2)$ task), and check if the graph is still connected (an $O(N_{edge})$ task). However, an algorithm exists to analyze tri-connected components of a graph in linear time using a data structure called an SPQR tree. This algorithm was adapted into a function which receives the bi-connected graph in Fig.~\ref{fig:network simplification}(c) and returns the lists of nodes in each tri-connected component of the graph. As shown in Fig.~\ref{fig:network simplification}(d), the largest tri-connected component is retained, and any smaller components are replaced with a single edge. The circuit rank of this final graph is the number of elastically active chains in a simulation volume. Note that the actual simulation networks contain thousands of beads, and even the final network can contain dozens of nodes. Due to the periodic, three-dimensional nature of the simulation, the networks are in general non-planar and difficult to visually interpret.

\begin{figure*}
    \includegraphics[width=\textwidth]{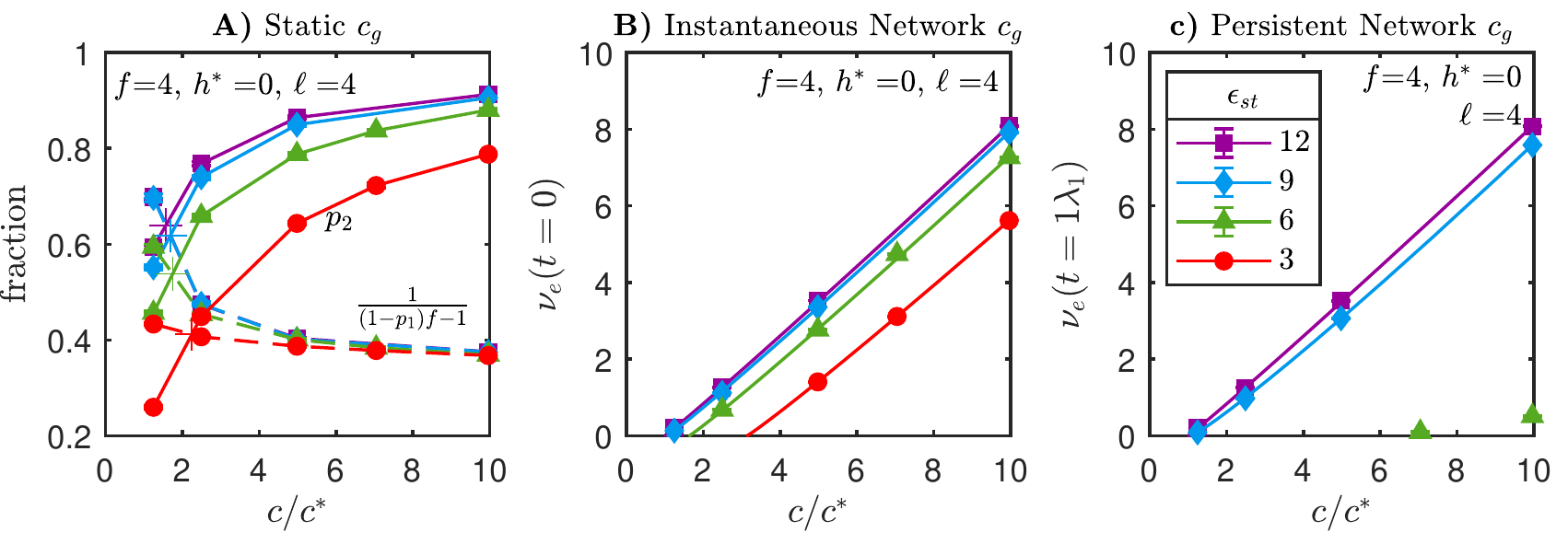}
     \caption{\small{Extraction of \cg{} from measurements by (a) finding the intersection (crosses) of the interchain conversion $p_2$ with the quantity $1/[(1-p_1)f-1]$, where $p_1$ is the fraction of intra-chain associations and $f$ is the number of stickers on a chain; or fitting with the form $\nu_e(c)=A(c/c_g-1)^z$ at (b) $t=0$, or at (c) $t=\lambda_1$ where $\lambda_1$ is the slowest Rouse mode for a 24 bead chain. After dissociations reduce $\nu_e$ to zero, fit is possible.}}
    \label{fig:nue vs c}
     \vspace{-10pt}
\end{figure*}

By applying this algorithm to simulation data and taking a time average over pairs of snapshots with the same $\Delta t=t_2-t_1$, a curve is constructed for the concentration of elastically active chains which have persisted for a length of time. The $\Delta t=0$ limit of this measurement is the equilibrium concentration of elastically active chains. A key realisation is that in order for an elastic element to exhibit a stress response, it must have maintained its connection to the network since the strain was applied. Fig.~\ref{fig:elastic emergence} shows the resulting curves for the concentration of persistent elastically active chains, $\nu_e(\Delta t)$, as well as $G(t)$. The immediate observation is that increasing polymer concentration increases the instantaneous concentration of elastically active chains, even at lower sticker strengths. But there is an ``association turnover'' with weak stickers, meaning that while a percolated network exists at any moment, the bonds within it are constantly dissociating, relaxing, and forming new associations in a relaxed configuration. This causes the elastic plateau observed for stronger stickers to recede to shorter times instead of dropping to lower heights. Fig.~\ref{fig:elastic emergence}(a) emphasizes the ``evanescent'' behavior, meaning that the elastic response of the network decays faster than the segmental relaxations of even single polymer chains, rendering it invisible, buried within the polymer relaxation spectrum.

Figs.~\ref{fig:nue vs c}~(b) and~(c) display the dependence of $\nu_e$ on concentration, sticker strength, and time scale. These measurements of the concentration of elastically active chains make it possible to estimate \cg{} as a function of \est{} by using $\nu_e$ as a proxy for $G_\epsilon$ and fitting $\nu_e(c)$ with a power law, $\nu_e(0)=A(c/c_g-1)^z$. Further, by considering the values of $\nu_e$ at different time scales, we can estimate the observed \cg{} if rheological measurements were limited to a particular time scale. The investigated time scales are chosen relative to the longest Rouse relaxation time for a 24 bead chain. The reasoning for this choice is that a process faster than this time scale will be difficult to distinguish from the polymer modes. In different complex fluids there might be other relaxation processes in play which might be competing with the dissociation mechanism. The points of intersection of the different curves in Figs.~\ref{fig:nue vs c}~(b) and~(c) with the $x$-axis (corresponding to  $\nu_e = 0$), represent the different estimates of \cg{}/$c^*$ in each case. For reference, a procedure for estimating \cg{}/$c^*$ from purely static measurements is also included in Figs.~\ref{fig:nue vs c}~(a), using the criterion adapted by Dobrynin~\cite{Dob} from the work of Flory and Stockmayer that the fraction of inter-chain associations $p_2$ at the gel point is $p_2=1/(1-p_1)f-1$, where $p_1$ is the fraction of intra-chain associations and $f$ is the number of attractive groups (stickers) on a chain. This criterion is applied because it is a method of extracting an estimate of \cg{} from static properties using very few concentration samples. 

Some notes must be made about the values of \cg{} extracted in Figs.~\ref{fig:nue vs c}~(b) and~(c).  The ability to extract a value of \cg{} is contingent on at least three different values of \cbyc{} with a non-zero value of $\nu_e$ in order to constrain the fit. With so few data points (due to limitations of computational resources), meaningful uncertainty estimates on these fit parameters are not available. The purpose of this exercise is to demonstrate that \cg{} may be identified systematically, rather than to resolve its value precisely. It is also notable that once the observation time scale passes beyond the dissociation time scale, $\nu_e$ drops to 0 and the gel network is not rheologically apparent (\est~=~3 and 6 in Fig.~\ref{fig:nue vs c}(c)).

\subsection{Competition of Time Scales}
\label{sec:Competition of Time Scales}

\begin{figure*}
    \includegraphics[width=\textwidth]{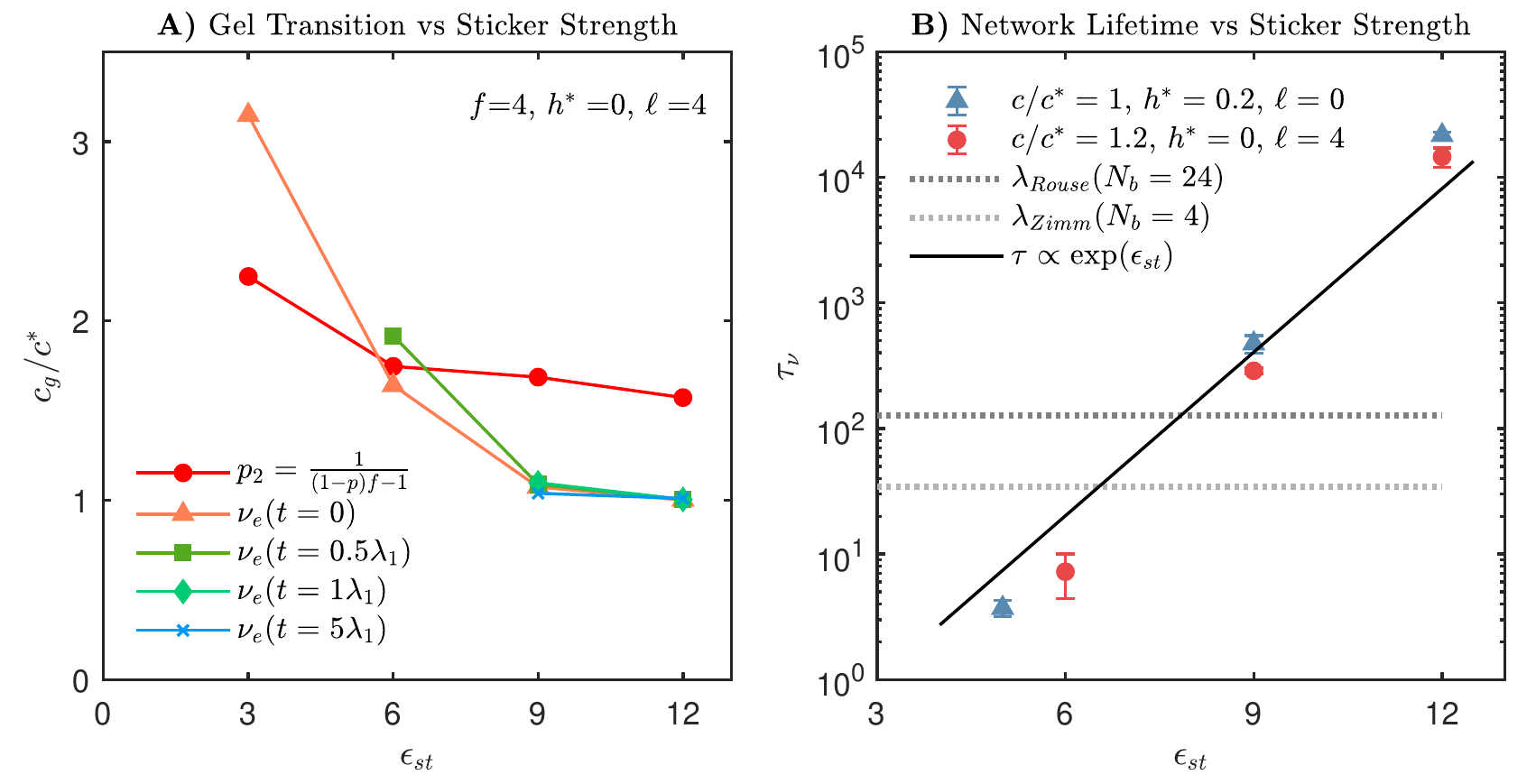}
         \vspace{-20pt}
     \caption{\small{(a) Dependence of \cg{} on \est. The curves compare between the apparent gel point when considering coarse static information (fractions of intra-chain $p_1$ and inter-chain $p_2$ associations), detailed instantaneous information ($\nu_e(t=0)$), or dynamic measurements at various time scales. (b) Dependence on \est{} of the time scale $\tau_\nu$ for the dissociation of elastically active chains. Horizontal lines indicate terminal relaxations predicted by Rouse and Zimm theory.}}
    \label{fig:cg vs epsilon}
     \vspace{-10pt}
\end{figure*}

Fig.~\ref{fig:cg vs epsilon}(a) collects the values of \cg{} as a function of \est{} extracted from static measurements and from $\nu_e$ data. The curves for different time scales suggest that, so long as an elastic network persists at all, the value of observed value of \cg{} is not significantly impacted by the observation time scale. Meanwhile, the absence of a \cg{} value for lower \est{} values, even when considering time scales much shorter than the single chain relaxation time, emphasizes the difficulty of rheologically detecting the presence of a network of weak stickers. Together, the curves in Fig.~\ref{fig:cg vs epsilon} show that, as one might intuitively expect, \cg{} generally decreases with \est{}, and that static or instantaneous measurements can detect a network at much lower sticker strength than dynamic measurements can, due to interference from other relaxation processes. Intuitively, even if the association strength was 0, a static analysis of the polymers would show apparent contact between beads labeled as ``stickers", which at high concentration would form a network. This is also evidenced by the associated cluster size distribution, available in the Supplementary Information, which exhibit a power law near the percolation threshold, and a bimodal distribution in the percolated state, even if the lifetime of the bonds is very sort. So, when stickers are very weak, a structural analysis of features such as cluster size or free chain concentration may not correlate precisely with the rheological signatures of gelation.

\begin{figure}[t]
    \includegraphics[width=\columnwidth]{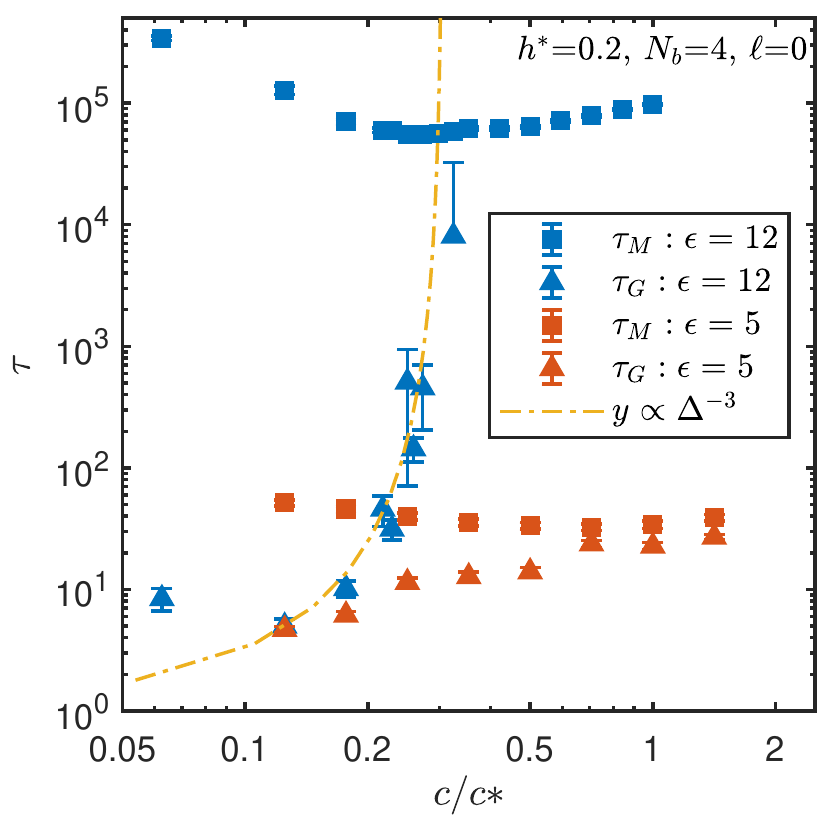}
     \caption{\small{The associative pair lifetime $\tau_M$ limits the growth of the longest observed rheological time scale $\tau_G$. When $\tau_M$ is short, the rapid growth of $\tau_G$ leading up to the gel transition is suppressed. A power law with exponent -3 represents scaling theory.}
    \label{fig:competition of time scales}}
     \vspace{-10pt}
\end{figure}

The crossover from the strong sticker regime with a clear rheological signature of gelation to the evanescent gel is understood by considering competing time scales.  Fig.~\ref{fig:cg vs epsilon}(b) shows the dependence of the network persistence time scale $\tau_\nu$ on \est. Two systems are represented with different chain length and hydrodynamic interactions, both showing a similar roughly exponential dependence on \est. The slightly super-exponential behavior is likely due to redundancies created in the elastic network when the concentraition of associated pairs is high, as the lifetimes of individual associated pairs don't show this exaggerated dependence on \est. The chain relaxation times $\lambda$ calculated using the Rouse and Zimm models are represented as horizontal lines to emphasize the ratio of these two time scales as \est{} is varied. The values of \est{} here correspond to those in the three panels of Fig~\ref{fig:elastic emergence}, demonstrating that the visibility of the elastic plateau is contingent on the mechanical rigidity of the network persisting longer than the single chain relaxation time. A similar comparison can be made between the longest observed relaxation time and the time scale for individual associations. Fig.~\ref{fig:competition of time scales} depicts the concentration dependence of two different relaxation time measurements for both strong and weak stickers. One time scale is the rheologically observed terminal relaxation time $\tau_G$. The other time scale is the renormalized bond lifetime $\tau_M$. This time scale is measured using the association matrix $M$. Specifically, just as $p_2$ is the fraction of stickers which are associated to a different chain, we define $M_2$ which includes only inter-chain associations. We then define an autocorrelation function
\begin{equation}
    C_M(\Delta t)=<M_2(t)\cdot M_2(t+\Delta t)>
\end{equation}
where $<A\cdot B>$ indicates an ensemble, time, and element-wise average of the element-wise product of $A$ and $B$. This autocorrelation function decays as a single exponential (plots are provided in the Supplementary Information), from which we extract the time scale $\tau_M$ using a single exponential fit. We exclude the intra-chain associations because there is a permanent probability for stickers on the same chain to re-associate, so the persistence of those bonds obscures the lifetime of the elastically relevant inter-chain bonds. $\tau_M$ therefore constitutes a robust measurement of intra-chain bond lifetime via the dissociation rate. Bond lifetimes for weak stickers could be measured by averaging the durations of individual association events, and such statistics are available in the Supplementary Information. However, the decay of the above autocorrelation captures the average lifetime even if it is longer than the observation window.

This association lifetime is compared to the longest relaxation time extracted from $G(t)$. Fig.~\ref{fig:competition of time scales} shows that the bond lifetimes for \est~=~12 are several orders of magnitude above the dilute terminal flow timescale. However, when \est~=~5, the bond lifetimes are unsurprisingly lower. The proximity of the bond lifetime to the single polymer relaxation time means that, even if there are instantaneously large associated clusters, the rapid growth of the cluster relaxation time is not observable because the clusters dissociate before that time scale is reached. With higher sticker strength, the divergence of the relaxation time for a cluster on approach to the gel transition is observable.

\subsection{Hyperscaling}

\begin{figure*}
    \includegraphics[width=\textwidth]{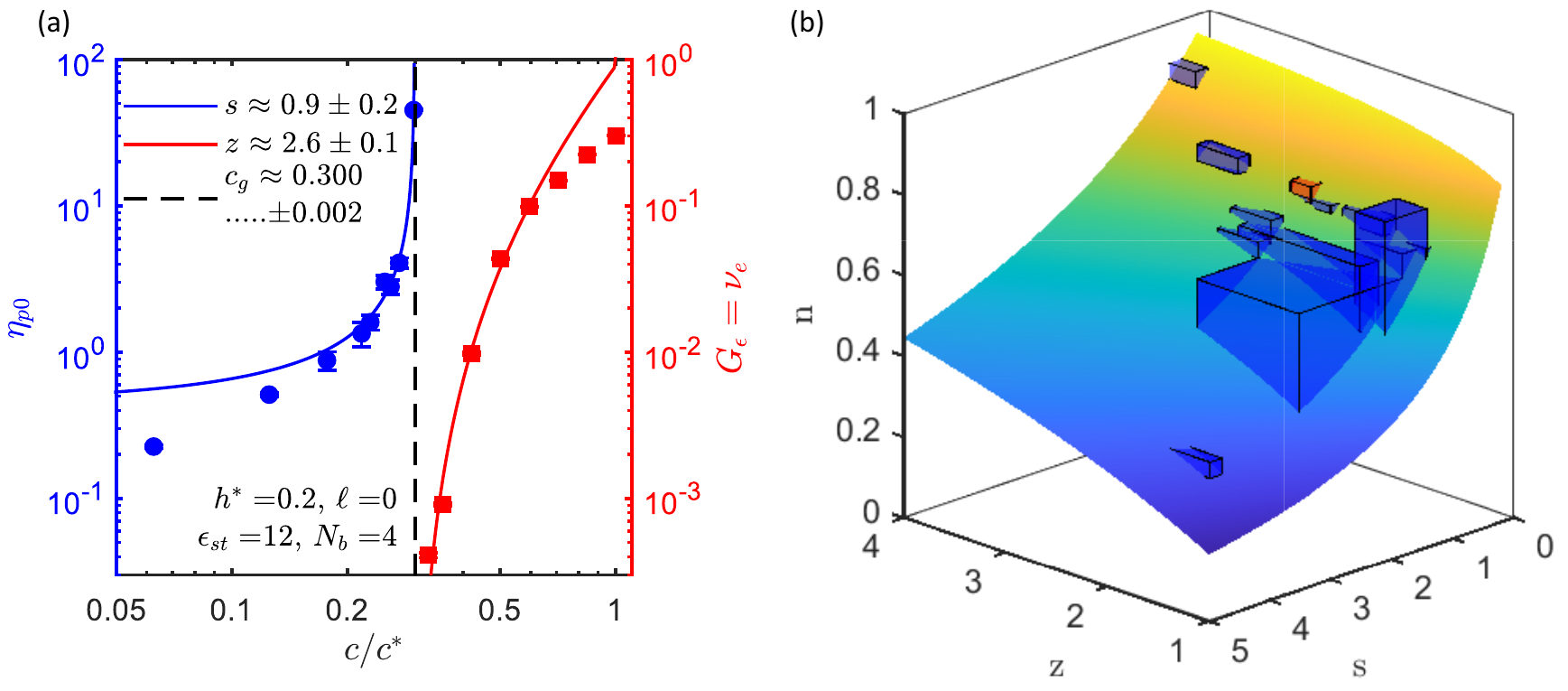}
     \caption{\small{(a) The polymer contribution to the zero-shear-rate viscosity $\eta_{p0}$ (blue circles) measured as the integral of $G(t)$ diverges on approach to \cg{} from below. Also shown is the gel elastic modulus $G_\epsilon$ (red squares) measured by equivalence to the concentration of elastically active chains $\nu_e$. (b) Comparison of the inter-relationship between different scaling exponents resolved in this work (red box), an array of experimental results for various gel-formers\cite{SumanJOR,doi:10.1021/ma00211a019,PhysRevLett.64.1457,PhysRevE.52.6271,doi:10.1021/ma034266c,cho2008dynamic,DAI20084012,doi:10.1021/acs.macromol.5b01922,TAN20085676,LU2006544,doi:10.1021/jp077685a} (blue boxes), and the hyperscaling relationship $n=z/(z+s)$ (surface). Box dimensions indicate confidence intervals.}}
    \label{fig:hyperscaling}
         \vspace{-10pt}
\end{figure*}

Precise determination of the scaling exponents $s$ and $z$ from the available data is challenging, because the scaling laws best describe only the data near the critical point. The power law tail on $G(t)$ is relatively straightforward to fit since we have resolved the tail over almost three decades in time, however because the scaling of $\eta_{p0}$ and $G_\epsilon$ are limited to the neighborhood of \cg, the exponents $s$ and $z$ are weakly determined by less than a decade of concentration values. There are also finite-size and run-time effects. On the sol side of the transition, capturing the divergence in the zero-shear-rate viscosity requires accounting for the longest relaxation mode, which is diverging. In simulations, viscosity calculation at the critical point could therefore easily be influenced by the finite duration of the simulation. Further, because we are modelling associations with finite lifetime (though approximately 100 times longer than our simulation duration on average at \est~=~12), the viscosity cannot truly diverge. Regarding the elastic modulus $G_\epsilon$ in the gel phase, near \cg, where the scaling is most accurate, the value of $G_\epsilon$ is infinitesimal, requiring many realizations to measure precisely. In addition, due to the finite system size, simulations will tend to over-count the number of apparently system-spanning elastic elements. Finally, as with the viscosity, dissociations will reduce the apparent elastic modulus of the gel, particularly near the critical point where relatively few dissociations can rupture the network.

To mitigate these challenges, we compare our results in Fig.~\ref{fig:hyperscaling} with the hyperscaling relationship $n = z/(z+s)$~\cite{SumanJOR} by performing simultaneous regression of $\eta_{p0}(c)$, $G_\epsilon(c)$, and $G(t)$ at \cbyc$=0.30$ to equations (2),(3), and (7) respectively, with the constraints that $n(z+s)=z$, and in the $G(t)$ fit that $G_\epsilon=0$ and $\alpha=-n$. Note that our $G(t)$ data exhibit a plateau at short $t$ due to the fastest relaxation mode of bead-spring chains, so a pure power law model~\cite{10.1063/5.0038830} such as equation (1) yields a poor fit and a distorted estimation of $n$ (fitting procedures are detailed in the Supplementary Information). This fact motivated our adoption of the Mittag-Leffler form to capture both the power law tail and the exponential head. In addition, to measure the elastic modulus in the gel phase, we have considered the concentration of elastically active chains as well as the long-time plateau in $G(t)$. This measurement agrees with the elastic modulus extracted from $G(t)$ at high concentration, while also revealing the modulus just above \cg, which is not distinguishable from the noise in $G(t)$ present at that concentration, as illustrated in Fig.~9. After all of these considerations, our constrained fit demonstrates agreement with the hyperscaling relationship $n = z/(z+s)$. At $c_g/c^*\approx 0.30$, (green circles in Fig.~3, the power law exponent for the tail is $n\approx0.73$. The power law fits to $\eta_{p0}$, $G_\epsilon$ in Fig.~13(a) show exponents $s\approx0.90$ and $z\approx2.6$. Additionally, another scaling exponent $\kappa=n/z\approx0.28$ (related to the shear modulus and crosslinking density $p$ as $[\partial\ln{G(t)}/\partial p]_{p=p_g}\sim t^\kappa$)\cite{Scanlan1991} is within the range 0.2--0.3 observed in experiment\cite{SumanJOR,Scanlan1991,doi:10.1021/acs.langmuir.7b00151,10.1122/1.4883675}. An illustrative comparison of experimental results\cite{doi:10.1021/ma00211a019,PhysRevLett.64.1457,PhysRevE.52.6271,doi:10.1021/ma034266c,cho2008dynamic,DAI20084012,doi:10.1021/acs.macromol.5b01922,TAN20085676,LU2006544,doi:10.1021/jp077685a} for the scaling exponents compiled by Suman and Joshi\cite{SumanJOR} is shown in Fig.~13(b).

\section{Conclusion}
\label{sec:Conclusion}

Dynamic signatures of gelation in associative polymer solutions have been investigated using Brownian dynamics simulations. Dynamic and linear viscoelastic properties like zero-shear-rate viscosity, storage ($G'$) and loss ($G''$) modulii, large scale relaxation time and bond lifetime are used to characterise the viscoelasticity of the associative polymer solutions in the pre and post-gel regimes.  Results obtained for the scaling of the zero-shear-rate viscosity and terminal relaxation time are compared with the scaling prediction of the mean-field theory for associative polymers when the energy barrier against dissociation is relatively low. For stronger stickers, this scaling prediction may still be followed, but the relaxation times in question are outside of our observational window. The concentration corresponding to the gelation crossover for \est~=~5, coincides with the maxima in free chain concentration, which is one of the static measures of gelation. For \est~=~12, several conventional rheological signatures of gelation were observed, including a diverging instantaneous viscosity, a growing plateau in the elastic modulus, power laws in $G(t), G'$, and $G''$, and the frequency independence of $\tan \delta$. The scaling exponents $n,s,z$, and $\kappa$ relating the viscosity, plateau modulus, and derivatives of $G$ at the gel point were found to be consistent with the hyperscaling relationships $n =z/(z+s)$ and $\kappa=n/z$.

In order to study the effect of HI on the dynamic properties in the pre-gel and post-gel regime we have compared the results from simulations carried out with and without HI. A key finding of this study is that the relaxation time as a function of the fraction of inter-chain associated stickers per chain follows the scaling prediction by the sticky Rouse model~\cite{RnSdynamics,RnS2001}, in the post gel regime. Moreover, the same asymptotic scaling is also obtained at the limit of high concentration in the post gel regime from simulations without HI, indicating the screening of hydrodynamic interaction due to formation of a dense network.

It has been verified by analyzing the structure and dynamics of the associated network directly that the viscoelastic response of physical gels at long times corresponds to the persistence of load-bearing structures within the associative network. By monitoring the status of the network in addition to the whole solution's stress response, we have shown that the network's response can be overwhelmed by the distribution of relaxation modes inherent in polymeric systems. This situation leads to a competition of time scales between the lifetime of the network and the terminal relaxation time of the unassociated polymer. If the association time scale is much greater, then the classical signatures of gelation established with chemical gels are present, at least approximately. If the association time scale is near or less than the single polymer's, then only the concentration scaling of terminal flow behavior can reveal the evanescent gel beneath the other relaxations.

This study was confined to very short chains, with $f=4$ stickers per chain and $\ell=0$ or 4 spacers between stickers.  Future studies should consider the effects of $f$ and $\ell$ on the gel transition systematically. An important consideration in modern applications is the nonlinear rheology of complex fluids. Associative polymers are known to have unique responses to high flow rates because the reversible bonds can be torn apart, yet recover quickly when the fluid is at rest. With this study's simulation framework in place, future studies should investigate the nonlinear behavior of these associative polymers, and whether the ``evanescent'' gel exhibits a more pronounced response in different flow conditions.

\section*{Acknowledgements}
This research was supported under Australian Research Council's Discovery Projects
funding scheme (project number DP190101825). It was undertaken with the assistance
of resources from the National Computational Infrastructure (NCI Australia), an NCRIS
enabled capability supported by the Australian Government. This work was also employed computational facilities at Monash University, the MASSIVE and MonARCH systems.

\bibliography{evanescent_gels}

\providecommand{\latin}[1]{#1}
\makeatletter
\providecommand{\doi}
  {\begingroup\let\do\@makeother\dospecials
  \catcode`\{=1 \catcode`\}=2 \doi@aux}
\providecommand{\doi@aux}[1]{\endgroup\texttt{#1}}
\makeatother
\providecommand*\mcitethebibliography{\thebibliography}
\csname @ifundefined\endcsname{endmcitethebibliography}
  {\let\endmcitethebibliography\endthebibliography}{}
\begin{mcitethebibliography}{81}
\providecommand*\natexlab[1]{#1}
\providecommand*\mciteSetBstSublistMode[1]{}
\providecommand*\mciteSetBstMaxWidthForm[2]{}
\providecommand*\mciteBstWouldAddEndPuncttrue
  {\def\EndOfBibitem{\unskip.}}
\providecommand*\mciteBstWouldAddEndPunctfalse
  {\let\EndOfBibitem\relax}
\providecommand*\mciteSetBstMidEndSepPunct[3]{}
\providecommand*\mciteSetBstSublistLabelBeginEnd[3]{}
\providecommand*\EndOfBibitem{}
\mciteSetBstSublistMode{f}
\mciteSetBstMaxWidthForm{subitem}{(\alph{mcitesubitemcount})}
\mciteSetBstSublistLabelBeginEnd
  {\mcitemaxwidthsubitemform\space}
  {\relax}
  {\relax}

\bibitem[Rubinstein and Colby(2003)Rubinstein, and Colby]{RubColby2003}
Rubinstein,~M.; Colby,~R.~H. \emph{Polymer Physics}; Oxford University Press,
  Oxford, 2003\relax
\mciteBstWouldAddEndPuncttrue
\mciteSetBstMidEndSepPunct{\mcitedefaultmidpunct}
{\mcitedefaultendpunct}{\mcitedefaultseppunct}\relax
\EndOfBibitem
\bibitem[Rubinstein and Dobrynin(1997)Rubinstein, and Dobrynin]{RubDob97}
Rubinstein,~M.; Dobrynin,~A. Solutions of Associative Polymers. \emph{Trends in
  Polymer Science} \textbf{1997}, \emph{5}, 181--186\relax
\mciteBstWouldAddEndPuncttrue
\mciteSetBstMidEndSepPunct{\mcitedefaultmidpunct}
{\mcitedefaultendpunct}{\mcitedefaultseppunct}\relax
\EndOfBibitem
\bibitem[Martin and Adolf(1991)Martin, and Adolf]{Martin1991-od}
Martin,~J.~E.; Adolf,~D. The sol-gel transition in chemical gels. \emph{Annu.
  Rev. Phys. Chem.} \textbf{1991}, \emph{42}, 311--339\relax
\mciteBstWouldAddEndPuncttrue
\mciteSetBstMidEndSepPunct{\mcitedefaultmidpunct}
{\mcitedefaultendpunct}{\mcitedefaultseppunct}\relax
\EndOfBibitem
\bibitem[Flory(1974)]{Flory1974-tm}
Flory,~P.~J. Introductory lecture. \emph{Faraday Discuss. Chem. Soc.}
  \textbf{1974}, \emph{57}, 7\relax
\mciteBstWouldAddEndPuncttrue
\mciteSetBstMidEndSepPunct{\mcitedefaultmidpunct}
{\mcitedefaultendpunct}{\mcitedefaultseppunct}\relax
\EndOfBibitem
\bibitem[Douglas(2018)]{Douglas2018-xh}
Douglas,~J.~F. Weak and strong gels and the emergence of the amorphous solid
  state. \emph{Gels} \textbf{2018}, \emph{4}, 19\relax
\mciteBstWouldAddEndPuncttrue
\mciteSetBstMidEndSepPunct{\mcitedefaultmidpunct}
{\mcitedefaultendpunct}{\mcitedefaultseppunct}\relax
\EndOfBibitem
\bibitem[Winter and Chambon(1986)Winter, and Chambon]{Winter:2000gw}
Winter,~H.~H.; Chambon,~F. {Analysis of Linear Viscoelasticity of a
  Crosslinking Polymer at the Gel Point}. \emph{J. Rheol.} \textbf{1986},
  \emph{30}, 367--382\relax
\mciteBstWouldAddEndPuncttrue
\mciteSetBstMidEndSepPunct{\mcitedefaultmidpunct}
{\mcitedefaultendpunct}{\mcitedefaultseppunct}\relax
\EndOfBibitem
\bibitem[Nicolai \latin{et~al.}(1997)Nicolai, Randrianantoandro, Prochazka, and
  Durand]{Nicolai1997-su}
Nicolai,~T.; Randrianantoandro,~H.; Prochazka,~F.; Durand,~D. Viscoelastic
  relaxation of polyurethane at different stages of the gel formation. 2.
  {Sol-Gel} transition dynamics. \emph{Macromolecules} \textbf{1997},
  \emph{30}, 5897--5904\relax
\mciteBstWouldAddEndPuncttrue
\mciteSetBstMidEndSepPunct{\mcitedefaultmidpunct}
{\mcitedefaultendpunct}{\mcitedefaultseppunct}\relax
\EndOfBibitem
\bibitem[Parada and Zhao(2018)Parada, and Zhao]{Parada2018-fx}
Parada,~G.~A.; Zhao,~X. Ideal reversible polymer networks. \emph{Soft Matter}
  \textbf{2018}, \emph{14}, 5186--5196\relax
\mciteBstWouldAddEndPuncttrue
\mciteSetBstMidEndSepPunct{\mcitedefaultmidpunct}
{\mcitedefaultendpunct}{\mcitedefaultseppunct}\relax
\EndOfBibitem
\bibitem[Chassenieux \latin{et~al.}(2011)Chassenieux, Nicolai, and
  Benyahia]{Chassenieux2011-fv}
Chassenieux,~C.; Nicolai,~T.; Benyahia,~L. Rheology of associative polymer
  solutions. \emph{Curr. Opin. Colloid Interface Sci.} \textbf{2011},
  \emph{16}, 18--26\relax
\mciteBstWouldAddEndPuncttrue
\mciteSetBstMidEndSepPunct{\mcitedefaultmidpunct}
{\mcitedefaultendpunct}{\mcitedefaultseppunct}\relax
\EndOfBibitem
\bibitem[Yan \latin{et~al.}(2004)Yan, Springsteen, Deeter, and
  Wang]{Yan2004-zl}
Yan,~J.; Springsteen,~G.; Deeter,~S.; Wang,~B. The relationship among pKa, pH,
  and binding constants in the interactions between boronic acids and
  diols---it is not as simple as it appears. \emph{Tetrahedron} \textbf{2004},
  \emph{60}, 11205--11209\relax
\mciteBstWouldAddEndPuncttrue
\mciteSetBstMidEndSepPunct{\mcitedefaultmidpunct}
{\mcitedefaultendpunct}{\mcitedefaultseppunct}\relax
\EndOfBibitem
\bibitem[Furikado \latin{et~al.}(2014)Furikado, Nagahata, Okamoto, Sugaya,
  Iwatsuki, Inamo, Takagi, Odani, and Ishihara]{Furikado2014-vr}
Furikado,~Y.; Nagahata,~T.; Okamoto,~T.; Sugaya,~T.; Iwatsuki,~S.; Inamo,~M.;
  Takagi,~H.~D.; Odani,~A.; Ishihara,~K. Universal reaction mechanism of
  boronic acids with diols in aqueous solution: kinetics and the basic concept
  of a conditional formation constant. \emph{Chemistry} \textbf{2014},
  \emph{20}, 13194--13202\relax
\mciteBstWouldAddEndPuncttrue
\mciteSetBstMidEndSepPunct{\mcitedefaultmidpunct}
{\mcitedefaultendpunct}{\mcitedefaultseppunct}\relax
\EndOfBibitem
\bibitem[Webber \latin{et~al.}(2016)Webber, Appel, Meijer, and
  Langer]{Webber2016-hj}
Webber,~M.~J.; Appel,~E.~A.; Meijer,~E.~W.; Langer,~R. Supramolecular
  biomaterials. \emph{Nat. Mater.} \textbf{2016}, \emph{15}, 13--26\relax
\mciteBstWouldAddEndPuncttrue
\mciteSetBstMidEndSepPunct{\mcitedefaultmidpunct}
{\mcitedefaultendpunct}{\mcitedefaultseppunct}\relax
\EndOfBibitem
\bibitem[Kloxin and Bowman(2013)Kloxin, and Bowman]{Kloxin2013-ci}
Kloxin,~C.~J.; Bowman,~C.~N. Covalent adaptable networks: smart, reconfigurable
  and responsive network systems. \emph{Chem. Soc. Rev.} \textbf{2013},
  \emph{42}, 7161--7173\relax
\mciteBstWouldAddEndPuncttrue
\mciteSetBstMidEndSepPunct{\mcitedefaultmidpunct}
{\mcitedefaultendpunct}{\mcitedefaultseppunct}\relax
\EndOfBibitem
\bibitem[Wang and Heilshorn(2015)Wang, and Heilshorn]{Wang2015-kv}
Wang,~H.; Heilshorn,~S.~C. Adaptable hydrogel networks with reversible linkages
  for tissue engineering. \emph{Adv. Mater.} \textbf{2015}, \emph{27},
  3717--3736\relax
\mciteBstWouldAddEndPuncttrue
\mciteSetBstMidEndSepPunct{\mcitedefaultmidpunct}
{\mcitedefaultendpunct}{\mcitedefaultseppunct}\relax
\EndOfBibitem
\bibitem[Voorhaar and Hoogenboom(2016)Voorhaar, and
  Hoogenboom]{Voorhaar:2016jt}
Voorhaar,~L.; Hoogenboom,~R. {Supramolecular polymer networks: hydrogels and
  bulk materials}. \emph{Chem. Soc. Rev.} \textbf{2016}, \emph{45},
  4013--4031\relax
\mciteBstWouldAddEndPuncttrue
\mciteSetBstMidEndSepPunct{\mcitedefaultmidpunct}
{\mcitedefaultendpunct}{\mcitedefaultseppunct}\relax
\EndOfBibitem
\bibitem[Zhang \latin{et~al.}(2018)Zhang, Chen, and Colby]{Zhang2018-qb}
Zhang,~Z.; Chen,~Q.; Colby,~R.~H. Dynamics of associative polymers. \emph{Soft
  Matter} \textbf{2018}, \emph{14}, 2961--2977\relax
\mciteBstWouldAddEndPuncttrue
\mciteSetBstMidEndSepPunct{\mcitedefaultmidpunct}
{\mcitedefaultendpunct}{\mcitedefaultseppunct}\relax
\EndOfBibitem
\bibitem[Zhang \latin{et~al.}(2017)Zhang, Huang, Weiss, and Chen]{Zhang2017-gs}
Zhang,~Z.; Huang,~C.; Weiss,~R.~A.; Chen,~Q. Association energy in strongly
  associative polymers. \emph{J. Rheol.} \textbf{2017}, \emph{61},
  1199--1207\relax
\mciteBstWouldAddEndPuncttrue
\mciteSetBstMidEndSepPunct{\mcitedefaultmidpunct}
{\mcitedefaultendpunct}{\mcitedefaultseppunct}\relax
\EndOfBibitem
\bibitem[Stukalin \latin{et~al.}(2013)Stukalin, Cai, Kumar, Leibler, and
  Rubinstein]{Stukalin2013-hu}
Stukalin,~E.~B.; Cai,~L.-H.; Kumar,~N.~A.; Leibler,~L.; Rubinstein,~M.
  Self-healing of unentangled polymer networks with reversible bonds.
  \emph{Macromolecules} \textbf{2013}, \emph{46}, 7525--7541\relax
\mciteBstWouldAddEndPuncttrue
\mciteSetBstMidEndSepPunct{\mcitedefaultmidpunct}
{\mcitedefaultendpunct}{\mcitedefaultseppunct}\relax
\EndOfBibitem
\bibitem[Herbst \latin{et~al.}(2013)Herbst, Döhler, Michael, and
  Binder]{Herbst2013-gi}
Herbst,~F.; Döhler,~D.; Michael,~P.; Binder,~W.~H. Self-Healing Polymers via
  Supramolecular Forces. \emph{Macromol. Rapid Commun.} \textbf{2013},
  \emph{34}, 203--220\relax
\mciteBstWouldAddEndPuncttrue
\mciteSetBstMidEndSepPunct{\mcitedefaultmidpunct}
{\mcitedefaultendpunct}{\mcitedefaultseppunct}\relax
\EndOfBibitem
\bibitem[Chaudhuri \latin{et~al.}(2016)Chaudhuri, Gu, Klumpers, Darnell,
  Bencherif, Weaver, Huebsch, Lee, Lippens, Duda, and Mooney]{Chaudhuri2016-lf}
Chaudhuri,~O.; Gu,~L.; Klumpers,~D.; Darnell,~M.; Bencherif,~S.~A.;
  Weaver,~J.~C.; Huebsch,~N.; Lee,~H.-P.; Lippens,~E.; Duda,~G.~N.;
  Mooney,~D.~J. Hydrogels with tunable stress relaxation regulate stem cell
  fate and activity. \emph{Nat. Mater.} \textbf{2016}, \emph{15},
  326--334\relax
\mciteBstWouldAddEndPuncttrue
\mciteSetBstMidEndSepPunct{\mcitedefaultmidpunct}
{\mcitedefaultendpunct}{\mcitedefaultseppunct}\relax
\EndOfBibitem
\bibitem[Holten-Andersen \latin{et~al.}(2011)Holten-Andersen, Harrington,
  Birkedal, Lee, Messersmith, Lee, and Waite]{Holten-Andersen2011-ke}
Holten-Andersen,~N.; Harrington,~M.~J.; Birkedal,~H.; Lee,~B.~P.;
  Messersmith,~P.~B.; Lee,~K. Y.~C.; Waite,~J.~H. pH-induced metal-ligand
  cross-links inspired by mussel yield self-healing polymer networks with
  near-covalent elastic moduli. \emph{Proc. Natl. Acad. Sci. U. S. A.}
  \textbf{2011}, \emph{108}, 2651--2655\relax
\mciteBstWouldAddEndPuncttrue
\mciteSetBstMidEndSepPunct{\mcitedefaultmidpunct}
{\mcitedefaultendpunct}{\mcitedefaultseppunct}\relax
\EndOfBibitem
\bibitem[Graham(1987)]{Graham1987-mp}
Graham,~N.~B. In \emph{Hydrogels in Medicine and Pharmacy}; Peppas,~N.~A., Ed.;
  CRC Press: Boca Raton, FL, 1987; Vol.~2\relax
\mciteBstWouldAddEndPuncttrue
\mciteSetBstMidEndSepPunct{\mcitedefaultmidpunct}
{\mcitedefaultendpunct}{\mcitedefaultseppunct}\relax
\EndOfBibitem
\bibitem[Fullenkamp \latin{et~al.}(2013)Fullenkamp, He, Barrett, Burghardt, and
  Messersmith]{Fullenkamp2013-da}
Fullenkamp,~D.~E.; He,~L.; Barrett,~D.~G.; Burghardt,~W.~R.; Messersmith,~P.~B.
  Mussel-inspired histidine-based transient network metal coordination
  hydrogels. \emph{Macromolecules} \textbf{2013}, \emph{46}, 1167--1174\relax
\mciteBstWouldAddEndPuncttrue
\mciteSetBstMidEndSepPunct{\mcitedefaultmidpunct}
{\mcitedefaultendpunct}{\mcitedefaultseppunct}\relax
\EndOfBibitem
\bibitem[Tang \latin{et~al.}(2018)Tang, Ma, Tu, Wang, Lin, and
  Anseth]{Tang2018-vm}
Tang,~S.; Ma,~H.; Tu,~H.-C.; Wang,~H.-R.; Lin,~P.-C.; Anseth,~K.~S. Adaptable
  Fast Relaxing Boronate-Based Hydrogels for Probing Cell–Matrix
  Interactions. \emph{Adv. Sci.} \textbf{2018}, \emph{5}, 1800638\relax
\mciteBstWouldAddEndPuncttrue
\mciteSetBstMidEndSepPunct{\mcitedefaultmidpunct}
{\mcitedefaultendpunct}{\mcitedefaultseppunct}\relax
\EndOfBibitem
\bibitem[Lee \latin{et~al.}(2010)Lee, Chung, Yeo, Ahn, Lee, Messersmith, and
  Park]{Lee2010-yl}
Lee,~Y.; Chung,~H.~J.; Yeo,~S.; Ahn,~C.-H.; Lee,~H.; Messersmith,~P.~B.;
  Park,~T.~G. Thermo-sensitive, injectable, and tissue adhesive sol--gel
  transition hyaluronic acid/pluronic composite hydrogels prepared from
  bio-inspired catechol-thiol reaction. \emph{Soft Matter} \textbf{2010},
  \emph{6}, 977\relax
\mciteBstWouldAddEndPuncttrue
\mciteSetBstMidEndSepPunct{\mcitedefaultmidpunct}
{\mcitedefaultendpunct}{\mcitedefaultseppunct}\relax
\EndOfBibitem
\bibitem[Winter(2016)]{WinterGel}
Winter,~H.~H. \emph{Encyclopedia of Polymer Science and Technology}; American
  Cancer Society, 2016; pp 1--15\relax
\mciteBstWouldAddEndPuncttrue
\mciteSetBstMidEndSepPunct{\mcitedefaultmidpunct}
{\mcitedefaultendpunct}{\mcitedefaultseppunct}\relax
\EndOfBibitem
\bibitem[Bromberg(1998)]{Bromberg}
Bromberg,~L. Scaling of Rheological Properties of Hydrogels from Associating
  Polymers. \emph{Macromolecules} \textbf{1998}, \emph{31}, 6148--6156\relax
\mciteBstWouldAddEndPuncttrue
\mciteSetBstMidEndSepPunct{\mcitedefaultmidpunct}
{\mcitedefaultendpunct}{\mcitedefaultseppunct}\relax
\EndOfBibitem
\bibitem[Thomas \latin{et~al.}(2009)Thomas, DePuit, and Khomami]{Dennis09}
Thomas,~D.~G.; DePuit,~R.~J.; Khomami,~B. Dynamic simulations of individual
  macromolecules in oscillatory shear flow. \emph{J. Rheol.} \textbf{2009},
  \emph{53}, 275--291\relax
\mciteBstWouldAddEndPuncttrue
\mciteSetBstMidEndSepPunct{\mcitedefaultmidpunct}
{\mcitedefaultendpunct}{\mcitedefaultseppunct}\relax
\EndOfBibitem
\bibitem[Wilson and Baljon(2017)Wilson, and Baljon]{polym9110556}
Wilson,~M.~A.; Baljon,~A. R.~C. Microstructural Origins of Nonlinear Response
  in Associating Polymers under Oscillatory Shear. \emph{Polymers}
  \textbf{2017}, \emph{9}\relax
\mciteBstWouldAddEndPuncttrue
\mciteSetBstMidEndSepPunct{\mcitedefaultmidpunct}
{\mcitedefaultendpunct}{\mcitedefaultseppunct}\relax
\EndOfBibitem
\bibitem[Brassinne \latin{et~al.}(2017)Brassinne, Cadix, Wilson, and van
  Ruymbeke]{Ruyumbeke2017}
Brassinne,~J.; Cadix,~A.; Wilson,~J.; van Ruymbeke,~E. Dissociating sticker
  dynamics from chain relaxation in supramolecular polymer networks—The
  importance of free partner! \emph{J. Rheol.} \textbf{2017}, \emph{61},
  1123--1134\relax
\mciteBstWouldAddEndPuncttrue
\mciteSetBstMidEndSepPunct{\mcitedefaultmidpunct}
{\mcitedefaultendpunct}{\mcitedefaultseppunct}\relax
\EndOfBibitem
\bibitem[Suman and Joshi(2020)Suman, and Joshi]{SumanJOR}
Suman,~K.; Joshi,~Y.~M. On the universality of the scaling relations during
  sol-gel transition. \emph{J. Rheol.} \textbf{2020}, \emph{64}, 863--877\relax
\mciteBstWouldAddEndPuncttrue
\mciteSetBstMidEndSepPunct{\mcitedefaultmidpunct}
{\mcitedefaultendpunct}{\mcitedefaultseppunct}\relax
\EndOfBibitem
\bibitem[Ozaki \latin{et~al.}(2017)Ozaki, Narita, Koga, and Indei]{Indei17}
Ozaki,~H.; Narita,~T.; Koga,~T.; Indei,~T. Theoretical Analysis of Critical
  Flowable Physical Gel Cross-Linked by Metal Ions and
  Polyacrylamide-Derivative Associating Polymers Containing Imidazole Groups.
  \emph{Polymers} \textbf{2017}, \emph{9}\relax
\mciteBstWouldAddEndPuncttrue
\mciteSetBstMidEndSepPunct{\mcitedefaultmidpunct}
{\mcitedefaultendpunct}{\mcitedefaultseppunct}\relax
\EndOfBibitem
\bibitem[Mewis \latin{et~al.}(2001)Mewis, Kaffashi, Vermant, and
  Butera]{Mewis2001}
Mewis,~J.; Kaffashi,~B.; Vermant,~J.; Butera,~R.~J. Determining Relaxation
  Modes in Flowing Associative Polymers Using Superposition Flows.
  \emph{Macromolecules} \textbf{2001}, \emph{34}, 1376--1383\relax
\mciteBstWouldAddEndPuncttrue
\mciteSetBstMidEndSepPunct{\mcitedefaultmidpunct}
{\mcitedefaultendpunct}{\mcitedefaultseppunct}\relax
\EndOfBibitem
\bibitem[Holten-Andersen \latin{et~al.}(2014)Holten-Andersen, Jaishankar,
  Harrington, Fullenkamp, DiMarco, He, McKinley, Messersmith, and
  Lee]{Andersen2014}
Holten-Andersen,~N.; Jaishankar,~A.; Harrington,~M.~J.; Fullenkamp,~D.~E.;
  DiMarco,~G.; He,~L.; McKinley,~G.~H.; Messersmith,~P.~B.; Lee,~K. Y.~C.
  Metal-coordination: using one of nature{'}s tricks to control soft material
  mechanics. \emph{J. Mater. Chem. B} \textbf{2014}, \emph{2}, 2467--2472\relax
\mciteBstWouldAddEndPuncttrue
\mciteSetBstMidEndSepPunct{\mcitedefaultmidpunct}
{\mcitedefaultendpunct}{\mcitedefaultseppunct}\relax
\EndOfBibitem
\bibitem[Wagner and McKinley(2017)Wagner, and McKinley]{Wagner2017}
Wagner,~C.~E.; McKinley,~G.~H. Age-dependent capillary thinning dynamics of
  physically-associated salivary mucin networks. \emph{J. Rheol.}
  \textbf{2017}, \emph{61}, 1309--1326\relax
\mciteBstWouldAddEndPuncttrue
\mciteSetBstMidEndSepPunct{\mcitedefaultmidpunct}
{\mcitedefaultendpunct}{\mcitedefaultseppunct}\relax
\EndOfBibitem
\bibitem[Rubinstein and Semenov(1998)Rubinstein, and Semenov]{RnSdynamics}
Rubinstein,~M.; Semenov,~A.~N. Thermoreversible Gelation in Solutions of
  Associative Polymers. 2. Linear Dynamics. \emph{Macromolecules}
  \textbf{1998}, \emph{31}, 1386--1397\relax
\mciteBstWouldAddEndPuncttrue
\mciteSetBstMidEndSepPunct{\mcitedefaultmidpunct}
{\mcitedefaultendpunct}{\mcitedefaultseppunct}\relax
\EndOfBibitem
\bibitem[Rubinstein and Semenov(2001)Rubinstein, and Semenov]{RnS2001}
Rubinstein,~M.; Semenov,~A.~N. Dynamics of Entangled Solutions of Associative
  Polymers. \emph{Macromolecules} \textbf{2001}, \emph{34}, 1058--1068\relax
\mciteBstWouldAddEndPuncttrue
\mciteSetBstMidEndSepPunct{\mcitedefaultmidpunct}
{\mcitedefaultendpunct}{\mcitedefaultseppunct}\relax
\EndOfBibitem
\bibitem[Stauffer \latin{et~al.}(1982)Stauffer, Coniglio, and
  Adam]{10.1007/3-540-11471-8_4}
Stauffer,~D.; Coniglio,~A.; Adam,~M. Gelation and critical phenomena. Polymer
  Networks. Berlin, Heidelberg, 1982; pp 103--158\relax
\mciteBstWouldAddEndPuncttrue
\mciteSetBstMidEndSepPunct{\mcitedefaultmidpunct}
{\mcitedefaultendpunct}{\mcitedefaultseppunct}\relax
\EndOfBibitem
\bibitem[Stauffer and Aharony(1992)Stauffer, and Aharony]{Stauffer}
Stauffer,~D.; Aharony,~A. \emph{Introduction to Percolation Theory}; Taylor and
  Francis, London, 1992\relax
\mciteBstWouldAddEndPuncttrue
\mciteSetBstMidEndSepPunct{\mcitedefaultmidpunct}
{\mcitedefaultendpunct}{\mcitedefaultseppunct}\relax
\EndOfBibitem
\bibitem[Winter and Mours(1997)Winter, and Mours]{Winter1997}
Winter,~H.~H.; Mours,~M. \emph{Neutron Spin Echo Spectroscopy Viscoelasticity
  Rheology}; Springer Berlin Heidelberg: Berlin, Heidelberg, 1997; pp
  165--234\relax
\mciteBstWouldAddEndPuncttrue
\mciteSetBstMidEndSepPunct{\mcitedefaultmidpunct}
{\mcitedefaultendpunct}{\mcitedefaultseppunct}\relax
\EndOfBibitem
\bibitem[Muthukumar(1989)]{Muthukumar1989-kk}
Muthukumar,~M. Screening effect on viscoelasticity near the gel point.
  \emph{Macromolecules} \textbf{1989}, \emph{22}, 4656--4658\relax
\mciteBstWouldAddEndPuncttrue
\mciteSetBstMidEndSepPunct{\mcitedefaultmidpunct}
{\mcitedefaultendpunct}{\mcitedefaultseppunct}\relax
\EndOfBibitem
\bibitem[Kroll and Croll(2015)Kroll, and Croll]{KROLL201582}
Kroll,~D.; Croll,~S. Influence of crosslinking functionality, temperature and
  conversion on heterogeneities in polymer networks. \emph{Polymer}
  \textbf{2015}, \emph{79}, 82--90\relax
\mciteBstWouldAddEndPuncttrue
\mciteSetBstMidEndSepPunct{\mcitedefaultmidpunct}
{\mcitedefaultendpunct}{\mcitedefaultseppunct}\relax
\EndOfBibitem
\bibitem[Santra \latin{et~al.}(2021)Santra, Dünweg, and
  Ravi~Prakash]{AritraStatJoR}
Santra,~A.; Dünweg,~B.; Ravi~Prakash,~J. Universal scaling and
  characterization of gelation in associative polymer solutions. \emph{J.
  Rheol.} \textbf{2021}, \emph{65}, 549--581\relax
\mciteBstWouldAddEndPuncttrue
\mciteSetBstMidEndSepPunct{\mcitedefaultmidpunct}
{\mcitedefaultendpunct}{\mcitedefaultseppunct}\relax
\EndOfBibitem
\bibitem[Anderson \latin{et~al.}(2020)Anderson, Glaser, and
  Glotzer]{ANDERSON2020109363}
Anderson,~J.~A.; Glaser,~J.; Glotzer,~S.~C. HOOMD-blue: A Python package for
  high-performance molecular dynamics and hard particle Monte Carlo
  simulations. \emph{Comput. Mater. Sci.} \textbf{2020}, \emph{173},
  109363\relax
\mciteBstWouldAddEndPuncttrue
\mciteSetBstMidEndSepPunct{\mcitedefaultmidpunct}
{\mcitedefaultendpunct}{\mcitedefaultseppunct}\relax
\EndOfBibitem
\bibitem[Howard \latin{et~al.}(2019)Howard, Statt, Madutsa, Truskett, and
  Panagiotopoulos]{HOWARD2019139}
Howard,~M.~P.; Statt,~A.; Madutsa,~F.; Truskett,~T.~M.; Panagiotopoulos,~A.~Z.
  Quantized bounding volume hierarchies for neighbor search in molecular
  simulations on graphics processing units. \emph{Comput. Mater. Sci.}
  \textbf{2019}, \emph{164}, 139--146\relax
\mciteBstWouldAddEndPuncttrue
\mciteSetBstMidEndSepPunct{\mcitedefaultmidpunct}
{\mcitedefaultendpunct}{\mcitedefaultseppunct}\relax
\EndOfBibitem
\bibitem[Stoltz \latin{et~al.}(2006)Stoltz, de~Pablo, and Graham]{Stoltz2006}
Stoltz,~C.; de~Pablo,~J.; Graham,~M. Concentration dependence of shear and
  extensional rheology of polymer solutions: {B}rownian dynamics simulations.
  \emph{J. Rheol.} \textbf{2006}, \emph{50}, 137--167\relax
\mciteBstWouldAddEndPuncttrue
\mciteSetBstMidEndSepPunct{\mcitedefaultmidpunct}
{\mcitedefaultendpunct}{\mcitedefaultseppunct}\relax
\EndOfBibitem
\bibitem[Soddemann \latin{et~al.}(2001)Soddemann, D{\"u}nweg, and
  Kremer]{soddemann2001generic}
Soddemann,~T.; D{\"u}nweg,~B.; Kremer,~K. A generic computer model for
  amphiphilic systems. \emph{Eur. Phys. J. E Soft Matter} \textbf{2001},
  \emph{6}, 409--419\relax
\mciteBstWouldAddEndPuncttrue
\mciteSetBstMidEndSepPunct{\mcitedefaultmidpunct}
{\mcitedefaultendpunct}{\mcitedefaultseppunct}\relax
\EndOfBibitem
\bibitem[Santra \latin{et~al.}(2019)Santra, Kumari, Padinhateeri, D\"{u}nweg,
  and Prakash]{Aritra2019}
Santra,~A.; Kumari,~K.; Padinhateeri,~R.; D\"{u}nweg,~B.; Prakash,~J.~R.
  Universality of the collapse transition of sticky polymers. \emph{Soft
  Matter} \textbf{2019}, \emph{15}, 7876--7887\relax
\mciteBstWouldAddEndPuncttrue
\mciteSetBstMidEndSepPunct{\mcitedefaultmidpunct}
{\mcitedefaultendpunct}{\mcitedefaultseppunct}\relax
\EndOfBibitem
\bibitem[Yamakawa(1971)]{Yamakawa1971}
Yamakawa,~H. \emph{Modern Theory of Polymer Solutions}; Harper and Row: New
  York, 1971\relax
\mciteBstWouldAddEndPuncttrue
\mciteSetBstMidEndSepPunct{\mcitedefaultmidpunct}
{\mcitedefaultendpunct}{\mcitedefaultseppunct}\relax
\EndOfBibitem
\bibitem[\"{O}ttinger(1996)]{Ottinger1996}
\"{O}ttinger,~H.~C. \emph{Stochastic Processes in Polymeric Fluids};
  Springer-Verlag: Berlin, 1996\relax
\mciteBstWouldAddEndPuncttrue
\mciteSetBstMidEndSepPunct{\mcitedefaultmidpunct}
{\mcitedefaultendpunct}{\mcitedefaultseppunct}\relax
\EndOfBibitem
\bibitem[Fiore \latin{et~al.}(2017)Fiore, Usabiaga, Donev, and Swan]{Fiore2017}
Fiore,~A.~M.; Usabiaga,~F.~B.; Donev,~A.; Swan,~J.~W. Rapid sampling of
  stochastic displacements in Brownian dynamics simulations. \emph{J. Chem.
  Phys.} \textbf{2017}, \emph{146}, 124116\relax
\mciteBstWouldAddEndPuncttrue
\mciteSetBstMidEndSepPunct{\mcitedefaultmidpunct}
{\mcitedefaultendpunct}{\mcitedefaultseppunct}\relax
\EndOfBibitem
\bibitem[Thurston(1974)]{THURSTON1974569}
Thurston,~G. Exact and approximate eigenvalues and intrinsic functions for the
  Gaussian chain theory. \emph{Polymer} \textbf{1974}, \emph{15},
  569--572\relax
\mciteBstWouldAddEndPuncttrue
\mciteSetBstMidEndSepPunct{\mcitedefaultmidpunct}
{\mcitedefaultendpunct}{\mcitedefaultseppunct}\relax
\EndOfBibitem
\bibitem[Djabourov \latin{et~al.}(2013)Djabourov, Nishinari, and
  Ross-Murphy]{djabourov_nishinari_ross-murphy_2013}
Djabourov,~M.; Nishinari,~K.; Ross-Murphy,~S.~B. \emph{Physical Gels from
  Biological and Synthetic Polymers}; Cambridge University Press, 2013\relax
\mciteBstWouldAddEndPuncttrue
\mciteSetBstMidEndSepPunct{\mcitedefaultmidpunct}
{\mcitedefaultendpunct}{\mcitedefaultseppunct}\relax
\EndOfBibitem
\bibitem[Brunsveld \latin{et~al.}(2001)Brunsveld, Folmer, Meijer, and
  Sijbesma]{doi:10.1021/cr990125q}
Brunsveld,~L.; Folmer,~B. J.~B.; Meijer,~E.~W.; Sijbesma,~R.~P. Supramolecular
  Polymers. \emph{Chem. Rev.} \textbf{2001}, \emph{101}, 4071--4098\relax
\mciteBstWouldAddEndPuncttrue
\mciteSetBstMidEndSepPunct{\mcitedefaultmidpunct}
{\mcitedefaultendpunct}{\mcitedefaultseppunct}\relax
\EndOfBibitem
\bibitem[Wu \latin{et~al.}(2023)Wu, Yang, and Chen]{Wu2023-sg}
Wu,~S.; Yang,~H.; Chen,~Q. Improving stretchability of associative polymers
  through tuning density of the secondary interactions. \emph{J. Rheol.}
  \textbf{2023}, \emph{67}, 293--304\relax
\mciteBstWouldAddEndPuncttrue
\mciteSetBstMidEndSepPunct{\mcitedefaultmidpunct}
{\mcitedefaultendpunct}{\mcitedefaultseppunct}\relax
\EndOfBibitem
\bibitem[Manousiouthakis and Deem(1999)Manousiouthakis, and
  Deem]{manousiouthakis1999strict}
Manousiouthakis,~V.~I.; Deem,~M.~W. Strict detailed balance is unnecessary in
  Monte Carlo simulation. \emph{J. Chem. Phys.} \textbf{1999}, \emph{110},
  2753--2756\relax
\mciteBstWouldAddEndPuncttrue
\mciteSetBstMidEndSepPunct{\mcitedefaultmidpunct}
{\mcitedefaultendpunct}{\mcitedefaultseppunct}\relax
\EndOfBibitem
\bibitem[Norris(1997)]{norris1998markov}
Norris,~J.~R. \emph{Markov Chains}; Cambridge Series in Statistical and
  Probabilistic Mathematics; Cambridge University Press, 1997\relax
\mciteBstWouldAddEndPuncttrue
\mciteSetBstMidEndSepPunct{\mcitedefaultmidpunct}
{\mcitedefaultendpunct}{\mcitedefaultseppunct}\relax
\EndOfBibitem
\bibitem[Podlubny(1999)]{Podlubny1999-ba}
Podlubny,~I. Fractional-order systems and $PI^{\lambda}\,D^{\mu}$-controllers.
  \emph{IEEE Trans. Automat. Contr.} \textbf{1999}, \emph{44}, 208--214\relax
\mciteBstWouldAddEndPuncttrue
\mciteSetBstMidEndSepPunct{\mcitedefaultmidpunct}
{\mcitedefaultendpunct}{\mcitedefaultseppunct}\relax
\EndOfBibitem
\bibitem[Schiessel \latin{et~al.}(1995)Schiessel, Metzler, Blumen, and
  Nonnenmacher]{Schiessel1995-ss}
Schiessel,~H.; Metzler,~R.; Blumen,~A.; Nonnenmacher,~T.~F. Generalized
  viscoelastic models: their fractional equations with solutions. \emph{J.
  Phys. A Math. Gen.} \textbf{1995}, \emph{28}, 6567--6584\relax
\mciteBstWouldAddEndPuncttrue
\mciteSetBstMidEndSepPunct{\mcitedefaultmidpunct}
{\mcitedefaultendpunct}{\mcitedefaultseppunct}\relax
\EndOfBibitem
\bibitem[Metzler and Klafter(2002)Metzler, and Klafter]{Metzler2002-pn}
Metzler,~R.; Klafter,~J. From stretched exponential to inverse power-law:
  fractional dynamics, {Cole--Cole} relaxation processes, and beyond. \emph{J.
  Non Cryst. Solids} \textbf{2002}, \emph{305}, 81--87\relax
\mciteBstWouldAddEndPuncttrue
\mciteSetBstMidEndSepPunct{\mcitedefaultmidpunct}
{\mcitedefaultendpunct}{\mcitedefaultseppunct}\relax
\EndOfBibitem
\bibitem[Jaishankar and McKinley(2013)Jaishankar, and
  McKinley]{Jaishankar2013-uo}
Jaishankar,~A.; McKinley,~G.~H. Power-law rheology in the bulk and at the
  interface: quasi-properties and fractional constitutive equations.
  \emph{Proc. Math. Phys. Eng. Sci.} \textbf{2013}, \emph{469}, 20120284\relax
\mciteBstWouldAddEndPuncttrue
\mciteSetBstMidEndSepPunct{\mcitedefaultmidpunct}
{\mcitedefaultendpunct}{\mcitedefaultseppunct}\relax
\EndOfBibitem
\bibitem[Jaishankar and McKinley(2014)Jaishankar, and
  McKinley]{Jaishankar2014-hg}
Jaishankar,~A.; McKinley,~G.~H. A fractional {K-BKZ} constitutive formulation
  for describing the nonlinear rheology of multiscale complex fluids. \emph{J.
  Rheol.} \textbf{2014}, \emph{58}, 1751--1788\relax
\mciteBstWouldAddEndPuncttrue
\mciteSetBstMidEndSepPunct{\mcitedefaultmidpunct}
{\mcitedefaultendpunct}{\mcitedefaultseppunct}\relax
\EndOfBibitem
\bibitem[Huang \latin{et~al.}(2010)Huang, Winkler, Sutmann, and
  Gompper]{Huang2010}
Huang,~C.~C.; Winkler,~R.~G.; Sutmann,~G.; Gompper,~G. Semidilute Polymer
  Solutions at Equilibrium and under Shear Flow. \emph{Macromolecules}
  \textbf{2010}, \emph{43}, 10107--10116\relax
\mciteBstWouldAddEndPuncttrue
\mciteSetBstMidEndSepPunct{\mcitedefaultmidpunct}
{\mcitedefaultendpunct}{\mcitedefaultseppunct}\relax
\EndOfBibitem
\bibitem[de~Gennes(1979)]{deGennes}
de~Gennes,~P.~G. \emph{Scaling Concepts in Polymer Physics}; Cornell University
  Press, Ithaca, 1979\relax
\mciteBstWouldAddEndPuncttrue
\mciteSetBstMidEndSepPunct{\mcitedefaultmidpunct}
{\mcitedefaultendpunct}{\mcitedefaultseppunct}\relax
\EndOfBibitem
\bibitem[Sing and Alexander-Katz(2011)Sing, and Alexander-Katz]{SingKatz2011}
Sing,~C.~E.; Alexander-Katz,~A. Equilibrium Structure and Dynamics of
  Self-Associating Single Polymers. \emph{Macromolecules} \textbf{2011},
  \emph{44}, 6962--6971\relax
\mciteBstWouldAddEndPuncttrue
\mciteSetBstMidEndSepPunct{\mcitedefaultmidpunct}
{\mcitedefaultendpunct}{\mcitedefaultseppunct}\relax
\EndOfBibitem
\bibitem[Dobrynin(2004)]{Dob}
Dobrynin,~A.~V. Phase Diagram of Solutions of Associative Polymers.
  \emph{Macromolecules} \textbf{2004}, \emph{37}, 3881--3893\relax
\mciteBstWouldAddEndPuncttrue
\mciteSetBstMidEndSepPunct{\mcitedefaultmidpunct}
{\mcitedefaultendpunct}{\mcitedefaultseppunct}\relax
\EndOfBibitem
\bibitem[Hodgson and Amis(1990)Hodgson, and Amis]{doi:10.1021/ma00211a019}
Hodgson,~D.~F.; Amis,~E.~J. Dynamic viscoelastic characterization of sol-gel
  reactions. \emph{Macromolecules} \textbf{1990}, \emph{23}, 2512--2519\relax
\mciteBstWouldAddEndPuncttrue
\mciteSetBstMidEndSepPunct{\mcitedefaultmidpunct}
{\mcitedefaultendpunct}{\mcitedefaultseppunct}\relax
\EndOfBibitem
\bibitem[Axelos and Kolb(1990)Axelos, and Kolb]{PhysRevLett.64.1457}
Axelos,~M. A.~V.; Kolb,~M. Crosslinked biopolymers: Experimental evidence for
  scalar percolation theory. \emph{Phys. Rev. Lett.} \textbf{1990}, \emph{64},
  1457--1460\relax
\mciteBstWouldAddEndPuncttrue
\mciteSetBstMidEndSepPunct{\mcitedefaultmidpunct}
{\mcitedefaultendpunct}{\mcitedefaultseppunct}\relax
\EndOfBibitem
\bibitem[Lusignan \latin{et~al.}(1995)Lusignan, Mourey, Wilson, and
  Colby]{PhysRevE.52.6271}
Lusignan,~C.~P.; Mourey,~T.~H.; Wilson,~J.~C.; Colby,~R.~H. Viscoelasticity of
  randomly branched polymers in the critical percolation class. \emph{Phys.
  Rev. E} \textbf{1995}, \emph{52}, 6271--6280\relax
\mciteBstWouldAddEndPuncttrue
\mciteSetBstMidEndSepPunct{\mcitedefaultmidpunct}
{\mcitedefaultendpunct}{\mcitedefaultseppunct}\relax
\EndOfBibitem
\bibitem[Guo \latin{et~al.}(2003)Guo, Colby, Lusignan, and
  Howe]{doi:10.1021/ma034266c}
Guo,~L.; Colby,~R.~H.; Lusignan,~C.~P.; Howe,~A.~M. Physical Gelation of
  Gelatin Studied with Rheo-Optics. \emph{Macromolecules} \textbf{2003},
  \emph{36}, 10009--10020\relax
\mciteBstWouldAddEndPuncttrue
\mciteSetBstMidEndSepPunct{\mcitedefaultmidpunct}
{\mcitedefaultendpunct}{\mcitedefaultseppunct}\relax
\EndOfBibitem
\bibitem[Cho and Heuzey(2008)Cho, and Heuzey]{cho2008dynamic}
Cho,~J.; Heuzey,~M.-C. Dynamic scaling for gelation of a thermosensitive
  chitosan-$\beta$-glycerophosphate hydrogel. \emph{Colloid Polym. Sci.}
  \textbf{2008}, \emph{286}, 427--434\relax
\mciteBstWouldAddEndPuncttrue
\mciteSetBstMidEndSepPunct{\mcitedefaultmidpunct}
{\mcitedefaultendpunct}{\mcitedefaultseppunct}\relax
\EndOfBibitem
\bibitem[Dai \latin{et~al.}(2008)Dai, Liu, Liu, and Tong]{DAI20084012}
Dai,~L.; Liu,~X.; Liu,~Y.; Tong,~Z. Concentration dependence of critical
  exponents for gelation in gellan gum aqueous solutions upon cooling.
  \emph{Eur. Polym. J.} \textbf{2008}, \emph{44}, 4012--4019\relax
\mciteBstWouldAddEndPuncttrue
\mciteSetBstMidEndSepPunct{\mcitedefaultmidpunct}
{\mcitedefaultendpunct}{\mcitedefaultseppunct}\relax
\EndOfBibitem
\bibitem[Liu \latin{et~al.}(2015)Liu, Chan, and
  Li]{doi:10.1021/acs.macromol.5b01922}
Liu,~S.; Chan,~W.~L.; Li,~L. Rheological Properties and Scaling Laws of
  k-Carrageenan in Aqueous Solution. \emph{Macromolecules} \textbf{2015},
  \emph{48}, 7649--7657\relax
\mciteBstWouldAddEndPuncttrue
\mciteSetBstMidEndSepPunct{\mcitedefaultmidpunct}
{\mcitedefaultendpunct}{\mcitedefaultseppunct}\relax
\EndOfBibitem
\bibitem[Tan \latin{et~al.}(2008)Tan, Pan, and Pan]{TAN20085676}
Tan,~L.; Pan,~D.; Pan,~N. Gelation behavior of polyacrylonitrile solution in
  relation to aging process and gel concentration. \emph{Polymer}
  \textbf{2008}, \emph{49}, 5676--5682\relax
\mciteBstWouldAddEndPuncttrue
\mciteSetBstMidEndSepPunct{\mcitedefaultmidpunct}
{\mcitedefaultendpunct}{\mcitedefaultseppunct}\relax
\EndOfBibitem
\bibitem[Lu \latin{et~al.}(2006)Lu, Liu, and Tong]{LU2006544}
Lu,~L.; Liu,~X.; Tong,~Z. Critical exponents for sol–gel transition in
  aqueous alginate solutions induced by cupric cations. \emph{Carbohydr.
  Polym.} \textbf{2006}, \emph{65}, 544--551\relax
\mciteBstWouldAddEndPuncttrue
\mciteSetBstMidEndSepPunct{\mcitedefaultmidpunct}
{\mcitedefaultendpunct}{\mcitedefaultseppunct}\relax
\EndOfBibitem
\bibitem[Lue and Zhang(2008)Lue, and Zhang]{doi:10.1021/jp077685a}
Lue,~A.; Zhang,~L. Investigation of the Scaling Law on Cellulose Solution
  Prepared at Low Temperature. \emph{J. Phys. Chem. B} \textbf{2008},
  \emph{112}, 4488--4495\relax
\mciteBstWouldAddEndPuncttrue
\mciteSetBstMidEndSepPunct{\mcitedefaultmidpunct}
{\mcitedefaultendpunct}{\mcitedefaultseppunct}\relax
\EndOfBibitem
\bibitem[Suman \latin{et~al.}(2021)Suman, Shanbhag, and
  Joshi]{10.1063/5.0038830}
Suman,~K.; Shanbhag,~S.; Joshi,~Y.~M. {Phenomenological model of
  viscoelasticity for systems undergoing sol–gel transition}. \emph{Phys.
  Fluids} \textbf{2021}, \emph{33}, 033103\relax
\mciteBstWouldAddEndPuncttrue
\mciteSetBstMidEndSepPunct{\mcitedefaultmidpunct}
{\mcitedefaultendpunct}{\mcitedefaultseppunct}\relax
\EndOfBibitem
\bibitem[Scanlan and Winter(1991)Scanlan, and Winter]{Scanlan1991}
Scanlan,~J.~C.; Winter,~H.~H. The evolution of viscoelasticity near the gel
  point of end-linking poly(dimethylsiloxane)s. \emph{Makromol. Chem.,
  Macromol. Symp.} \textbf{1991}, \emph{45}, 11--21\relax
\mciteBstWouldAddEndPuncttrue
\mciteSetBstMidEndSepPunct{\mcitedefaultmidpunct}
{\mcitedefaultendpunct}{\mcitedefaultseppunct}\relax
\EndOfBibitem
\bibitem[Jatav and Joshi(2017)Jatav, and
  Joshi]{doi:10.1021/acs.langmuir.7b00151}
Jatav,~S.; Joshi,~Y.~M. Phase Behavior of Aqueous Suspension of Laponite: New
  Insights with Microscopic Evidence. \emph{Langmuir} \textbf{2017}, \emph{33},
  2370--2377\relax
\mciteBstWouldAddEndPuncttrue
\mciteSetBstMidEndSepPunct{\mcitedefaultmidpunct}
{\mcitedefaultendpunct}{\mcitedefaultseppunct}\relax
\EndOfBibitem
\bibitem[Negi \latin{et~al.}(2014)Negi, Redmon, Ramakrishnan, and
  Osuji]{10.1122/1.4883675}
Negi,~A.~S.; Redmon,~C.~G.; Ramakrishnan,~S.; Osuji,~C.~O. {Viscoelasticity of
  a colloidal gel during dynamical arrest: Evolution through the critical gel
  and comparison with a soft colloidal glass}. \emph{J. Rheol.} \textbf{2014},
  \emph{58}, 1557--1579\relax
\mciteBstWouldAddEndPuncttrue
\mciteSetBstMidEndSepPunct{\mcitedefaultmidpunct}
{\mcitedefaultendpunct}{\mcitedefaultseppunct}\relax
\EndOfBibitem
\end{mcitethebibliography}


\begin{thebibliography}{21}%
\makeatletter
\providecommand \@ifxundefined [1]{%
 \@ifx{#1\undefined}
}%
\providecommand \@ifnum [1]{%
 \ifnum #1\expandafter \@firstoftwo
 \else \expandafter \@secondoftwo
 \fi
}%
\providecommand \@ifx [1]{%
 \ifx #1\expandafter \@firstoftwo
 \else \expandafter \@secondoftwo
 \fi
}%
\providecommand \natexlab [1]{#1}%
\providecommand \enquote  [1]{``#1''}%
\providecommand \bibnamefont  [1]{#1}%
\providecommand \bibfnamefont [1]{#1}%
\providecommand \citenamefont [1]{#1}%
\providecommand \href@noop [0]{\@secondoftwo}%
\providecommand \href [0]{\begingroup \@sanitize@url \@href}%
\providecommand \@href[1]{\@@startlink{#1}\@@href}%
\providecommand \@@href[1]{\endgroup#1\@@endlink}%
\providecommand \@sanitize@url [0]{\catcode `\\12\catcode `\$12\catcode
  `\&12\catcode `\#12\catcode `\^12\catcode `\_12\catcode `\%12\relax}%
\providecommand \@@startlink[1]{}%
\providecommand \@@endlink[0]{}%
\providecommand \url  [0]{\begingroup\@sanitize@url \@url }%
\providecommand \@url [1]{\endgroup\@href {#1}{\urlprefix }}%
\providecommand \urlprefix  [0]{URL }%
\providecommand \Eprint [0]{\href }%
\providecommand \doibase [0]{https://doi.org/}%
\providecommand \selectlanguage [0]{\@gobble}%
\providecommand \bibinfo  [0]{\@secondoftwo}%
\providecommand \bibfield  [0]{\@secondoftwo}%
\providecommand \translation [1]{[#1]}%
\providecommand \BibitemOpen [0]{}%
\providecommand \bibitemStop [0]{}%
\providecommand \bibitemNoStop [0]{.\EOS\space}%
\providecommand \EOS [0]{\spacefactor3000\relax}%
\providecommand \BibitemShut  [1]{\csname bibitem#1\endcsname}%
\let\auto@bib@innerbib\@empty
\bibitem [{\citenamefont {Bird}\ \emph {et~al.}(1987)\citenamefont {Bird},
  \citenamefont {Curtiss}, \citenamefont {Armstrong},\ and\ \citenamefont
  {Hassager}}]{Bird1987v1}%
  \BibitemOpen
  \bibfield  {author} {\bibinfo {author} {\bibfnamefont {R.~B.}\ \bibnamefont
  {Bird}}, \bibinfo {author} {\bibfnamefont {C.~F.}\ \bibnamefont {Curtiss}},
  \bibinfo {author} {\bibfnamefont {R.~C.}\ \bibnamefont {Armstrong}},\ and\
  \bibinfo {author} {\bibfnamefont {O.}~\bibnamefont {Hassager}},\ }\href@noop
  {} {\emph {\bibinfo {title} {Dynamics of polymeric liquids}}},\ Vol.~\bibinfo
  {volume} {1}\ (\bibinfo  {publisher} {John Wiley and Sons, New York},\
  \bibinfo {year} {1987})\BibitemShut {NoStop}%
\bibitem [{\citenamefont {Stoltz}(2006)}]{Stoltzthesis}%
  \BibitemOpen
  \bibfield  {author} {\bibinfo {author} {\bibfnamefont {C.}~\bibnamefont
  {Stoltz}},\ }\href@noop {} {\emph {\bibinfo {title} {Simulation of Dilute
  Polymer and Polyelectrolyte Solutions: Concentration Effects (Ph.D.
  thesis)}}}\ (\bibinfo  {publisher} {The University of Wisconsin},\ \bibinfo
  {year} {2006})\BibitemShut {NoStop}%
\bibitem [{\citenamefont {Lee}\ and\ \citenamefont {Kremer}(2009)}]{LeeKremer}%
  \BibitemOpen
  \bibfield  {author} {\bibinfo {author} {\bibfnamefont {W.~B.}\ \bibnamefont
  {Lee}}\ and\ \bibinfo {author} {\bibfnamefont {K.}~\bibnamefont {Kremer}},\
  }\bibfield  {title} {\enquote {\bibinfo {title} {Entangled polymer melts:
  Relation between plateau modulus and stress autocorrelation function},}\
  }\href@noop {} {\bibfield  {journal} {\bibinfo  {journal} {Macromolecules}\
  }\textbf {\bibinfo {volume} {42}},\ \bibinfo {pages} {6270--6276} (\bibinfo
  {year} {2009})}\BibitemShut {NoStop}%
\bibitem [{\citenamefont {Mours}\ and\ \citenamefont
  {Winter}(2012)}]{MoursWinter}%
  \BibitemOpen
  \bibfield  {author} {\bibinfo {author} {\bibfnamefont {M.}~\bibnamefont
  {Mours}}\ and\ \bibinfo {author} {\bibfnamefont {H.}~\bibnamefont {Winter}},\
  }\bibfield  {title} {\enquote {\bibinfo {title} {Mechanical spectroscopy of
  polymers},}\ }\href@noop {} {\bibfield  {journal} {\bibinfo  {journal}
  {Experimental Methods in Polymer Science: Modern Methods in Polymer Research
  and Technology}\ ,\ \bibinfo {pages} {495--546}} (\bibinfo {year}
  {2012})}\BibitemShut {NoStop}%
\bibitem [{\citenamefont {Wittmer}\ \emph {et~al.}(2015)\citenamefont
  {Wittmer}, \citenamefont {Xu}, \citenamefont {Benzerara},\ and\ \citenamefont
  {Baschnagel}}]{WittmerMolPy}%
  \BibitemOpen
  \bibfield  {author} {\bibinfo {author} {\bibfnamefont {J.}~\bibnamefont
  {Wittmer}}, \bibinfo {author} {\bibfnamefont {H.}~\bibnamefont {Xu}},
  \bibinfo {author} {\bibfnamefont {O.}~\bibnamefont {Benzerara}},\ and\
  \bibinfo {author} {\bibfnamefont {J.}~\bibnamefont {Baschnagel}},\ }\bibfield
   {title} {\enquote {\bibinfo {title} {Fluctuation-dissipation relation
  between shear stress relaxation modulus and shear stress autocorrelation
  function revisited},}\ }\href@noop {} {\bibfield  {journal} {\bibinfo
  {journal} {Molecular Physics}\ }\textbf {\bibinfo {volume} {113}},\ \bibinfo
  {pages} {2881--2893} (\bibinfo {year} {2015})}\BibitemShut {NoStop}%
\bibitem [{\citenamefont {Wittmer}, \citenamefont {Xu},\ and\ \citenamefont
  {Baschnagel}(2015)}]{WittmerPRE}%
  \BibitemOpen
  \bibfield  {author} {\bibinfo {author} {\bibfnamefont {J.}~\bibnamefont
  {Wittmer}}, \bibinfo {author} {\bibfnamefont {H.}~\bibnamefont {Xu}},\ and\
  \bibinfo {author} {\bibfnamefont {J.}~\bibnamefont {Baschnagel}},\ }\bibfield
   {title} {\enquote {\bibinfo {title} {Shear-stress relaxation and ensemble
  transformation of shear-stress autocorrelation functions},}\ }\href@noop {}
  {\bibfield  {journal} {\bibinfo  {journal} {Physical Review E}\ }\textbf
  {\bibinfo {volume} {91}} (\bibinfo {year} {2015})}\BibitemShut {NoStop}%
\bibitem [{\citenamefont {Pan}\ \emph {et~al.}(2014)\citenamefont {Pan},
  \citenamefont {Ahirwal}, \citenamefont {Nguyen}, \citenamefont {Sunthar},
  \citenamefont {Sridhar},\ and\ \citenamefont {Prakash}}]{Sharad2014}%
  \BibitemOpen
  \bibfield  {author} {\bibinfo {author} {\bibfnamefont {S.}~\bibnamefont
  {Pan}}, \bibinfo {author} {\bibfnamefont {D.}~\bibnamefont {Ahirwal}},
  \bibinfo {author} {\bibfnamefont {D.~A.}\ \bibnamefont {Nguyen}}, \bibinfo
  {author} {\bibfnamefont {P.}~\bibnamefont {Sunthar}}, \bibinfo {author}
  {\bibfnamefont {T.}~\bibnamefont {Sridhar}},\ and\ \bibinfo {author}
  {\bibfnamefont {J.~R.}\ \bibnamefont {Prakash}},\ }\bibfield  {title}
  {\enquote {\bibinfo {title} {Viscosity radius of polymers in dilute
  solutions: {U}niversal behaviour from {DNA} rheology and {B}rownian dynamics
  simulations},}\ }\href@noop {} {\bibfield  {journal} {\bibinfo  {journal}
  {Macromolecules}\ }\textbf {\bibinfo {volume} {47}},\ \bibinfo {pages}
  {7548--7560} (\bibinfo {year} {2014})}\BibitemShut {NoStop}%
\bibitem [{\citenamefont {Fixman}(1981)}]{Fixman1981}%
  \BibitemOpen
  \bibfield  {author} {\bibinfo {author} {\bibfnamefont {M.}~\bibnamefont
  {Fixman}},\ }\bibfield  {title} {\enquote {\bibinfo {title} {{Inclusion of
  hydrodynamic interaction in polymer dynamical simulations}},}\ }\href@noop {}
  {\bibfield  {journal} {\bibinfo  {journal} {Macromolecules}\ }\textbf
  {\bibinfo {volume} {14}},\ \bibinfo {pages} {1710--1717} (\bibinfo {year}
  {1981})}\BibitemShut {NoStop}%
\bibitem [{\citenamefont {\"{O}ttinger}(1996)}]{Ottinger1996}%
  \BibitemOpen
  \bibfield  {author} {\bibinfo {author} {\bibfnamefont {H.~C.}\ \bibnamefont
  {\"{O}ttinger}},\ }\href@noop {} {\emph {\bibinfo {title} {Stochastic
  Processes in Polymeric Fluids}}}\ (\bibinfo  {publisher} {Springer-Verlag:
  Berlin},\ \bibinfo {year} {1996})\BibitemShut {NoStop}%
\bibitem [{\citenamefont {Lees}\ and\ \citenamefont
  {Edwards}(1972)}]{Lees_1972}%
  \BibitemOpen
  \bibfield  {author} {\bibinfo {author} {\bibfnamefont {A.~W.}\ \bibnamefont
  {Lees}}\ and\ \bibinfo {author} {\bibfnamefont {S.~F.}\ \bibnamefont
  {Edwards}},\ }\bibfield  {title} {\enquote {\bibinfo {title} {The computer
  study of transport processes under extreme conditions},}\ }\href@noop {}
  {\bibfield  {journal} {\bibinfo  {journal} {Journal of Physics C: Solid State
  Physics}\ }\textbf {\bibinfo {volume} {5}},\ \bibinfo {pages} {1921--1928}
  (\bibinfo {year} {1972})}\BibitemShut {NoStop}%
\bibitem [{\citenamefont {Jain}\ \emph {et~al.}(2015)\citenamefont {Jain},
  \citenamefont {Sasmal}, \citenamefont {Hartkamp}, \citenamefont {Todd},\ and\
  \citenamefont {Prakash}}]{JainSasmal2015}%
  \BibitemOpen
  \bibfield  {author} {\bibinfo {author} {\bibfnamefont {A.}~\bibnamefont
  {Jain}}, \bibinfo {author} {\bibfnamefont {C.}~\bibnamefont {Sasmal}},
  \bibinfo {author} {\bibfnamefont {R.}~\bibnamefont {Hartkamp}}, \bibinfo
  {author} {\bibfnamefont {B.~D.}\ \bibnamefont {Todd}},\ and\ \bibinfo
  {author} {\bibfnamefont {J.~R.}\ \bibnamefont {Prakash}},\ }\bibfield
  {title} {\enquote {\bibinfo {title} {Brownian dynamics simulations of planar
  mixed flows of polymer solutions at finite concentrations},}\ }\href@noop {}
  {\bibfield  {journal} {\bibinfo  {journal} {Chem. Eng. Sci.}\ }\textbf
  {\bibinfo {volume} {121}},\ \bibinfo {pages} {245--257} (\bibinfo {year}
  {2015})}\BibitemShut {NoStop}%
\bibitem [{\citenamefont {Myung}, \citenamefont {Winkler},\ and\ \citenamefont
  {Gompper}(2015)}]{Gompper2015}%
  \BibitemOpen
  \bibfield  {author} {\bibinfo {author} {\bibfnamefont {J.~S.}\ \bibnamefont
  {Myung}}, \bibinfo {author} {\bibfnamefont {R.~G.}\ \bibnamefont {Winkler}},\
  and\ \bibinfo {author} {\bibfnamefont {G.}~\bibnamefont {Gompper}},\
  }\bibfield  {title} {\enquote {\bibinfo {title} {Self-organization in
  suspensions of end-functionalized semiflexible polymers under shear flow},}\
  }\href@noop {} {\bibfield  {journal} {\bibinfo  {journal} {The Journal of
  Chemical Physics}\ }\textbf {\bibinfo {volume} {143}},\ \bibinfo {pages}
  {243117} (\bibinfo {year} {2015})}\BibitemShut {NoStop}%
\bibitem [{\citenamefont {Brassinne}\ \emph {et~al.}(2017)\citenamefont
  {Brassinne}, \citenamefont {Cadix}, \citenamefont {Wilson},\ and\
  \citenamefont {van Ruymbeke}}]{Ruyumbeke2017}%
  \BibitemOpen
  \bibfield  {author} {\bibinfo {author} {\bibfnamefont {J.}~\bibnamefont
  {Brassinne}}, \bibinfo {author} {\bibfnamefont {A.}~\bibnamefont {Cadix}},
  \bibinfo {author} {\bibfnamefont {J.}~\bibnamefont {Wilson}},\ and\ \bibinfo
  {author} {\bibfnamefont {E.}~\bibnamefont {van Ruymbeke}},\ }\bibfield
  {title} {\enquote {\bibinfo {title} {Dissociating sticker dynamics from chain
  relaxation in supramolecular polymer networks—the importance of free
  partner!}}\ }\href@noop {} {\bibfield  {journal} {\bibinfo  {journal}
  {Journal of Rheology}\ }\textbf {\bibinfo {volume} {61}},\ \bibinfo {pages}
  {1123--1134} (\bibinfo {year} {2017})}\BibitemShut {NoStop}%
\bibitem [{\citenamefont {Rubinstein}\ and\ \citenamefont
  {Semenov}(1998)}]{RnSdynamics}%
  \BibitemOpen
  \bibfield  {author} {\bibinfo {author} {\bibfnamefont {M.}~\bibnamefont
  {Rubinstein}}\ and\ \bibinfo {author} {\bibfnamefont {A.~N.}\ \bibnamefont
  {Semenov}},\ }\bibfield  {title} {\enquote {\bibinfo {title}
  {Thermoreversible gelation in solutions of associative polymers. 2. linear
  dynamics},}\ }\href@noop {} {\bibfield  {journal} {\bibinfo  {journal}
  {Macromolecules}\ }\textbf {\bibinfo {volume} {31}},\ \bibinfo {pages}
  {1386--1397} (\bibinfo {year} {1998})}\BibitemShut {NoStop}%
\bibitem [{\citenamefont {Huang}\ \emph {et~al.}(2010)\citenamefont {Huang},
  \citenamefont {Winkler}, \citenamefont {Sutmann},\ and\ \citenamefont
  {Gompper}}]{Huang2010}%
  \BibitemOpen
  \bibfield  {author} {\bibinfo {author} {\bibfnamefont {C.~C.}\ \bibnamefont
  {Huang}}, \bibinfo {author} {\bibfnamefont {R.~G.}\ \bibnamefont {Winkler}},
  \bibinfo {author} {\bibfnamefont {G.}~\bibnamefont {Sutmann}},\ and\ \bibinfo
  {author} {\bibfnamefont {G.}~\bibnamefont {Gompper}},\ }\bibfield  {title}
  {\enquote {\bibinfo {title} {Semidilute polymer solutions at equilibrium and
  under shear flow},}\ }\href@noop {} {\bibfield  {journal} {\bibinfo
  {journal} {Macromolecules}\ }\textbf {\bibinfo {volume} {43}},\ \bibinfo
  {pages} {10107--10116} (\bibinfo {year} {2010})}\BibitemShut {NoStop}%
\bibitem [{\citenamefont {Nafar~Sefiddashti}, \citenamefont {Edwards},\ and\
  \citenamefont {Khomami}(2015)}]{Nafar2015}%
  \BibitemOpen
  \bibfield  {author} {\bibinfo {author} {\bibfnamefont {M.~H.}\ \bibnamefont
  {Nafar~Sefiddashti}}, \bibinfo {author} {\bibfnamefont {B.~J.}\ \bibnamefont
  {Edwards}},\ and\ \bibinfo {author} {\bibfnamefont {B.}~\bibnamefont
  {Khomami}},\ }\bibfield  {title} {\enquote {\bibinfo {title} {Individual
  chain dynamics of a polyethylene melt undergoing steady shear flow},}\
  }\href@noop {} {\bibfield  {journal} {\bibinfo  {journal} {Journal of
  Rheology}\ }\textbf {\bibinfo {volume} {59}},\ \bibinfo {pages} {119--153}
  (\bibinfo {year} {2015})}\BibitemShut {NoStop}%
\bibitem [{\citenamefont {Santra}, \citenamefont {Dünweg},\ and\ \citenamefont
  {Ravi~Prakash}(2021)}]{AritraStatJoR}%
  \BibitemOpen
  \bibfield  {author} {\bibinfo {author} {\bibfnamefont {A.}~\bibnamefont
  {Santra}}, \bibinfo {author} {\bibfnamefont {B.}~\bibnamefont {Dünweg}},\
  and\ \bibinfo {author} {\bibfnamefont {J.}~\bibnamefont {Ravi~Prakash}},\
  }\bibfield  {title} {\enquote {\bibinfo {title} {Universal scaling and
  characterization of gelation in associative polymer solutions},}\ }\href@noop
  {} {\bibfield  {journal} {\bibinfo  {journal} {Journal of Rheology}\ }\textbf
  {\bibinfo {volume} {65}},\ \bibinfo {pages} {549--581} (\bibinfo {year}
  {2021})}\BibitemShut {NoStop}%
\bibitem [{\citenamefont {Stauffer}, \citenamefont {Coniglio},\ and\
  \citenamefont {Adam}(2007)}]{Stauffer2007}%
  \BibitemOpen
  \bibfield  {author} {\bibinfo {author} {\bibfnamefont {D.}~\bibnamefont
  {Stauffer}}, \bibinfo {author} {\bibfnamefont {A.}~\bibnamefont {Coniglio}},\
  and\ \bibinfo {author} {\bibfnamefont {M.}~\bibnamefont {Adam}},\ }\bibfield
  {title} {\enquote {\bibinfo {title} {Gelation and critical phenomena},}\
  }\href@noop {} {\bibfield  {journal} {\bibinfo  {journal} {Adv. Polym. Sci.}\
  }\textbf {\bibinfo {volume} {44}},\ \bibinfo {pages} {103--158} (\bibinfo
  {year} {2007})}\BibitemShut {NoStop}%
\bibitem [{\citenamefont {Ozaki}\ \emph {et~al.}(2017)\citenamefont {Ozaki},
  \citenamefont {Narita}, \citenamefont {Koga},\ and\ \citenamefont
  {Indei}}]{Indei17}%
  \BibitemOpen
  \bibfield  {author} {\bibinfo {author} {\bibfnamefont {H.}~\bibnamefont
  {Ozaki}}, \bibinfo {author} {\bibfnamefont {T.}~\bibnamefont {Narita}},
  \bibinfo {author} {\bibfnamefont {T.}~\bibnamefont {Koga}},\ and\ \bibinfo
  {author} {\bibfnamefont {T.}~\bibnamefont {Indei}},\ }\bibfield  {title}
  {\enquote {\bibinfo {title} {Theoretical analysis of critical flowable
  physical gel cross-linked by metal ions and polyacrylamide-derivative
  associating polymers containing imidazole groups},}\ }\href@noop {}
  {\bibfield  {journal} {\bibinfo  {journal} {Polymers}\ }\textbf {\bibinfo
  {volume} {9}} (\bibinfo {year} {2017})}\BibitemShut {NoStop}%
\bibitem [{\citenamefont {Bird}, \citenamefont {Curtiss},\ and\ \citenamefont
  {anf O.~Hassager}(1987)}]{Bird1987}%
  \BibitemOpen
  \bibfield  {author} {\bibinfo {author} {\bibfnamefont {R.~B.}\ \bibnamefont
  {Bird}}, \bibinfo {author} {\bibfnamefont {C.~F.}\ \bibnamefont {Curtiss}},\
  and\ \bibinfo {author} {\bibfnamefont {R.~C.~A.}\ \bibnamefont {anf
  O.~Hassager}},\ }\href@noop {} {\emph {\bibinfo {title} {Dynamics of
  polymeric liquids}}},\ Vol.~\bibinfo {volume} {2}\ (\bibinfo  {publisher}
  {John Wiley and Sons, New York},\ \bibinfo {year} {1987})\BibitemShut
  {NoStop}%
\bibitem [{\citenamefont {Suman}, \citenamefont {Shanbhag},\ and\ \citenamefont
  {Joshi}(2021)}]{10.1063/5.0038830}%
  \BibitemOpen
  \bibfield  {author} {\bibinfo {author} {\bibfnamefont {K.}~\bibnamefont
  {Suman}}, \bibinfo {author} {\bibfnamefont {S.}~\bibnamefont {Shanbhag}},\
  and\ \bibinfo {author} {\bibfnamefont {Y.~M.}\ \bibnamefont {Joshi}},\
  }\bibfield  {title} {\enquote {\bibinfo {title} {{Phenomenological model of
  viscoelasticity for systems undergoing sol–gel transition}},}\ }\href@noop
  {} {\bibfield  {journal} {\bibinfo  {journal} {Physics of Fluids}\ }\textbf
  {\bibinfo {volume} {33}},\ \bibinfo {pages} {033103} (\bibinfo {year}
  {2021})}\BibitemShut {NoStop}%
\end{thebibliography}%

\end{document}


\beginsupplement

\title{Supplementary Information for "Evanescent Gels: Competition Between Sticker Dynamics and Single Chain Relaxation"}

\author{Dominic Robe}
    \affiliation{Department of Chemical and Biological Engineering, Monash University, Melbourne, VIC, 3800, Australia}
    \affiliation{Current address: Department of Mechanical Engineering, University of Melbourne, Melbourne, VIC, 3052, Australia}
\author{Aritra Santra}
 \affiliation{Department of Chemical and Biological Engineering, Monash University, Melbourne, VIC, 3800, Australia}
 \affiliation{Current address: Department of Chemical Engineering, Indian Institute of Technology (Indian School of Mines), Dhanbad, Jharkhand,  826004, India}
\author{Gareth H. McKinley}
 \affiliation{Department of Mechanical Engineering, Massachusetts Institute of Technology, Cambridge, MA, 02139, United States}
\author{J. Ravi Prakash}
    \email{ravi.jagadeeshan@monash.edu}
    \affiliation{Department of Chemical and Biological Engineering, Monash University, Melbourne, VIC, 3800, Australia}

\maketitle




\section{\label{sec:Dynamic prop} Computation of dynamic properties}

In this section, we discuss the formulation to compute various dynamic properties from Brownian dynamics simulations and present some simulation data to demonstrate the calculations.

The dynamic properties such as relaxation modulus, zero-shear rate viscosity, dynamic moduli ($G'$ \& $G''$), investigated in this work, can be defined in terms of the components of stress-tensor~\cite{Bird1987v1} for the polymer solutions. In the absence of external forces, the stress tensor ($\bm{\sigma}^*$) (non-dimensionalized by $n_pk_BT$, where $n_p$ is the number of polymer chains per unit volume), for a multi-chain system, can be shown to be~\citep{Stoltzthesis}
\begin{equation}\label{DyEq:stresstensor}
\bm{\sigma}^* = \frac{\bm{\sigma}}{n_pk_BT} = \frac{1}{N_c}\left[\sum\limits_{\mu\,=\,1}^{N}\sum\limits_{\nu\,=\,1}^{N}\,\langle \mathbf{r}_{\mu\nu}\,\mathbf{F}_{\mu\nu}^{\text{SDK}}\rangle + \sum\limits^{N_c}\sum\limits_{\nu\,=\,1}^{N_b-1}\,\langle \mathbf{Q}_{\nu}\,\mathbf{F}^c(\mathbf{Q}_{\nu})\rangle\right]
\end{equation}
Here, $N_c$ is the total number of chains and $N = N_c\times N_b$ is the total number of beads in the system. In the above equation, the first summation represents all the excluded volume and associative interactions between backbone monomers and sticker monomers, where $\mathbf{r}_{\mu\nu} = \mathbf{r}_{\mu}-\mathbf{r}_{\nu}$ is the vector between bead $\mu$ and $\nu$ and $\mathbf{F}_{\mu\nu}^{\text{SDK}}$ is the force acting between the beads due to SDK potential. The second term represents the contribution from the spring force, $\mathbf{F}^c(\mathbf{Q}_{\nu})$, due to the connector vector $\mathbf{Q}_{\nu} = \mathbf{r}_{\nu+1}-\mathbf{r}_{\nu}$ between two adjacent beads along the backbone of a polymer chain.

Once the stress tensor is computed, we can easily estimate various dynamic properties and material functions for the polymer solutions. Here, we focus on the calculation of dynamic and linear viscoelastic properties in terms of relaxation modulus, shear viscosity and dynamic moduli. While these properties are typically computed from non-equilibrium simulations or experiments, there are sophisticated techniques for their calculation based on equilibrium simulations~\cite{Bird1987v1,LeeKremer,MoursWinter,WittmerMolPy}. We discuss some of these techniques in the subsequent sections.

\subsection{\label{sec:RelaxMod}Relaxation modulus}

\begin{figure}[b]
    \centering{\includegraphics*[width=12cm,height=!]{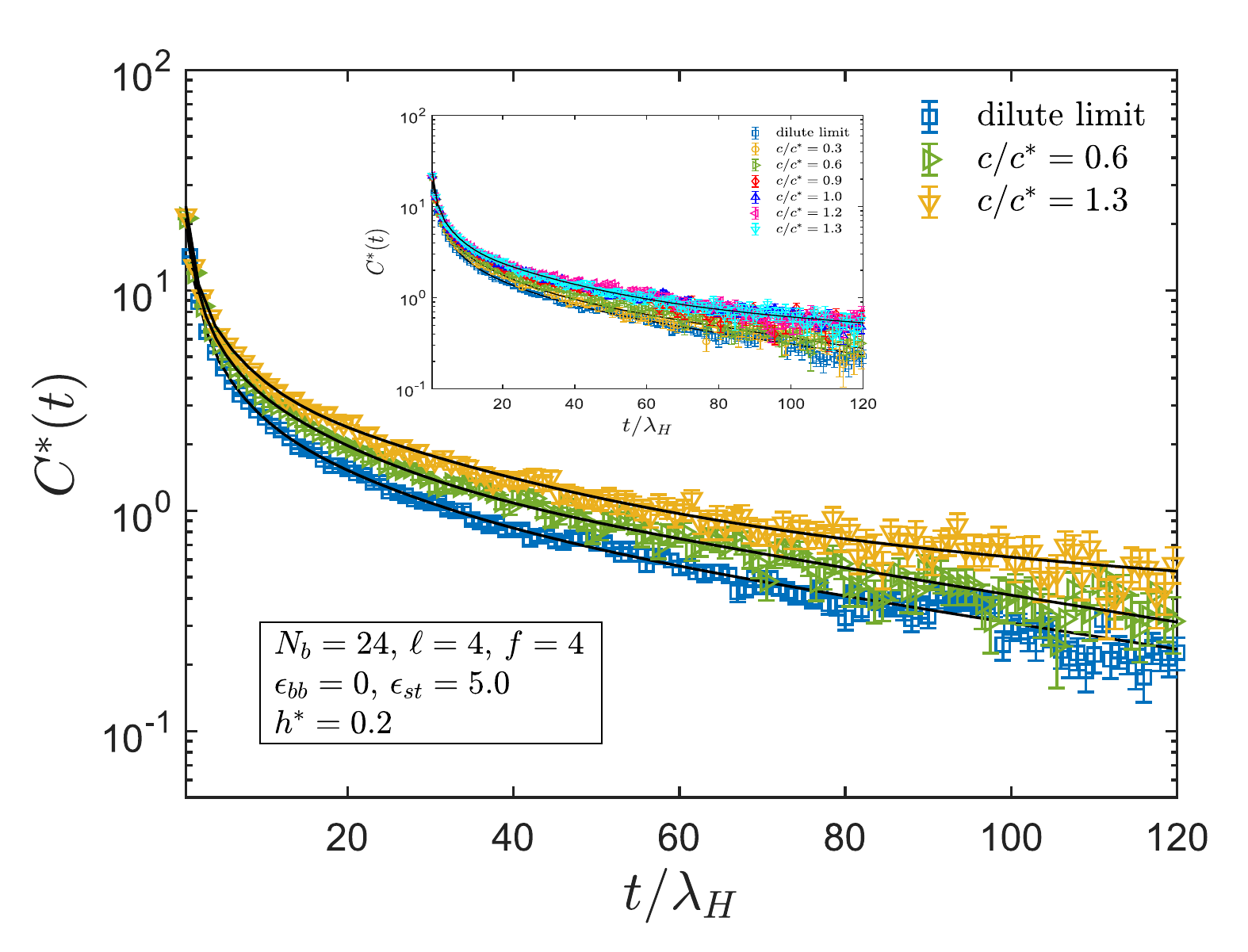}}
 \vskip-15pt
 \caption{Non-dimensionalized stress auto-correlation function, $C^*(t)$, for systems of associative polymers with $N_b=24$, $\ell=4$, $f$, $\epsilon_{bb}=0$ and $\epsilon_{st}=5.0$ at different values of scaled monomer concentration, $c/c^*$, in the dilute and semi-dilute regimes. The solid black lines are fit to the simulation data using a sum of exponentials. The inset shows the entire range of $c/c^*$ covered in the study.}
\label{fig:stressauto_expfit}
\vspace{-10pt}
\end{figure}

The relaxation modulus, $G(t)$, is related to the stress auto-correlation function, $C(t)$, by the following equations,
\begin{eqnarray}
G_{ij}(t) & = G_{\text{eq}} + C_{ij}(t) \label{DyEq:G(t)} \\
C_{ij}(t) &= \dfrac{V}{k_BT}\langle\sigma_{ij}(0)\,\sigma_{ij}(t)\rangle \label{DyEq:C(t)}
\end{eqnarray}
where $G_{ij}(t)$ and $C_{ij}(t)$ are the $ij^{\text{th}}$ component of the relaxation modulus and stress auto-correlation function, respectively, and $V$ is the total volume. The stress-autocorrelation function, $C(t)$, can be easily computed from equilibrium simulations using Eq.~(\ref{DyEq:C(t)}). $G_{\text{eq}}$ is the equilibrium modulus which takes a non-zero value for systems having an infinite relaxation time~\cite{WittmerMolPy,WittmerPRE}. For systems where $G_{\text{eq}}$ is not zero, the general protocol is to estimate the equilibrium modulus by doing actual step strain experiments or simulations and allowing the system to relax. However, in the limit of $G_{\text{eq}}=0$, the relaxation modulus is equal to the stress auto-correlation function. In such a case the relaxation modulus is exactly calculated from stress auto-correlation function. Additionally, the stress auto-correlation function may also be used for the determination of relaxation time scale, as explained in the subsequent section. Since, at equilibrium the stress-tensor, $\bm{\sigma}$, is isotropic, the relaxation modulus and stress auto-correlation function can be expressed as
\begin{align}
G(t) &= \frac{1}{3}(G_{xy}(t)+G_{xz}(t)+G_{yz}(t)) \\
C(t) &= \frac{1}{3}(C_{xy}(t)+C_{xz}(t)+C_{yz}(t))
\end{align}  
Finally, we define a non-dimensionalized relaxation modulus and stress auto-correlation function given by,
\begin{align}
\tilde{G}(t) &= \frac{G(t)}{n_pk_BT} \\
C^*(t) &= \frac{C(t)}{n_pk_BT}
\end{align}

In this work we have evaluated the stress auto-correlation function for a range of scaled monomer concentration, $c/c^*$, in the dilute and semi-dilute regimes as discussed in the main text. Here, we present the results obtained with chain length $N_b=24$, spacer length, $\ell=4$, $\epsilon_{bb}=0$ and $\epsilon_{st}=5.0$. As shown in Fig.~\ref{fig:stressauto_expfit}, the decay in the stress auto-correlation function, obtained from simulations, is generally fitted with a sum of exponentials~\cite{Sharad2014}, as given below, and all the subsequent calculations are carried out using the fit.
\begin{equation}
 C^*(t) = \sum\limits_{i=1}^n a_i\,\exp(b_i t)
\end{equation}      
Here, $a_i$ and $b_i$ are the fitting parameters and $n$ is the number of exponentials used to fit the auto-correlation function. All the stress auto-correlation functions evaluated here are typically fitted with $6$ to $7$ exponentials. From this fit, we can estimate the longest relaxation time by the relation, $\tau_{max} = -\max\lbrace 1/b_i\rbrace$. Several other methods to estimate the terminal and characteristic relaxation times are discussed further in subsequent sections. In the following sections we present the linear viscoselastic properties, such as zero-shear rate viscosity and dynamic moduli, obtained from the stress auto-correlation function and their scaling with concentration which has been used to characterize the dynamic signatures of gelation for weak stickers.

\begin{figure*}
    \includegraphics[width=\textwidth]{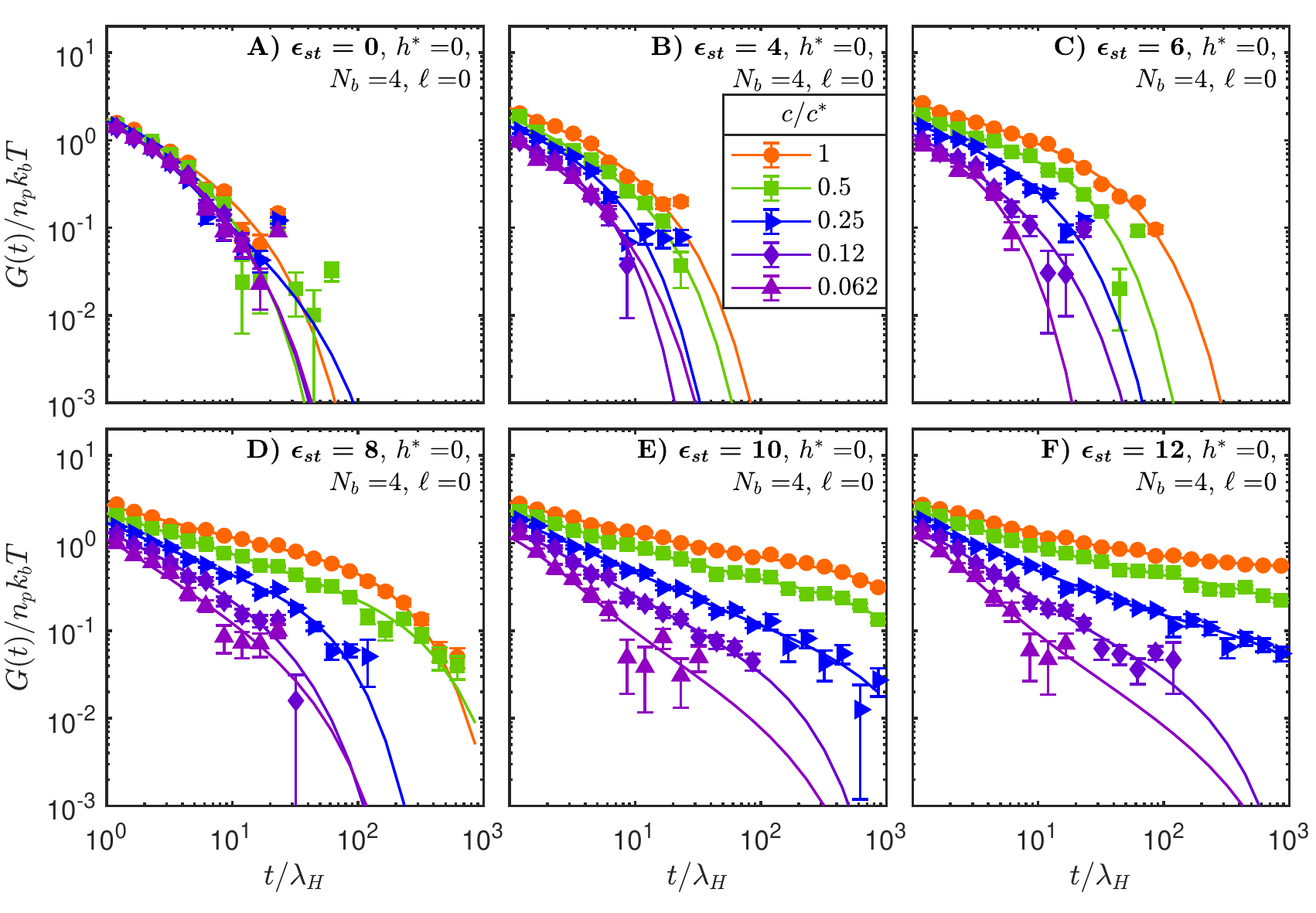}
    \caption{All $G(t)$ data for $\ell=0$, $h^*=0$.}
    \label{fig:Goft_Nb4_h0}
    \vspace{-10pt}
\end{figure*}

Alternatively, by using non-equilibrium simulations with shear schedule involving a step-strain of magnitude $\gamma$, we can measure $G(t)$ directly as
\begin{equation}
G(t) = \frac{1}{3\gamma}(\sigma_\mathrm{xy}(t)+\sigma_\mathrm{xz}(t)+\sigma_\mathrm{yz}(t)),
\end{equation}
where $t=0$ is defined as the end of the step. This protocol is more complicated to implement, but captures any equilibrium modulus. Results presented in this work applied a step strain $\gamma=0.2$, with strain rate $\dot{\gamma}=10^6$. Initial post-step stress was found to depend linearly on the strain at this and higher $\gamma$, but at higher $\gamma$ some associated pairs were broken apart during the strain. Figures \ref{fig:Goft_Nb4_h0}-\ref{fig:Goft_Nb24_h0} show comprehensive plots of the shear relaxation modulus $G(t)$ measured by this method under various conditions simulated in this work.

\begin{figure*}
    \includegraphics[width=0.95\textwidth]{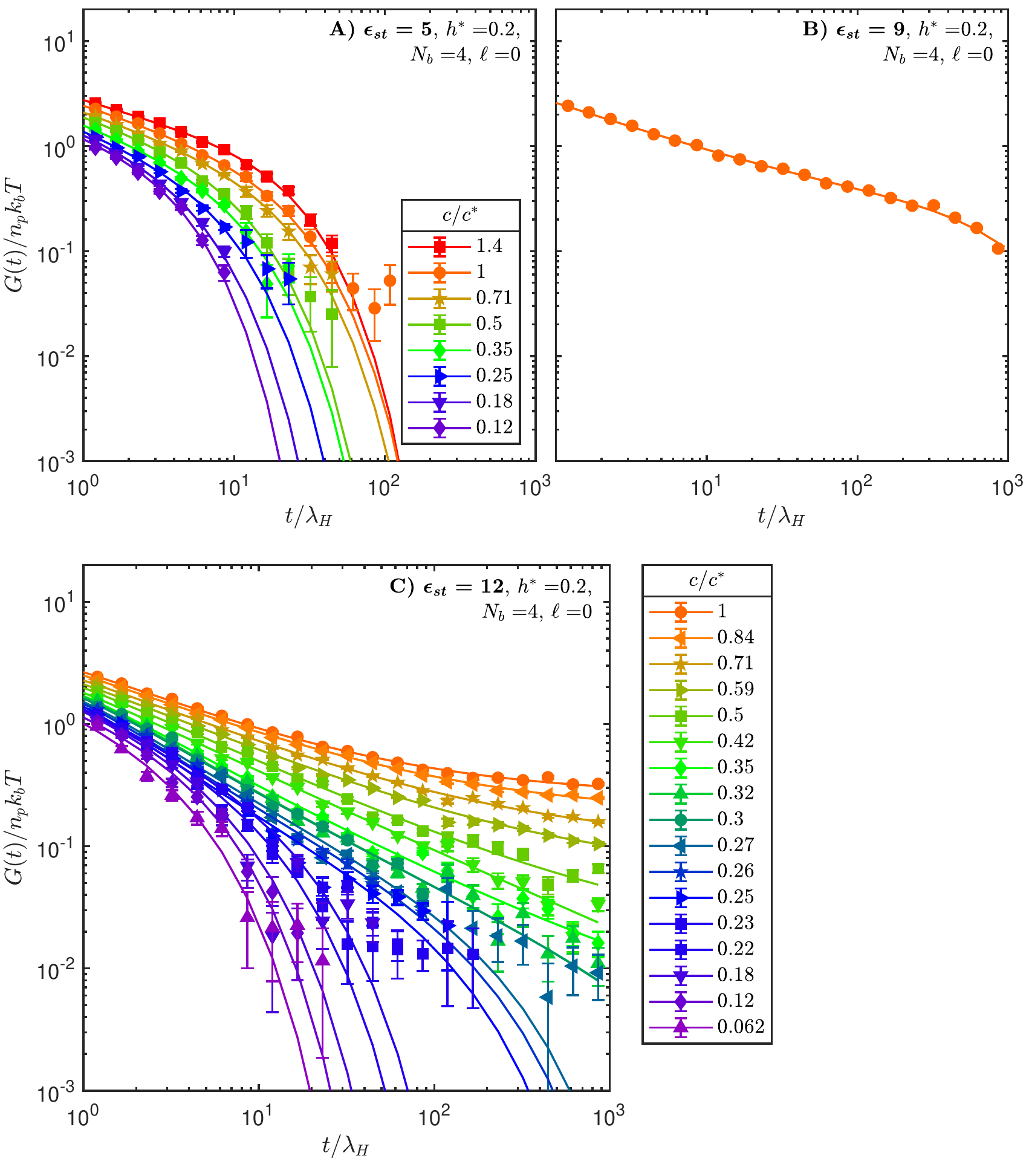}
    \caption{All $G(t)$ data for $\ell=0$, $h^*=0.2$. The concentration in (B) is $c/c^*=1$.}
    \vspace{-20pt}
    \end{figure*}

\begin{figure*}
    \includegraphics[width=0.95\textwidth]{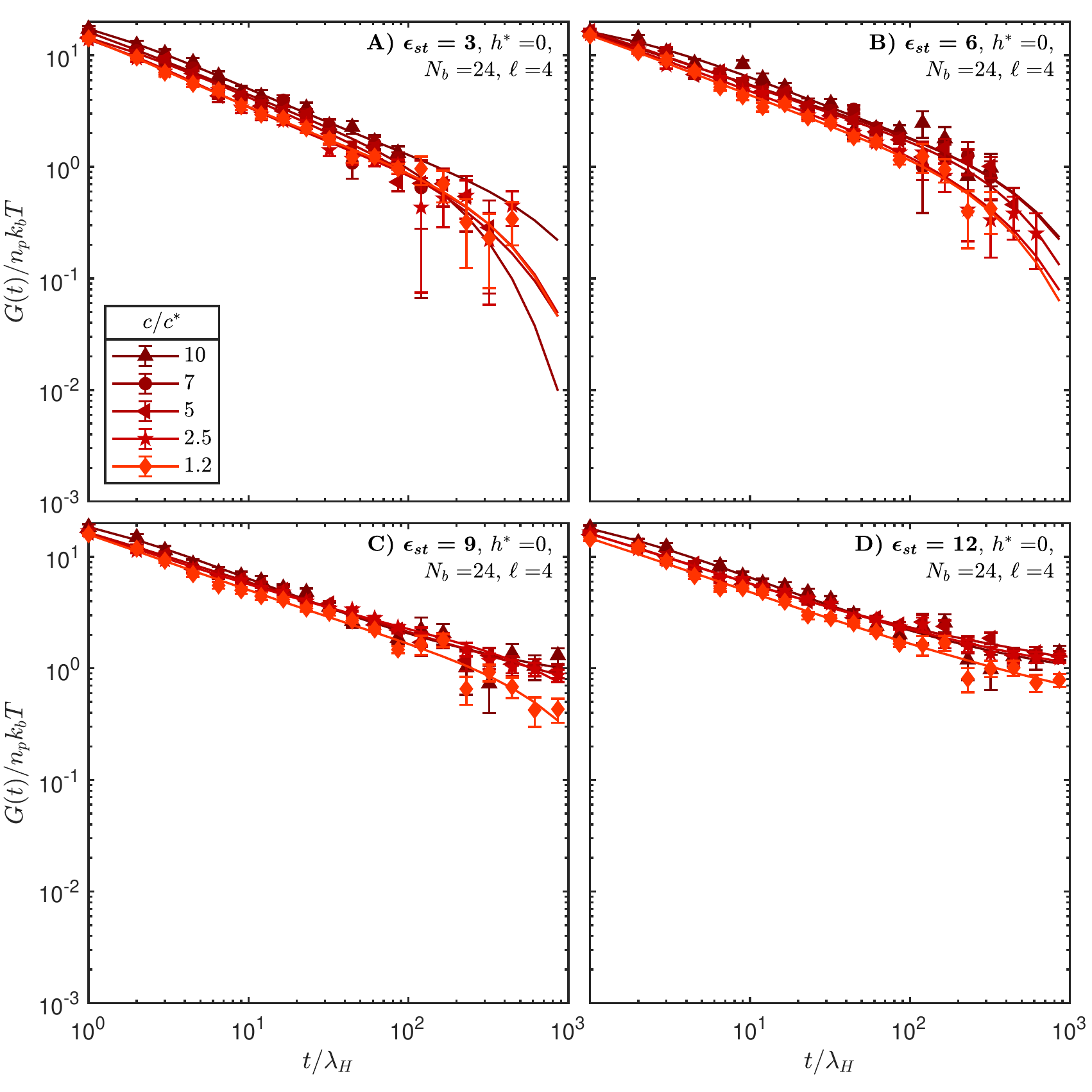}
    \caption{All $G(t)$ data using the HOOMD code for $\ell=4$, $h^*=0$.}
    \label{fig:Goft_Nb24_h0}
    \vspace{-23pt}
\end{figure*}

\subsection{\label{sec:eta_p0}Zero-shear rate viscosity}
An important viscometric function to understand the dynamics and viscoelasticity of polymer solutions is the polymeric component of shear viscosity, $\eta_p$, calculated from shear flow experiments or simulations. For a planar shear flow, the velocity gradient tensor, $\bm{\kappa} = (\bm{\nabla v})^{T}$, presented in Eq.~(4) in the main text, is defined as,
\begin{gather}
 \bm{\kappa}
 =
  \begin{bmatrix}
   0 & \dot{\gamma} & 0 \\
   0 & 0 & 0 \\
   0 & 0 & 0 
   \end{bmatrix}
\end{gather} 
Here $\dot{\gamma}$ is a constant shear rate. For such a system, the shear viscosity, $\eta_p$, is computed as
\begin{equation}
\eta_p = -\frac{\sigma_{xy}}{\dot{\gamma}}
\end{equation}
While the study of shear viscosity at moderately high shear rates are important in non-linear rheology, for linear viscoelasticity the focus is on polymeric component of the zero-shear rate viscosity, defined as $\eta_p^0 = \lim_{\dot{\gamma}\rightarrow 0} \eta_p$, which is typically estimated by measuring the shear viscosity of the polymer solution subjected to multiple small values of shear rate, followed by extrapolation to the zero-shear rate limit. Alternatively, for $G_{\text{eq}} = 0$, $\eta_p^0$ can be calculated from equilibrium simulations, by an application of the Green-Kubo relation~\cite{Fixman1981,Sharad2014,LeeKremer} to the stress auto-correlation function, as shown below,
\begin{equation}\label{DyEq:eta0}
\eta_p^{0^*} = \frac{\eta_p^0}{n_pk_BT\lambda_H} = \int_0^{\infty}C^*(t)\,dt
\end{equation} 
In the above equation, $\eta_p^{0^*}$ is the zero-shear rate viscosity non-dimensionalized by $n_pk_BT\lambda_H$, where $n_p$ is the number of polymer chains per unit volume and $\lambda_H=\zeta/4H$ is the typical time unit for Brownian dynamics simulations. 

\begin{figure}[tbh]
 \begin{center}
   \resizebox{11cm}{!}{\includegraphics*[width=4cm]{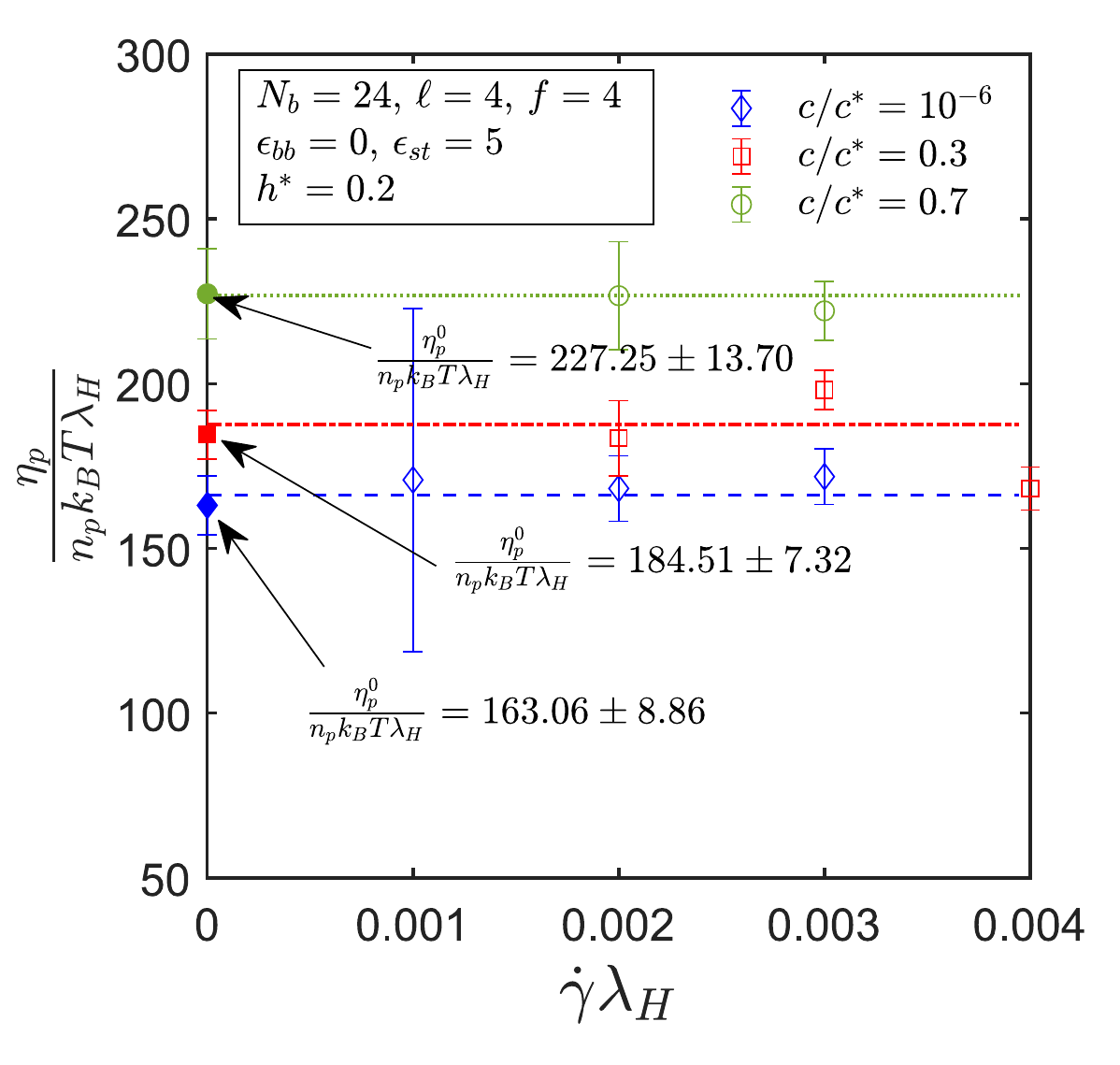}}
 \end{center}
 \vskip-15pt
 \caption{\small{Polymeric component of the non-dimensionalized shear viscosity, $\eta^*_p = \displaystyle\frac{\eta_p}{n_pk_BT\lambda_H}$, as a function of non-dimensional shear rate ($\dot{\gamma}\,\lambda_H$) for different values of monomer concentration, $c/c^*$. The open symbols are the values of $\eta^*_p$ calculated from shear flow simulations and filled symbols are the values of zero-shear rate viscosity, $\eta_p^{0^*}$, evaluated as the integral of $G(t)$}}
\label{fig:eta_v_gam}
\end{figure}

A more direct method of estimating $\eta_p^{0^*}$ is by performing shear flow simulations under very low shear rates and then extrapolating the value of the shear viscosity to the limit of zero-shear rate. Fig.~\ref{fig:eta_v_gam} presents the non-dimensional shear viscosity, $\eta^*_p$, as a function of dimensionless shear rate, $\dot{\gamma}\,\lambda_H$ (in a range of very small values of shear rate), at different $c/c^*$, where the extrapolated values of $\eta^*_p$ in the limit of $\dot{\gamma}\,\lambda_H \rightarrow 0$ are in good agreement with the zero-shear rate viscosity computed from Eq.~(\ref{DyEq:eta0}), which are represented by the filled symbols. By considering these few representative concentrations for the systems of associative polymer solutions, we have illustrated that both the methods produce the same zero-shear rate viscosity, which implies that the equilibrium modulus, $G_{\text{eq}}$, is essentially 0. 

\begin{figure}[b]
 \begin{center}
   \resizebox{13cm}{!}{\includegraphics*[width=4cm]{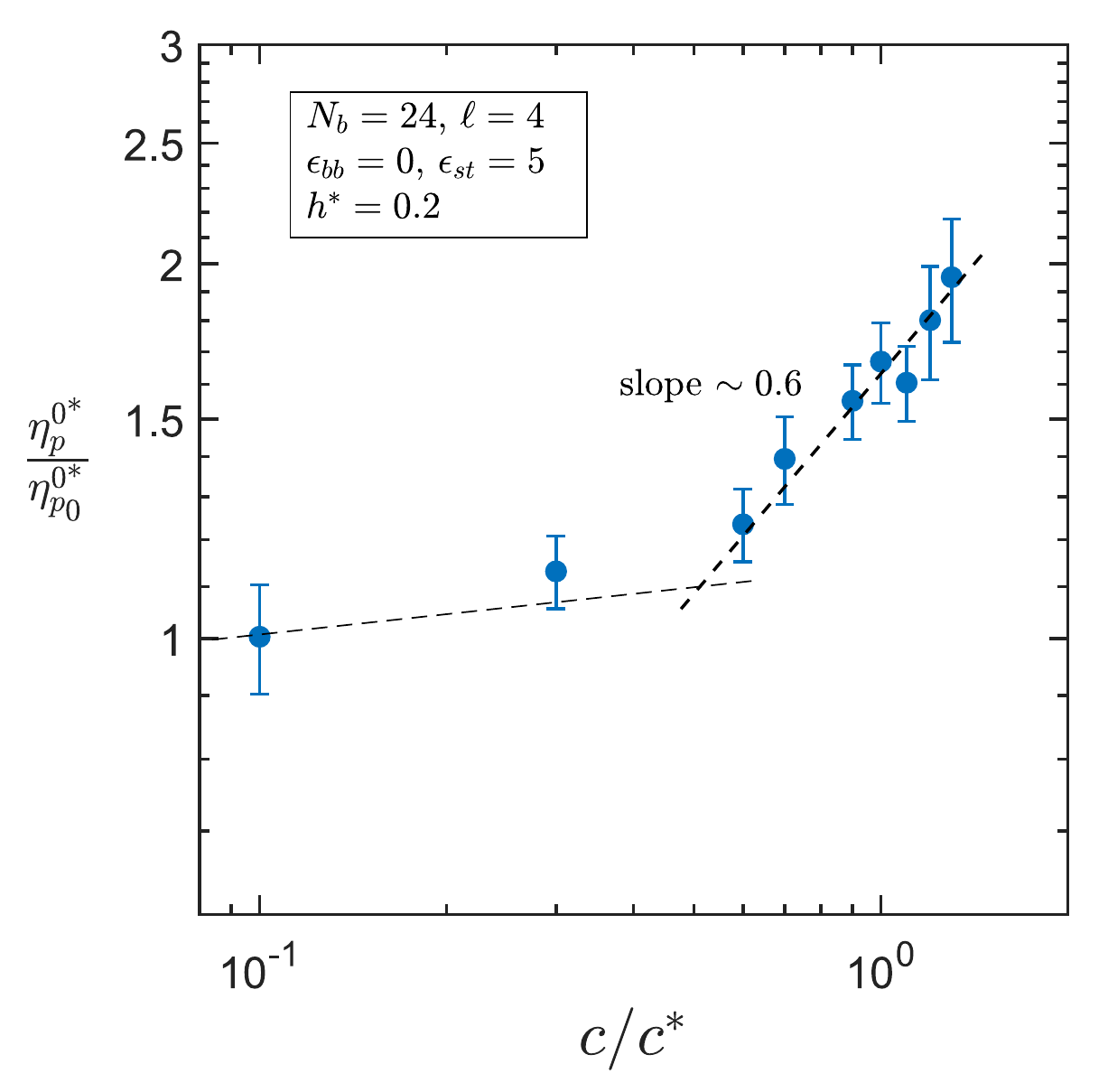}}
 \end{center}
 \vspace{-20pt}
 \caption{\small{Scaling of the ratio of zero-shear rate viscosity at finite concentration, $\eta_p^{0^*}$, to its value in the dilute limit, $\eta_{p_0}^{0^*}$, with the scaled concentration, $c/c^*$, for solutions of associative polymers with $N_b=24$, $\ell=4$, $f=4$, $\epsilon_{bb}=0$ and $\epsilon_{st}=5.0$. The symbols from simulations and the dashed lines with slope $0.6$ is obtained from least square fit.}}
\label{fig:eta0_v_c}
\end{figure}

Using the Green-Kubo relation to estimate the zero-shear rate viscosity, $\eta_p^{0^*}$, we now investigate the scaling of $\eta_p^{0^*}$ with concentration, $c/c^*$. In Fig.~\ref{fig:eta0_v_c}, the ratio of zero-shear rate viscosity at finite concentration, $\eta_p^{0^*}$, to its value in the dilute limit, $\eta_{p_0}^{0^*}$, is plotted against the scaled monomer concentration, $c/c^*$, for associative polymer solutions with chain length, $N_b=24$, spacer length $\ell=4$, number of stickers per chain $f=4$, $\epsilon_{bb}=0$ and $\epsilon_{st}=5.0$. The scaling of the normalised zero-shear rate viscosity, $\eta_p^{0^*}/\eta_{p_0}^{0^*}$, with scaled concentration indicates a cross-over at $c/c^*\approx 0.5$, beyond which the scaling exponent takes an asymptotic value of $0.6$. This cross-over scaling behaviour is identical to the scaling of the longest relaxation time for weak stickers, as shown in the main text. This is not surprising since an alternative method to estimate the large scale relaxation time ($\lambda_{\eta}$) is based on the polymeric contribution of the zero-shear rate viscosity ($\eta_p^0$), which is given by the following expression~\citep{Ottinger1996}
\begin{equation}
\lambda_{\eta} = \frac{M\,\eta_p^0}{cN_Ak_BT}
\end{equation}
where $M$ is the molecular weight of the polymers, $c$ is the monomer concentration. It noteworthy that normalising $\lambda_{\eta}$ at a finite concentration with its value in the dilute limit, $\lambda_{\eta}^0$, gives the ratio $\lambda_{\eta}/\lambda_{\eta}^0$, which is equivalent to the ratio $\eta_p^{0^*}/\eta_{p_0}^{0^*}$, used for the scaling of the zero-shear rate viscosity. Thus, the scaling of $\lambda_{\eta}/\lambda_{\eta}^0$ and  $\eta_p^{0^*}/\eta_{p_0}^{0^*}$ with concentration are identical.

\subsection{\label{sec:DyModuli}Dynamic moduli ($G'$ \& $G''$)}
The elastic and viscous response of a viscoelastic fluid is generally characterised by the storage ($G'$) and the loss ($G''$) moduli, which are together referred to as the dynamic moduli. These properties are typically calculated from oscillatory shear flow (OSF) experiments or simulations. In numerical simulations, an OSF is implemented by subjecting the simulation box to an oscillatory shear strain ($\gamma(t)$), using Lees-Edwards boundary conditions~\cite{Lees_1972,JainSasmal2015,Gompper2015}, such that
\begin{align}
\gamma(t) &= \gamma_0\,\sin(\omega\,t)\\
\dot{\gamma}(t) &= \gamma_0\,\omega\,\cos(\omega\,t)
\end{align} 
where $\gamma_0$ is the strain amplitude and $\omega$ is the frequency of oscillation. The response to this oscillatory strain input produces an oscillatory stress component $\sigma_{xy}(t)$, from which $G'$ and $G''$ are extracted as follows,
\begin{eqnarray}
\sigma_{xy}(t) &=& \sigma_0\,\sin(\omega\,t+\delta)\nonumber\\
    &=& \sigma_0\,\cos(\delta)\,\sin(\omega\,t) + \sigma_0\,\sin(\delta)\,\cos(\omega\,t)
\end{eqnarray}
\begin{equation}
G'(\omega) = -\frac{\sigma_0\,\cos(\delta)}{\gamma_0},\quad\quad G''(\omega) = -\frac{\sigma_0\,\sin(\delta)}{\gamma_0}
\end{equation}
Here $\sigma_0$ is the amplitude of the stress response, $\delta$ is the phase difference. It is important to note that $\delta = 0$ for a Newtonian liquid, whereas, $\delta = 90^{\circ}$ for a purely elastic solid. Since $G'$ and $G''$ determine the shear stress that is linearly dependent on the strain, these material functions are also linear viscoelastic properties. In the limit of very small strain amplitude ($\gamma_0 \ll 1$), $G'$ and $G''$ can be estimated by Fourier transformation of the stress autocorrelation function, $C(t)$,~\cite{MoursWinter,WittmerMolPy} as described below,
\begin{align}
G'(\omega) - G_{eq} &= \int\limits_0^{\infty}d(\omega t)C(t)\,\sin(\omega t) \label{DyEq:G'} \\
G''(\omega) &= \int\limits_0^{\infty}d(\omega t)\,C(t)\,\cos(\omega t) \label{DyEq:G''}
\end{align}
Finally, the dynamic moduli may be scaled by $n_pk_BT$ to define the dimensionless storage modulus ($G'^*$) and loss modulus ($G''^*$).

\begin{figure*}[t]
    \centerline{
    \begin{tabular}{c c}
        \includegraphics[width=74mm]{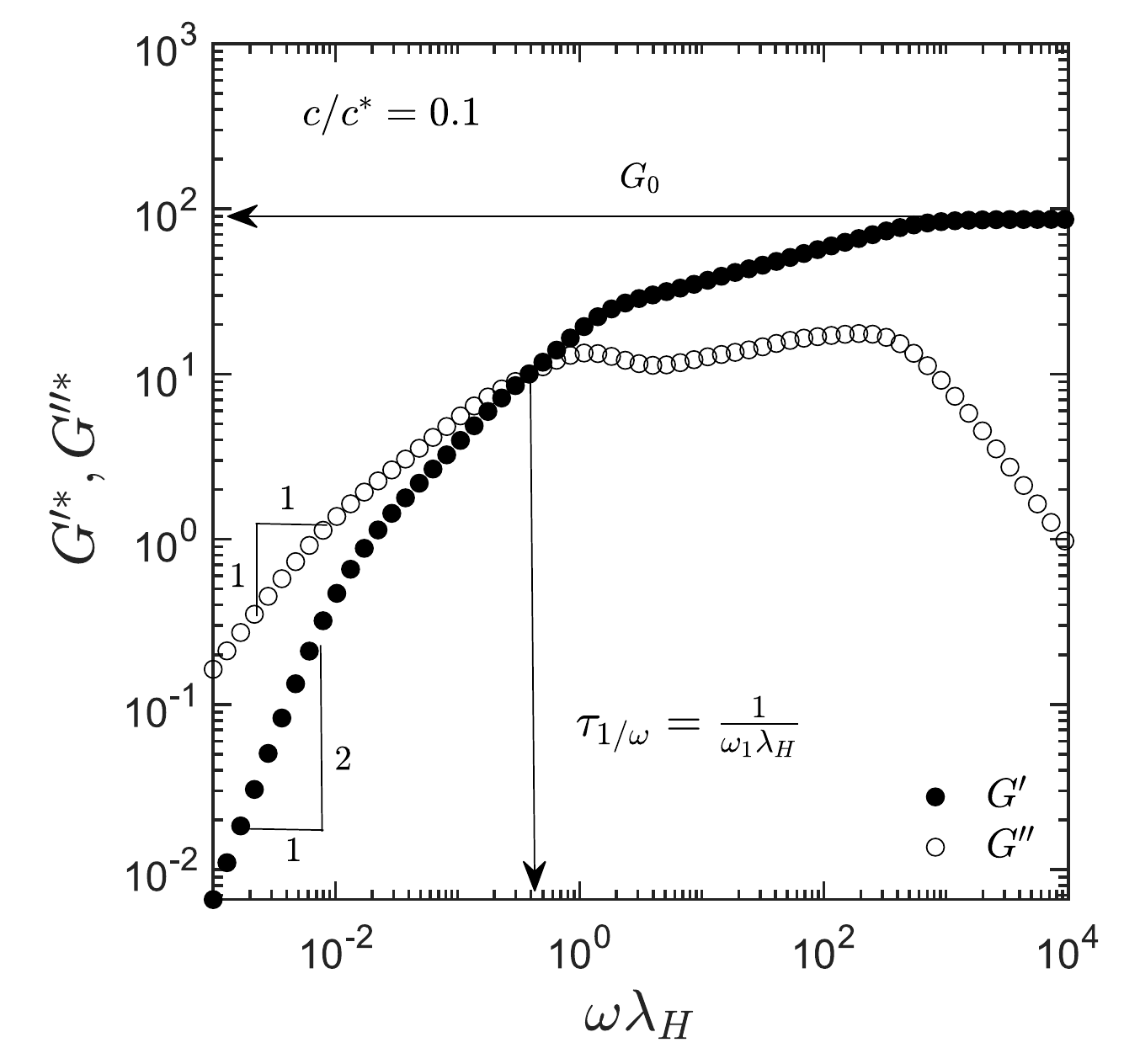} 
        & \includegraphics[width=74mm]{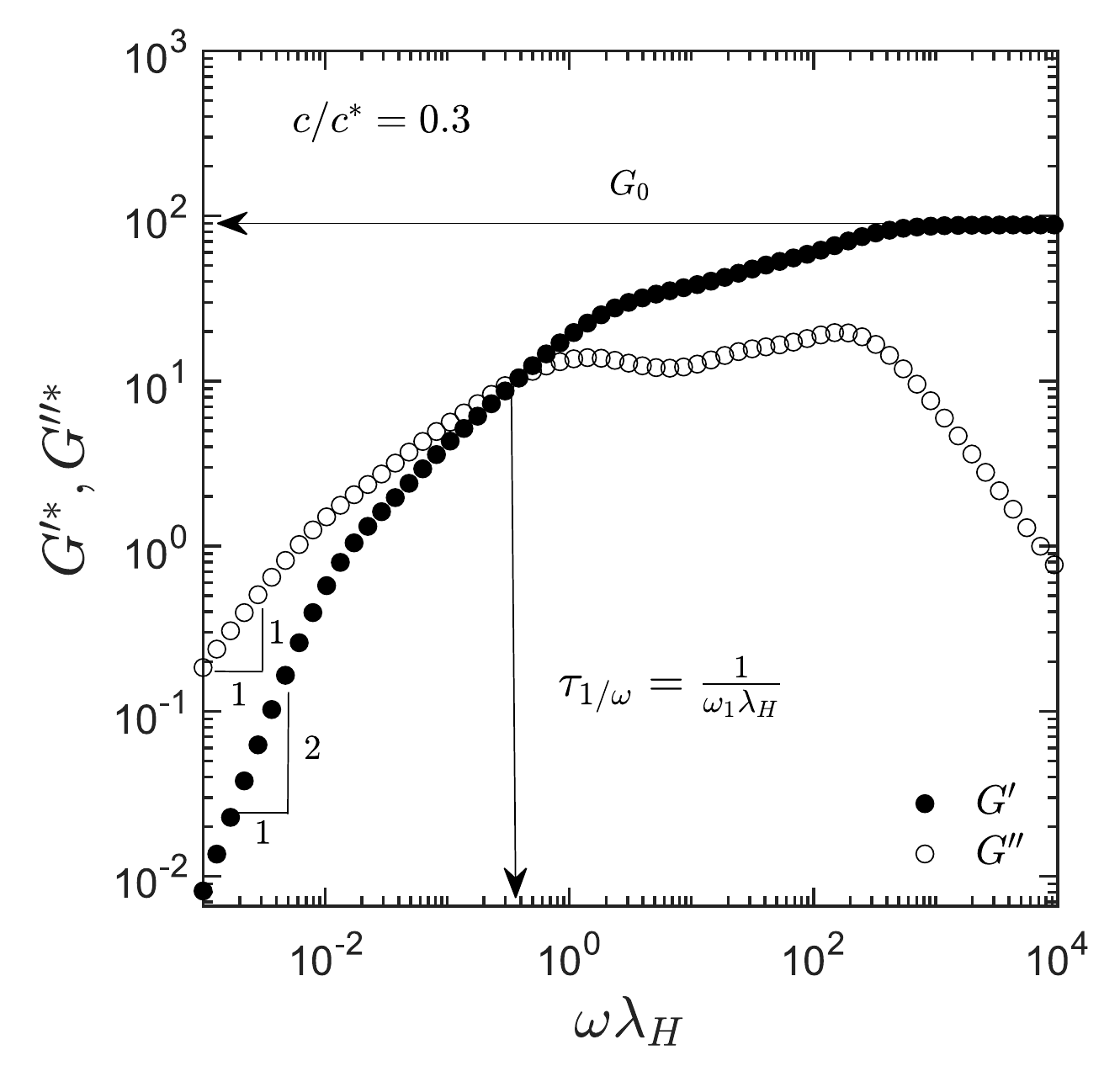} \\
        (a) & (b) \\
        \includegraphics[width=74mm]{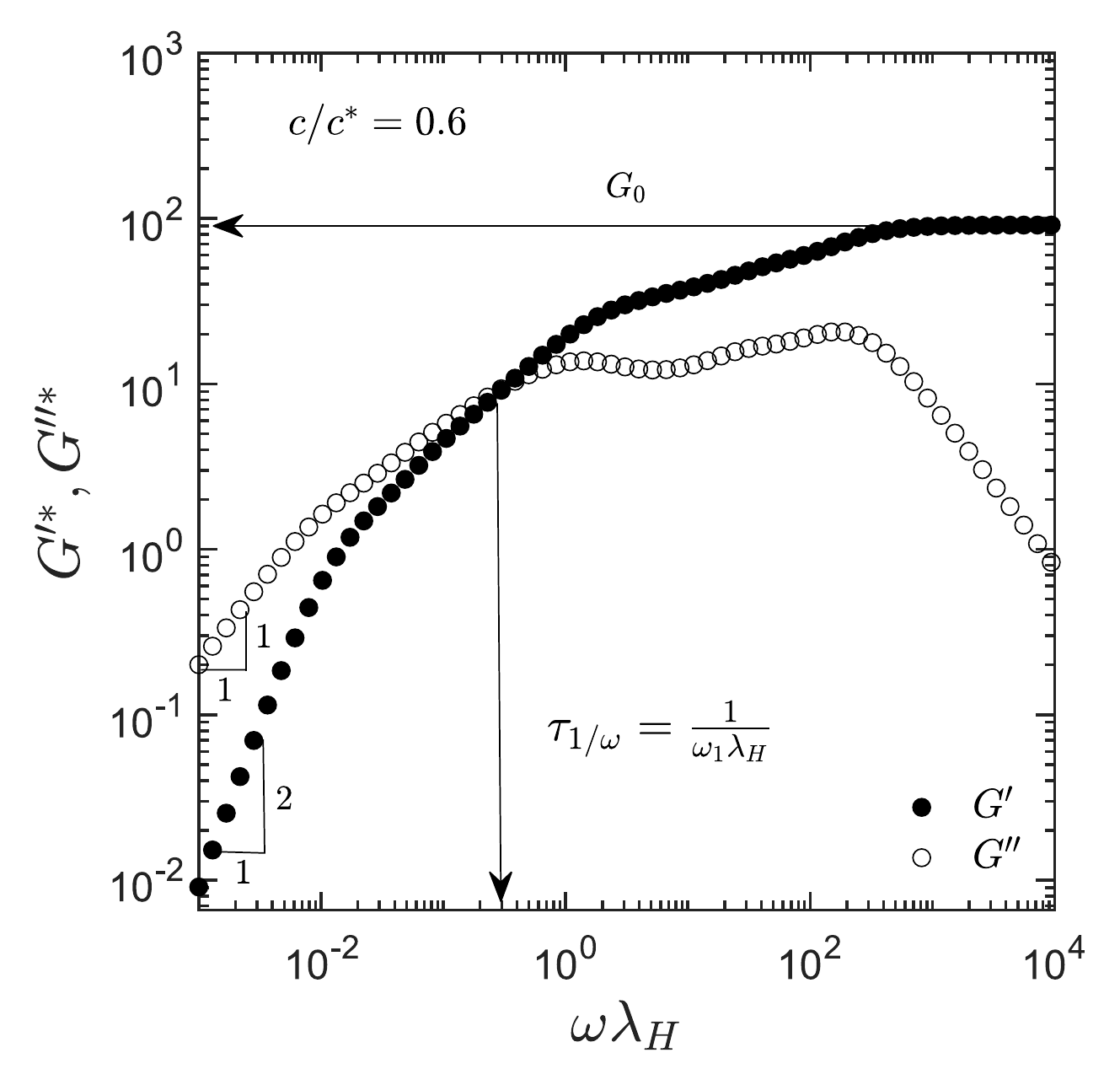} 
        & \includegraphics[width=74mm]{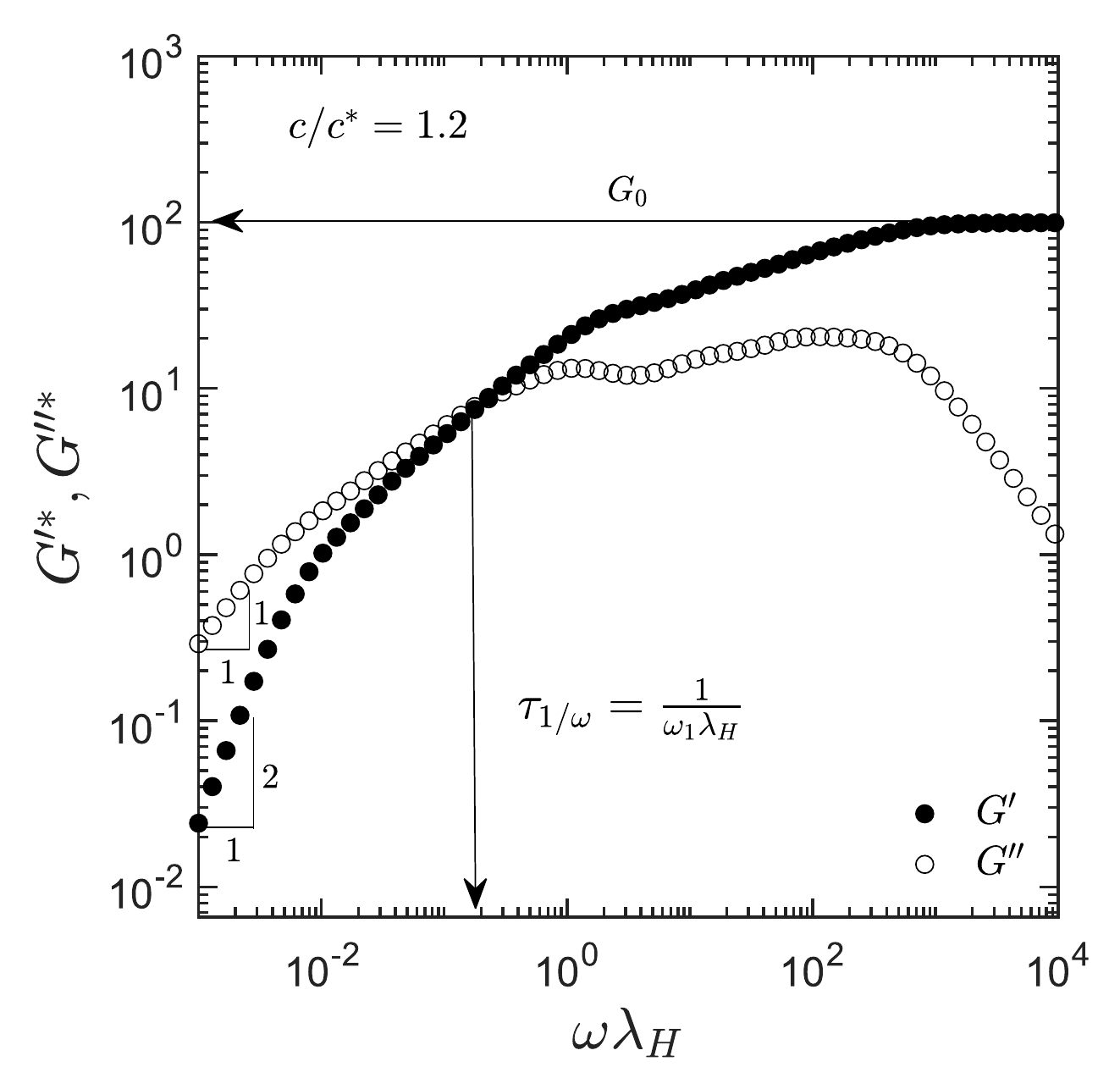} \\
        (c) & (d) \\
    \end{tabular}
    }
   \caption{\small{Non-dimensional storage ($G'^*$) and loss ($G''^*$) moduli of associative polymer solutions as a function of dimensionless frequency, $\omega\,\lambda_H$, for different values of $c/c^*$ with chain length $N_b=24$, spacer length $\ell=4$, number stickers per chain $f=4$, backbone monomer interaction strength $\epsilon_{bb}=0$ and sticker strength $\epsilon_{st}=5.0$. The filled and open circles in the plots are $G'$ and $G''$, respectively. The inverse of the frequency corresponding to the point of intersection of $G'$ and $G''$ curves represents the characteristic relaxation time $\tau_{1/\omega} = 1/(\omega_1\,\lambda_H)$. $G_0$ indicates the elastic modulus. The slopes of $G'^*$ and $G''^*$ in the terminal flow regime are 2 and 1, respectively, as indicated in the figures.} 
\label{fig:DyMod_SP}}
 \vspace{-20pt}
 \end{figure*}

\begin{figure}[t]
 \begin{center}
   \resizebox{12cm}{!}{\includegraphics*[width=4cm]{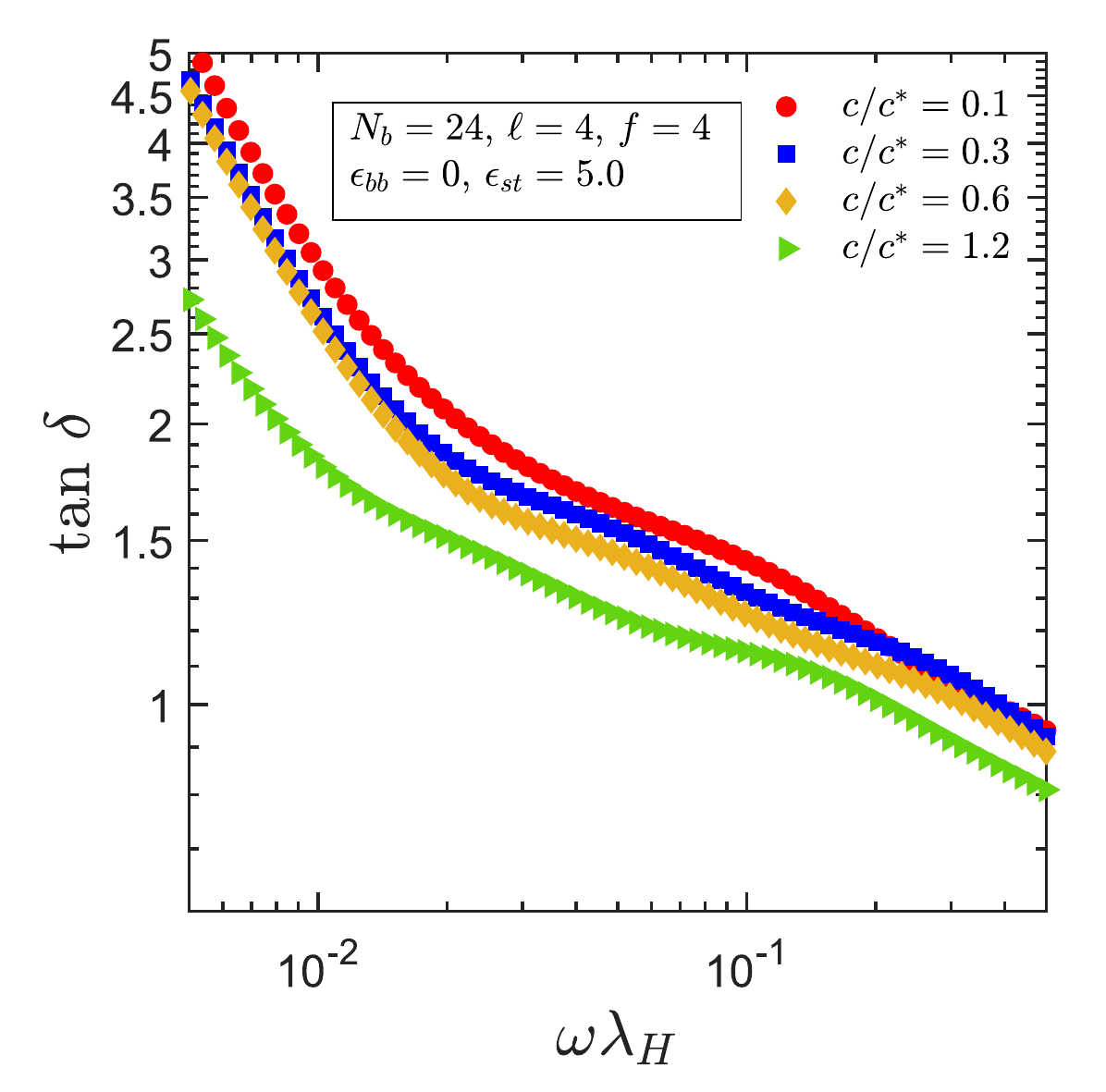}}
 \end{center}
 \vskip-15pt
 \caption{\small{Variation of the loss tangent, $\tan\,\delta$, with the dimensionless frequency, $\omega\,\lambda_H$, at different values of $c/c^*$ for associative polymer solutions with $N_b=24$, $\ell=4$, $f=4$, $\epsilon_{bb}=0$ and $\epsilon_{st}=5.0$.}}
\label{fig:tandelta}
 \vspace{-10pt}
\end{figure}

Using the above formulation, we compute $G'$ and $G''$ from the stress auto-correlation function for associative polymer solutions for a range of concentration, $c/c^*$, and investigate their frequency dependencies in the pre and post-gel regimes. Figs.~\ref{fig:DyMod_SP} display the non-dimensional storage ($G'^*$) and loss ($G''^*$) moduli as a function of frequency for different concentrations, $c/c^*$, with chain length $N_b=24$, spacer length $\ell=4$, backbone monomer interaction strength $\epsilon_{bb}=0$ and sticker strength $\epsilon_{st}=5.0$. The inverse of the frequency corresponding to the point of intersection of $G'^*$ and $G''^*$ is defined as a characteristic relaxation time of the polymer solutions, $\tau_{1/\omega} = 1/(\omega_1\,\lambda_H)$, where $\omega_1$ is the intersection frequency. For each concentration, $c/c^*$, the storage modulus, $G'^*$, in the limit of high frequency, saturates to a value $G_0$, which is identified as the elastic plateau modulus. In addition, the variation of loss tangent, defined as $\tan\,\delta = G''/G'$,  with frequency, $\omega\,\lambda_H$, is presented for different concentrations in Fig.~\ref{fig:tandelta}. For all the concentrations, shown in Figs.~\ref{fig:DyMod_SP}, there exists a unique point of interaction between $G'^*$ and $G''^*$, along with a terminal flow regime in the limit of low frequencies, where $G'\sim \omega^2$ and $G''\sim\omega$~\cite{Bird1987v1,Ruyumbeke2017}. Considering $c_g/c^*=0.5$ as the gel-point, these observations imply the existence of a finite terminal relaxation time in both the pre-gel and post-gel regimes, for associative polymer solutions with weak stickers. These results diverge from the typical power-law behaviour of $G'(\omega)$ and $G''(\omega)$, with no discernible relaxation time scale, as observed for the strong stickers and discussed in the main text. Besides, the loss tangent, $\tan\,\delta$, becomes independent of frequency at the gelation concentration for a typical gel-forming polymer solution, which is not observed for the present systems, as $\tan\,\delta$ decreases with  frequency at all the values of $c/c^*$, as shown in Fig.~\ref{fig:tandelta}.

\begin{figure}[tbh]
 \begin{center}
   \resizebox{12cm}{!}{\includegraphics*[width=4cm]{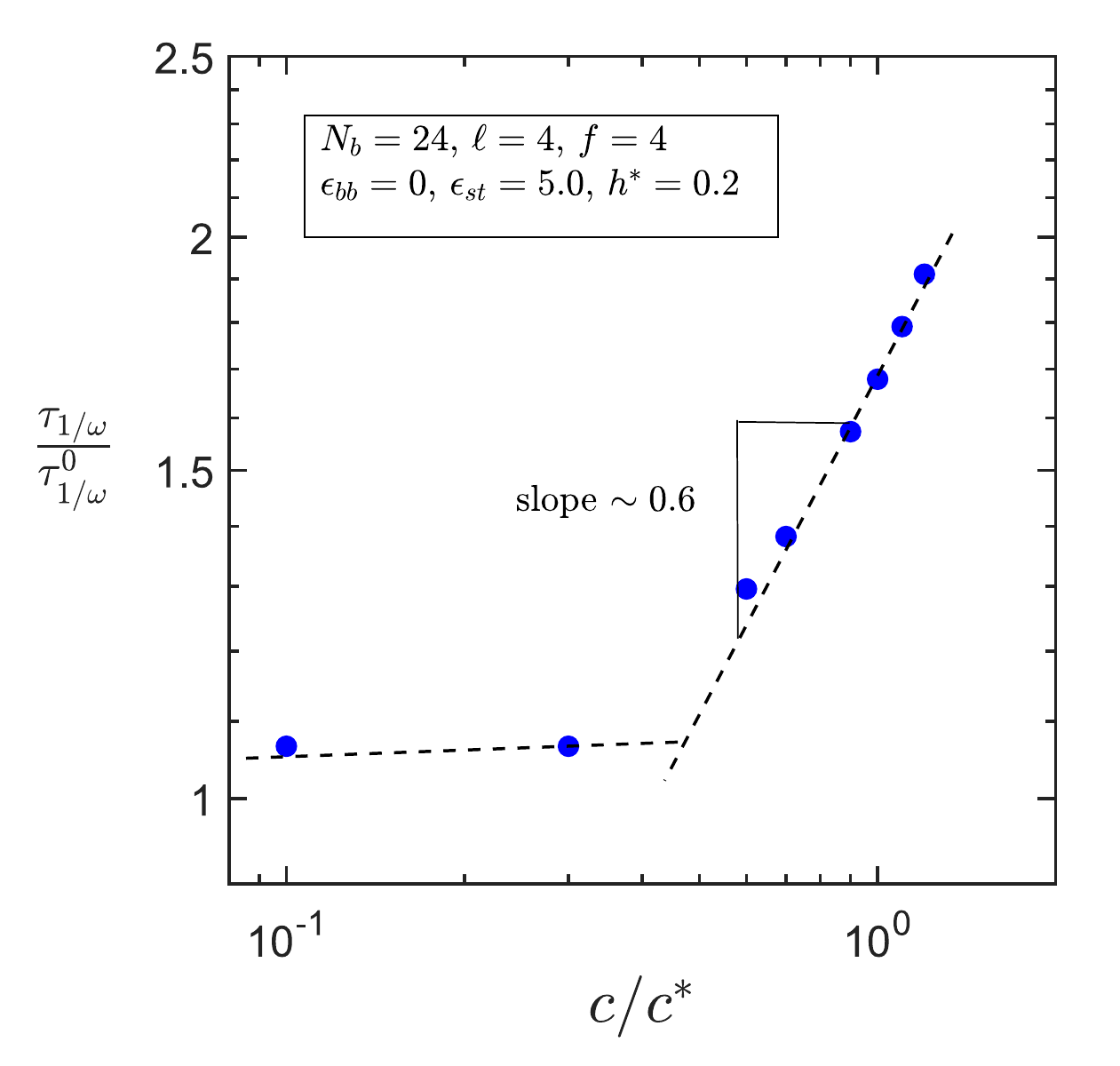}}
 \end{center}
 \vskip-15pt
 \caption{\small{Scaling of the ratio of characteristic relaxation time, $\tau_{1/\omega}$, to its value in the dilute limit, $\tau^0_{1/\omega}$, with the scaled concentration, $c/c^*$, for solutions of associative polymers with $N_b=24$, $\ell=4$, $f=4$, $\epsilon_{bb}=0$ and $\epsilon_{st}=5.0$. The symbols from simulations and the dashed lines with slope $0.6$ is the best fit to the data.}}
\label{fig:invomega_v_c}
 \vspace{-20pt}
\end{figure}

\begin{figure*}[t]
    \centerline{
    \begin{tabular}{c c}
        \includegraphics[width=85mm]{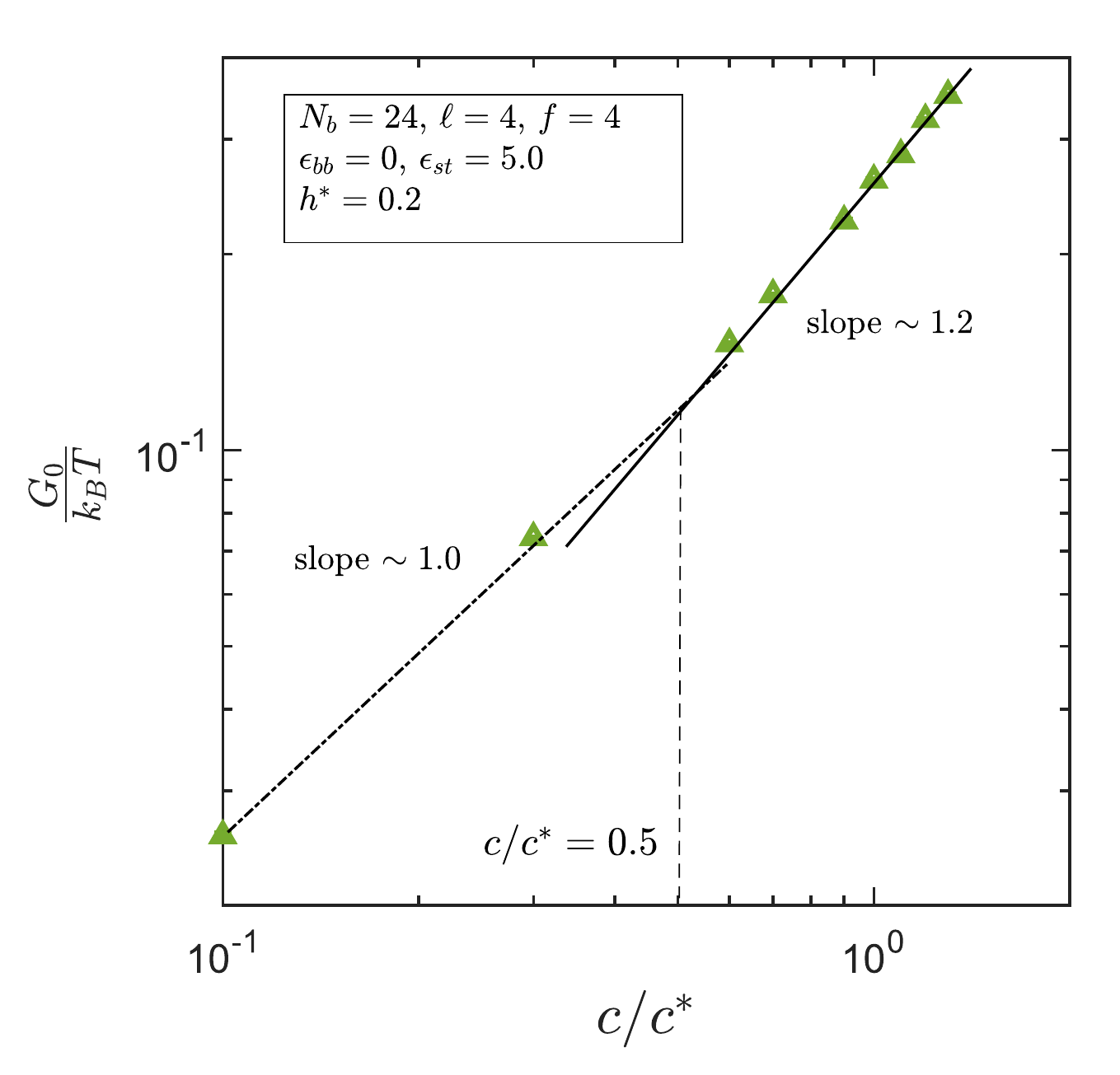} 
        & \includegraphics[width=85mm]{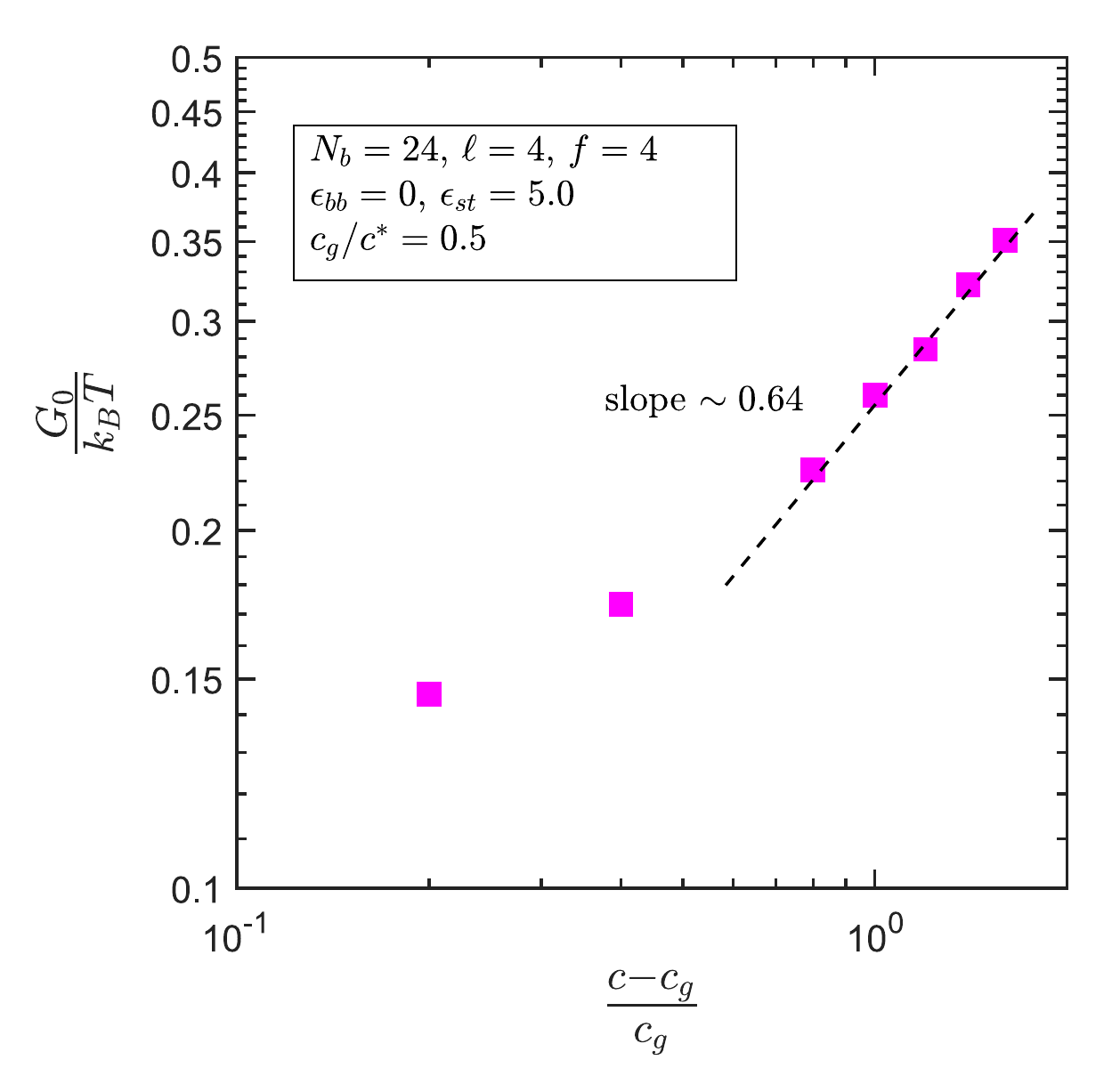} \\
        (a) & (b) \\
    \end{tabular}
    }
   \caption{\small{(a) Scaled elastic modulus, $G_0/k_BT$, versus concentration, $c/c^*$, for the solutions of associative polymers with $N_b=24$, $\ell=4$, $f=4$, $\epsilon_{bb}=0$, $\epsilon_{st}=5.0$. (b) Elastic modulus, $G_0/k_BT$, as a function of the relative distance from the gelation concentration $\Delta = (c-c_g)/c_g$, near the gel-point in the post gel regime, where the gelation concentration is considered to be $c_g/c^*=0.5$. Symbols are the simulation data and the solid and broken lines indicate the corresponding scaling exponents.} 
\label{fig:G0_v_c}}
 \vspace{-10pt}
\end{figure*}

Notably, the characteristic relaxation time, $\tau_{1/\omega}$, normalised with the relaxation time $\tau^0_{1/\omega}$ in the dilute limit, plotted as a function of scaled monomer concentration, $c/c^*$, shown in Fig.~\ref{fig:invomega_v_c}, exhibits a change in the scaling exponent at a value of $c/c^*\approx 0.5$, which is also the concentration ($c_{g_3}/c^*$) at which maxima in the free chain concentration and a crossover in the zero-shear rate viscosity scaling is observed. Surprisingly, the asymptotic value of the exponent following the crossover is $0.6$, which is the same as that for the crossover in the scaling of zero-shear rate viscosity. In the main text, $\tau_{1/\omega}/\tau^0_{1/\omega}$ is used as one of the measurements for the longest relaxation time scale and it is found to follow a universal curve. Additionally, in Fig.~\ref{fig:G0_v_c}~(a) we show a crossover in the scaling of the elastic plateau modulus, $G_0/k_BT$, with monomer concentration, at $c/c^*\approx 0.5$. For the elastic modulus the scaling exponent is found to change from $1$ to $1.2$, following the crossover. According to the prediction of mean-field theory for associative polymer solutions proposed by Rubinstein and Semenov~\cite{RnSdynamics}, the elastic modulus ($G_0$) in the post-gel regime, close to the gel-point, scales with the relative distance from  the gelation concentration, $c_g$, following the relation $G_0\sim \Delta^{3\mu}$, where $\Delta = (c-c_g)/c_g$ and exponent $\mu=0.85$ (i.e. $3\mu=2.55$). However, our simulations (see Fig.~\ref{fig:G0_v_c}~(b)) show a much weaker dependence of $G_0$ on $\Delta$, where the asymptotic value of the exponent in the post-gel regime, near the gel-point $c_g/c^* = 0.5$, is $0.64$. This difference in the scaling of $G_0$ between the simulations and theory may be a consequence of the formation of ``evanescent" gels with weak stickers, where the increase in concentration in the post gel regime does not cause a sharp increase in the elastic modulus.

\begin{figure}[t]
 \begin{center}
   \resizebox{11cm}{!}{\includegraphics*[width=4cm]{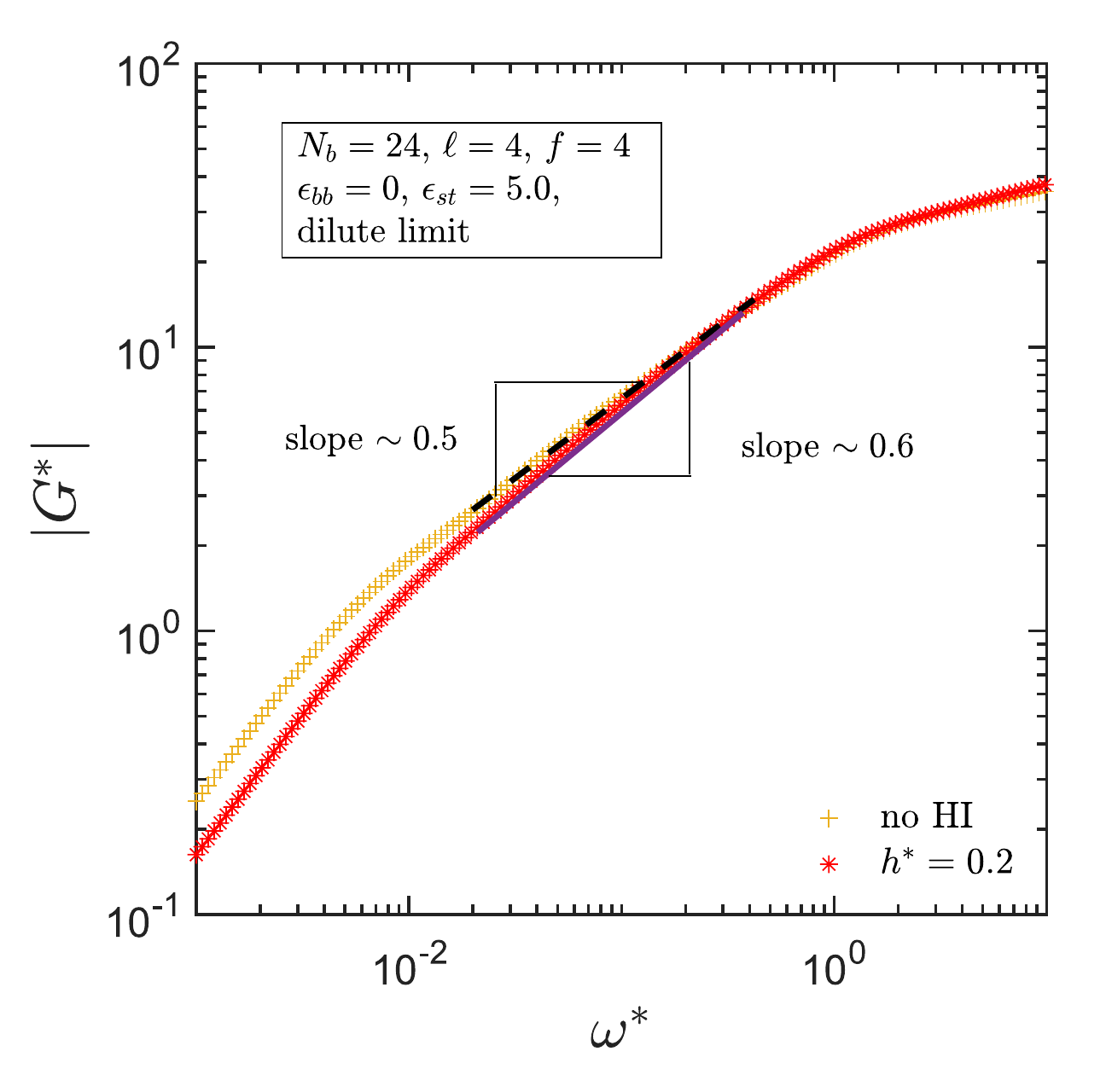}}
 \end{center}
 \vskip-15pt
 \caption{Effect of hydrodynamic interaction on the scaling of complex modulus.}
\label{fig:GstrSP}
 \vspace{-10pt}
\end{figure}
Figure \ref{fig:GstrSP} Demonstrates a fundamental rheological discrepancy between models with and without hydrodynamic interactions. When HI is neglected, there is a modest but non-negligible change in the distribution of relaxation times. This difference is represented here using the magnitude of the complex modulus. Since a modest change in the exponent of a power-law-like region of the dynamic moduli can significantly affect the point of intersection of $G'$ and $G''$, this may account for the separation observed in the main text of $\tau_\omega$ from the other time scales in systems without HI.

\subsection{End-to-end $\hat{R_e}$ auto-correlation}
\label{sec:Re auto}

\begin{figure*}[tbh]
    \centerline{
    \begin{tabular}{c c}
        \includegraphics[width=85mm]{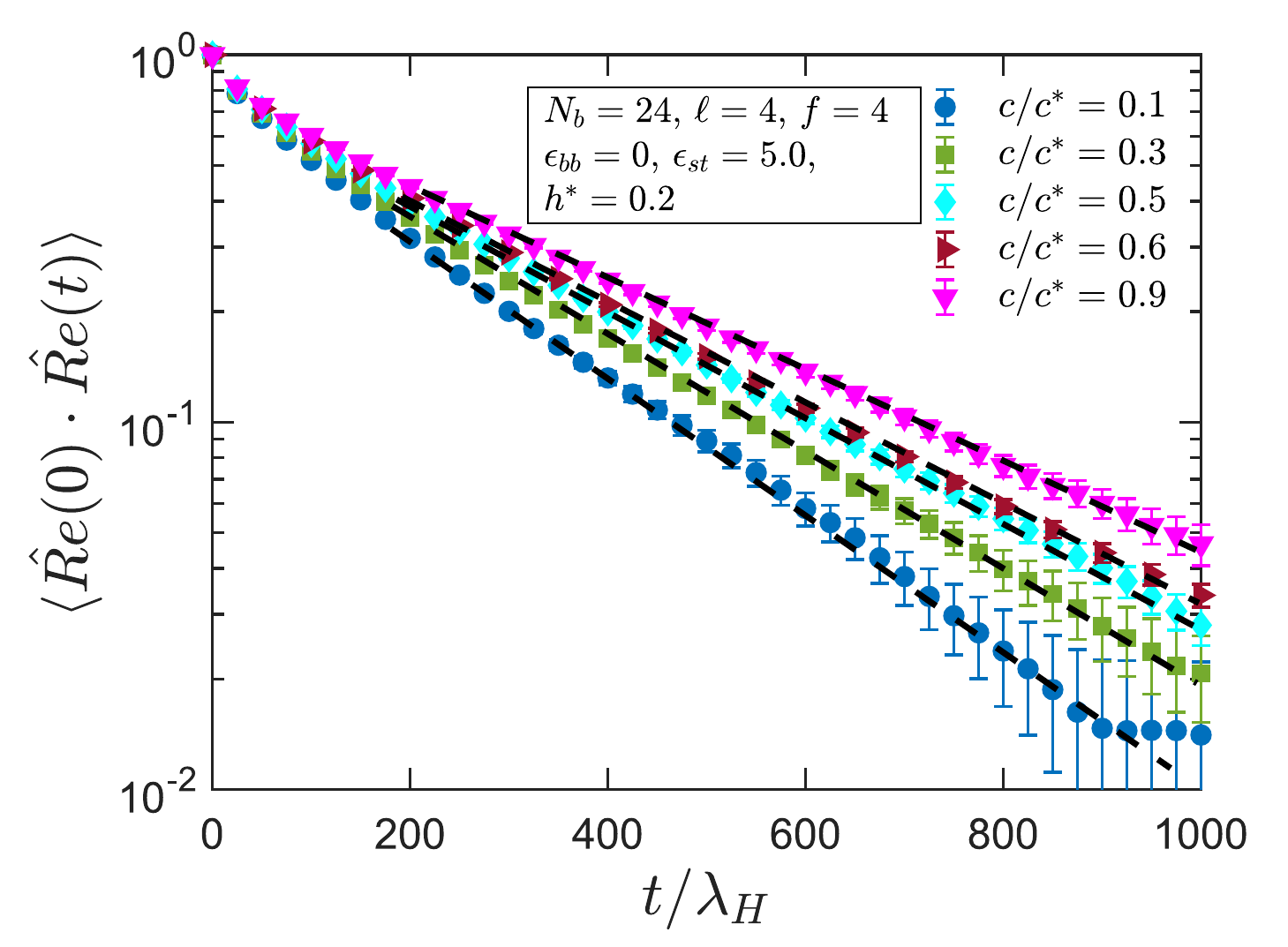} 
        & \includegraphics[width=85mm]{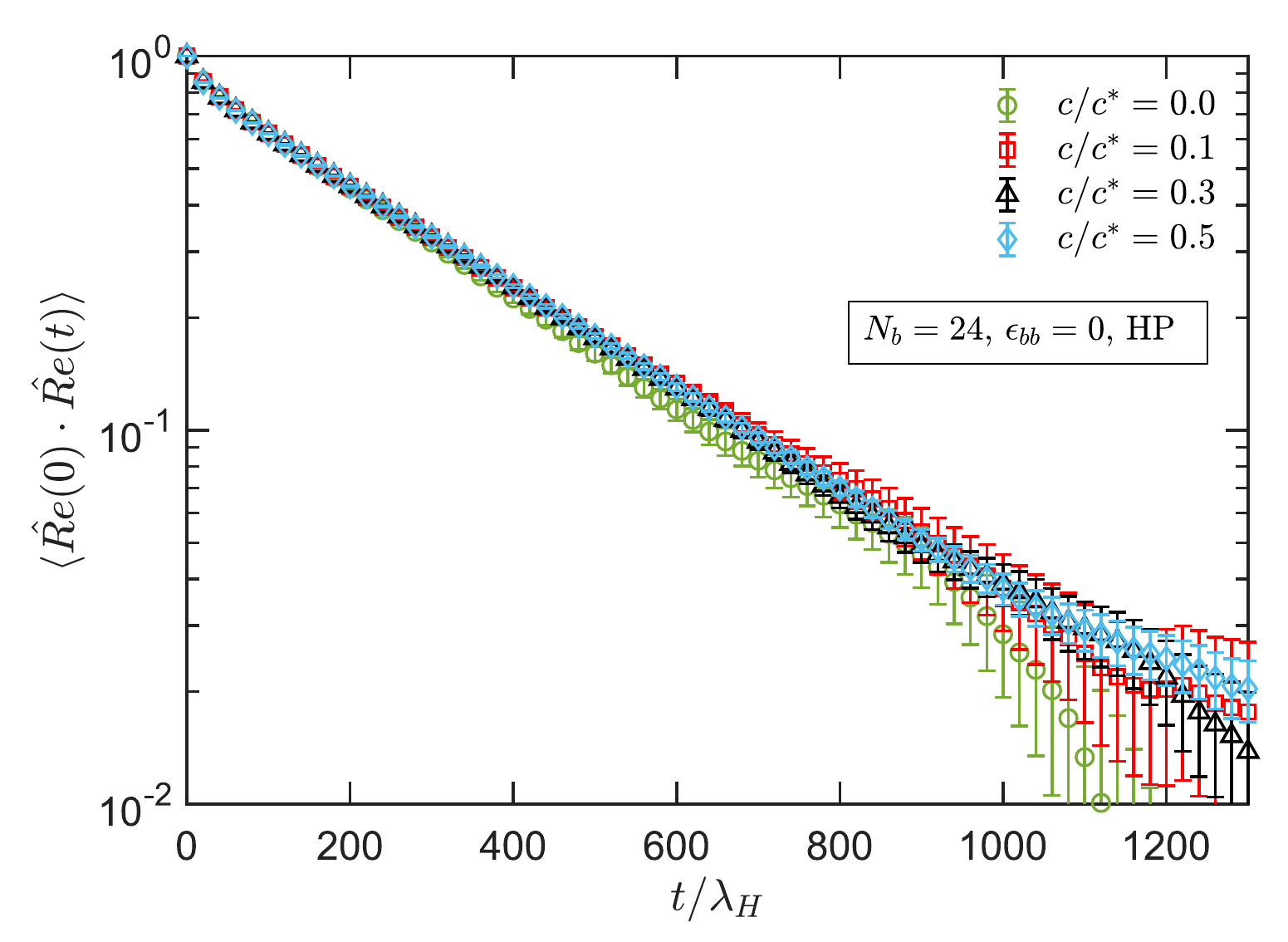} \\
        (a) & (b) \\
    \end{tabular}
    }
   \caption{\small{End-to-end unit vector auto-correlation function at different monomer concentrations, $c/c^*$, for (a) associative polymer solutions with $N_b=24$, $\ell=4$, $f=4$, $\epsilon_{bb}=0$ and $\epsilon_{st}=5.0$ and (b) homopolymer solutions in athermal solvent ($\epsilon_{bb}=0$) with chain length $N_b=24$. Symbols represents the simulation data and dashed lines are the exponential fit at different concentrations.} 
\label{fig:ReAutocorr}}
 \vspace{-20pt}
\end{figure*}

\begin{figure}[t]
    \includegraphics[width=\textwidth]{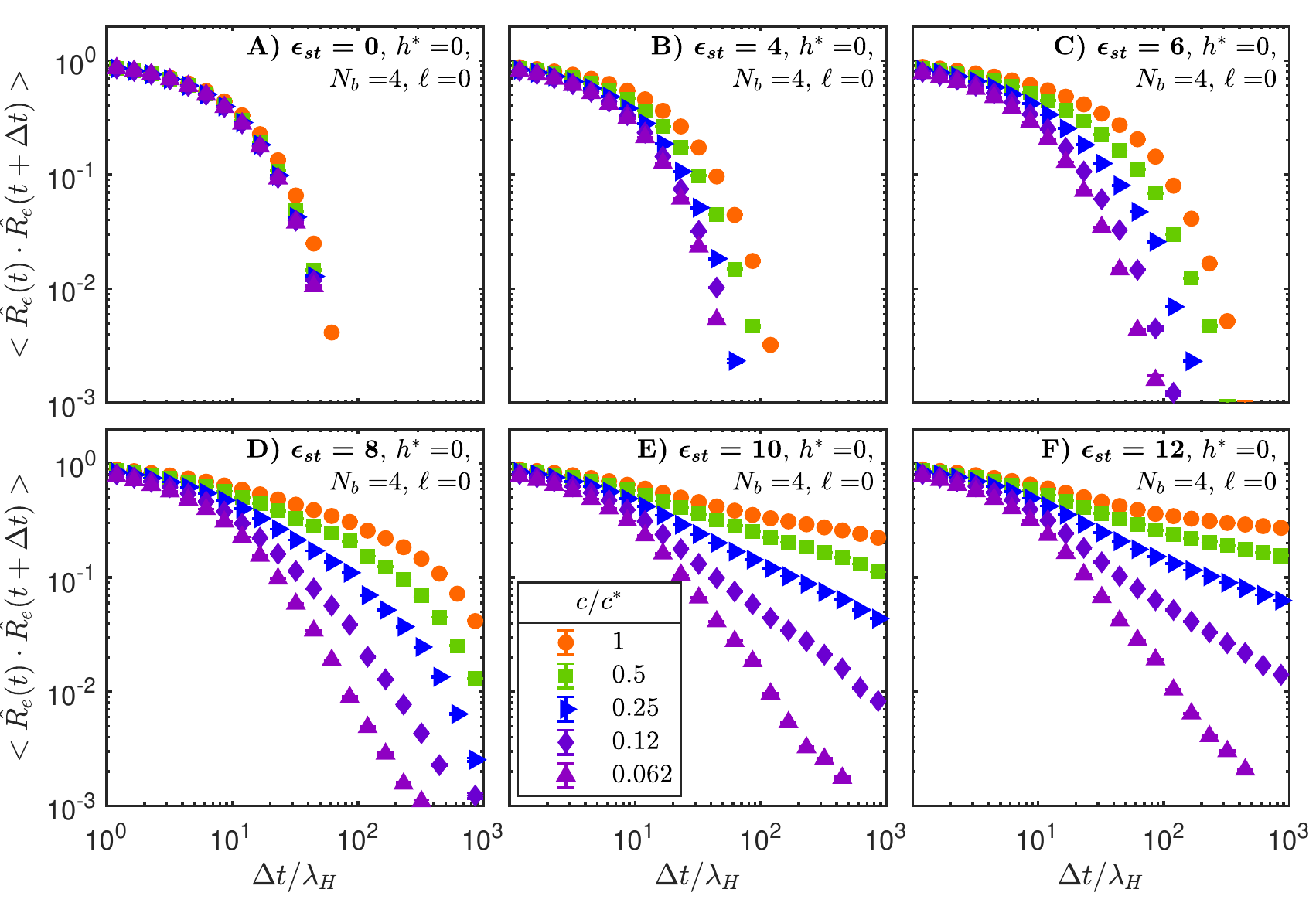}
    \caption{All $R_e$ ACF data for $\ell=0$, $h^*=0$. }
    \label{fig:ReAuto_highSt}
     \vspace{-15pt}
\end{figure}

In simulations, the terminal relaxation time of polymer solutions is typically evaluated by fitting the tail of the decay of stress auto-correlation, as discussed in the earlier section, or from end-to-end vector auto-correlation functions~\cite{Huang2010,Sharad2014,Nafar2015}. Here, we will discuss the method to extract the terminal relaxation time from end-to-end vector auto-correlation. Longest relaxation time scale can be estimated by fitting a single exponential to the decay of end-to-end unit vector auto-correlation function of the polymer chains, defined as follows
\begin{equation}
u(t) = \frac{1}{N_c}\sum\limits_{i=1}^{N_c}\langle \hat{\mathbf{R}}^i_{e}(0)\cdot \hat{\mathbf{R}}^i_{e}(t) \rangle
\end{equation}
where, $\hat{\mathbf{R}}^i_{e}$ is the end-to-end unit vector of the $i^{\text{th}}$ chain and $N_c$ is the number of chains in the system. Fig.~\ref{fig:ReAutocorr}~(a) presents the end-to-end unit vector auto-correlation function of associative polymer solutions at different concentrations, where the last $30$ to $40$ percent of the decay is fitted with an exponential of the form, $\hat{u}(t) = A\,\exp(-t/\tau)$, where $\tau$ gives the estimate for the longest relaxation time. It becomes clear from Fig.~\ref{fig:ReAutocorr}~(a) that with increase in concentration the rate of decay of the auto-correlation function for associative polymers becomes progressively slower, whereas, in case of homopolymer solutions, shown in Fig.~\ref{fig:ReAutocorr}~(b), the decay of the end-to-end unit vector auto-correlation function has a very weak dependence on concentration, $c/c^*$. The relaxation time extracted from the exponential fit is compared with other measurements of longest relaxation time and the data is found to collapse on a universal curve, as shown in the main text. Additionally, we also present the end-to-end unit vector auto-correlation functions for all other conditions studied, including with relatively higher sticker strength as shown in Figs.~\ref{fig:ReAuto_highSt}-\ref{fig:ReACF_Nb24_h0}.

\begin{figure}[t]
    \includegraphics[width=160mm]{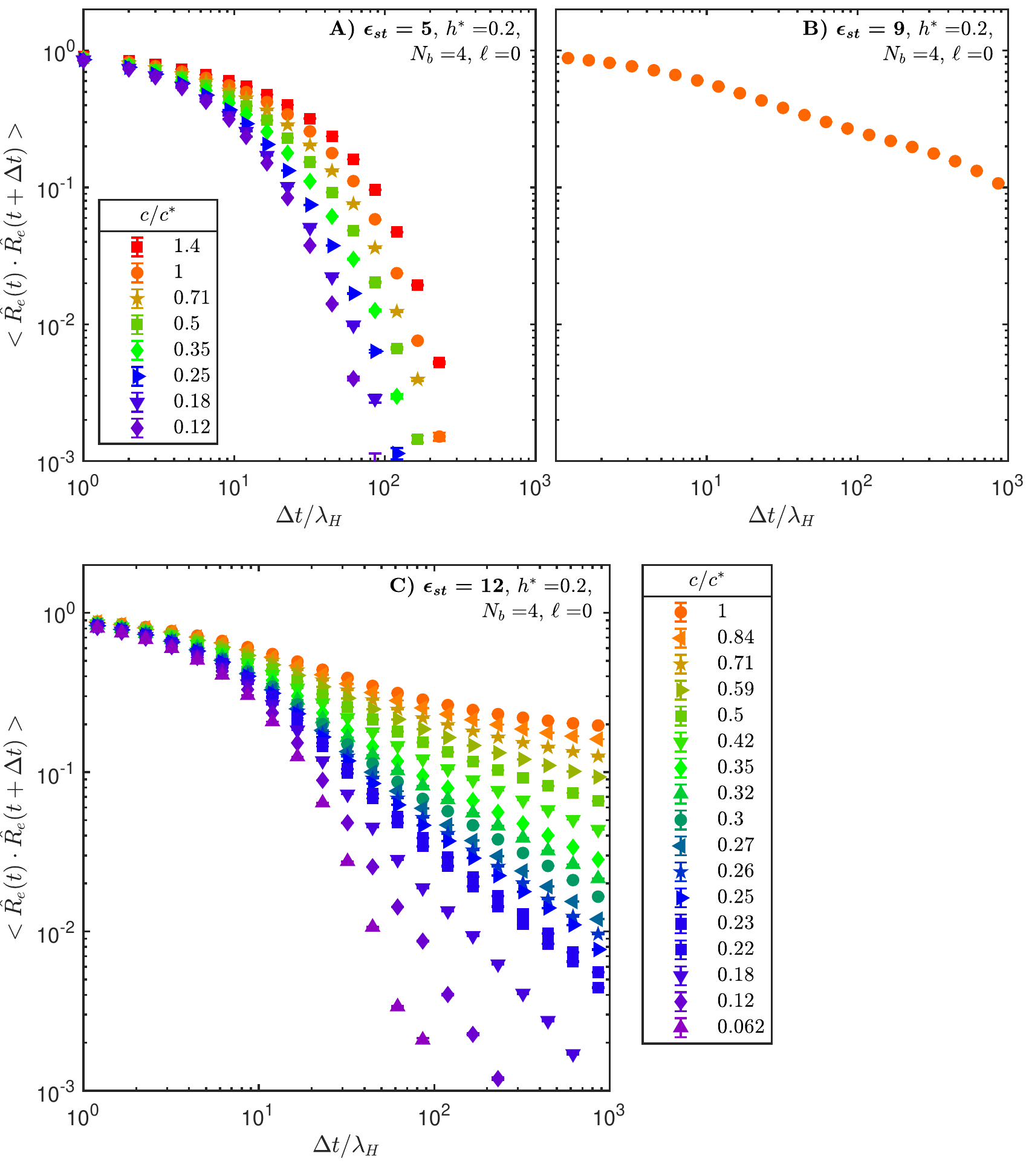}
    \caption{End-to-end unit vector auto-correlation function at values of sticker strength, $\epsilon_{st} = 5,\,9,\,12$, and different monomer concentrations, $c/c^*$, for associative polymer solutions with $N_b=4$, $\ell=0$ and HI parameter, $h^*=0.2$.}
     \vspace{-15pt}
\end{figure}
\begin{figure}[t]
    \includegraphics[width=\textwidth]{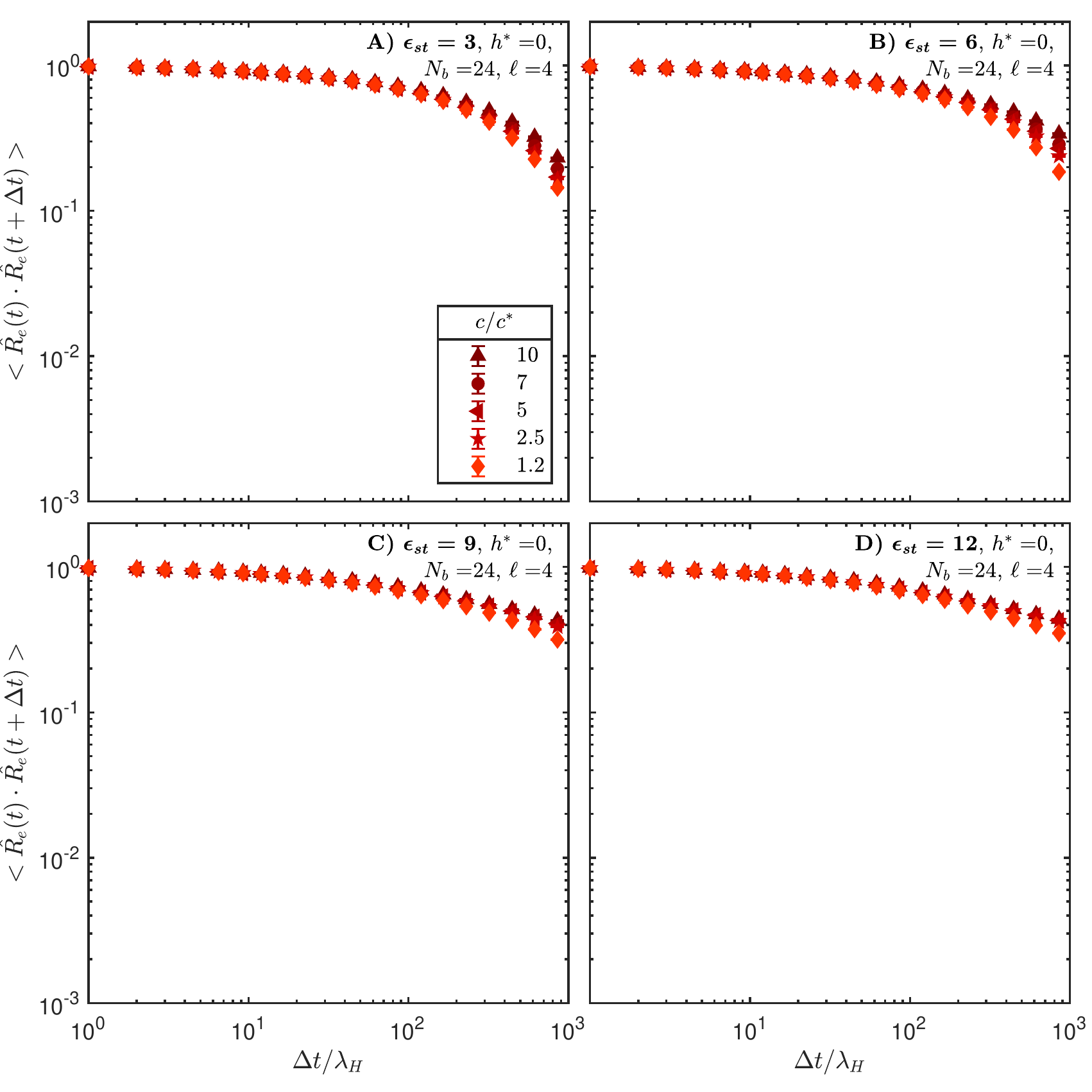}
    \caption{All $R_e$ ACF data using the HOOMD code for $\ell=4$, $h^*=0$.}
    \label{fig:ReACF_Nb24_h0}
     \vspace{-15pt}
\end{figure}

\subsection{Bond Lifetime}

\begin{figure}[t]
 \begin{center}
   \resizebox{12cm}{!}{\includegraphics*[width=4cm]{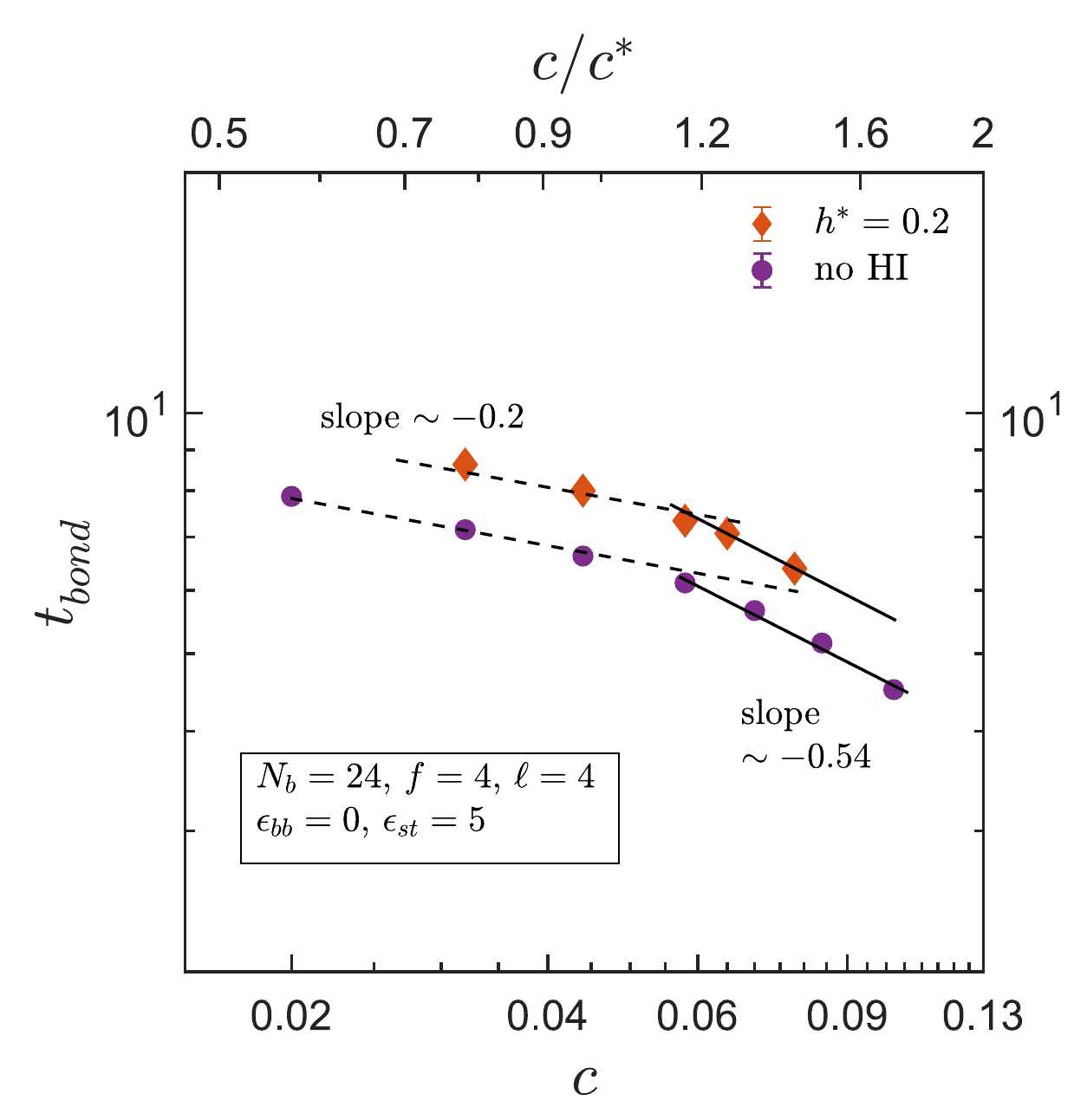}}
 \end{center}
 \vskip-15pt
 \caption{\small{Dimensionless mean bond lifetime, $t_{\text{bond}}$ as a function of concentration, $c/c^*$, for associative polymer solutions with $N_b=24$, $\ell=4$, $f=4$, $\epsilon_{bb}=0$ and $\epsilon_{st}=5.0$. The symbols are from simulations and the dashed line is best fit to the data.}}
\label{fig:tbond}
  \vspace{-20pt}
\end{figure}

Bond lifetimes are measured by extracting the durations of bond events from simulation trajectories. Alternatively, the autocorrelation of bonded pairs provides a renormalized lifetime which includes bind-unbind-rebind events. The bond lifetimes used to validate the Sticky Rouse Model, discussed in the main text, is measured by the average association duration, shown in Fig.~\ref{fig:tbond}, where the average bond lifetime ($t_{\text{bond}}$) of associated stickers is plotted as a function of concentration with and without HI for the case of $\epsilon_{st}=5.0$.
Here, $t_{\text{bond}}$ is computed by considering all possible associating pairs for the systems of associative polymer solutions at different concentrations. Fig.~\ref{fig:tbond} indicates that at a constant value of sticker strength, the bond lifetime decreases with concentration for both HI and no HI cases. Interestingly, in both HI and no HI cases there is a crossover in the scaling behaviour of the average bond lifetime with concentration, where the scaling exponent goes from a value of $-0.2$ to $-0.54$. In the main text, using these values for bond lifetime as function a concentration, we have validated the Sticky Rouse Model showing satisfactory data collapse. 

\begin{figure*}[t]
    \includegraphics[width=\textwidth]{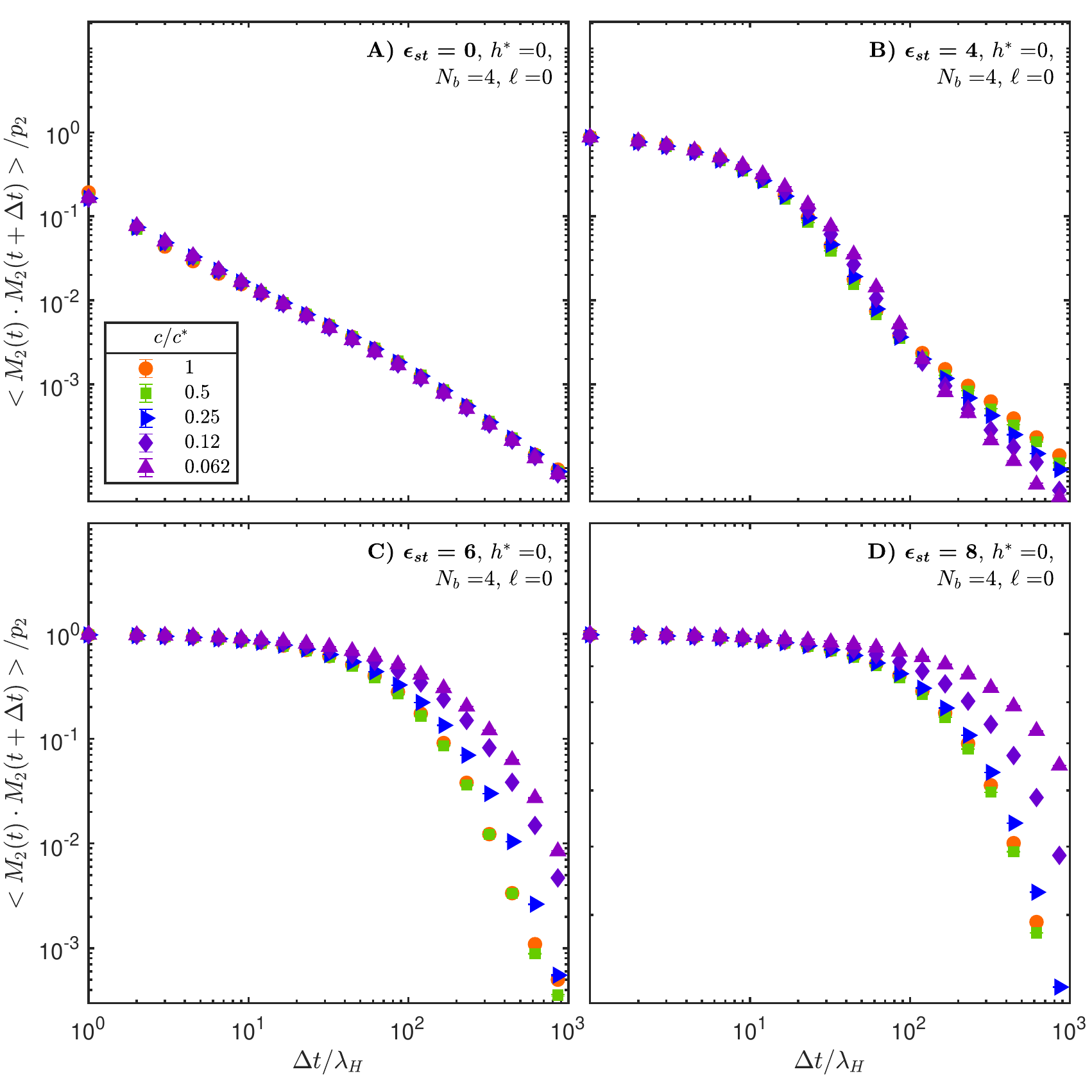}
    \caption{All $M_2$ ACF data for $\ell=0$, $h^*=0$.}
    \label{fig:M2acf_Nb4_h0}
    \vspace{-15pt}
\end{figure*}

Figures \ref{fig:M2acf_Nb4_h0}-\ref{fig:M2acf_Nb24_h0} present the auto-correlation function for the fraction of inter-chain associations at time $t$ which are also present at $t+\Delta t$. These data sets were fit with a single exponential for which the only parameter is the relaxation time $\tau_M$. Surprisingly, there appears to be a power-law tail in this auto-correlation function, which is robustly excluded by the single-parameter fit, as demonstrated in Fig.~\ref{fig:M2acf_Nb24_h0}A. Note that this tail seems largely independent of both concentration and sticker strength, suggesting generic characteristic of this measurement. This could be due to a mechanism such as sticker pairs dissociating, de-correlating completely, then subsequently re-associating due to the finite number of potential partners in the simulation volume.

\begin{figure*}[t]
    \includegraphics[width=\textwidth]{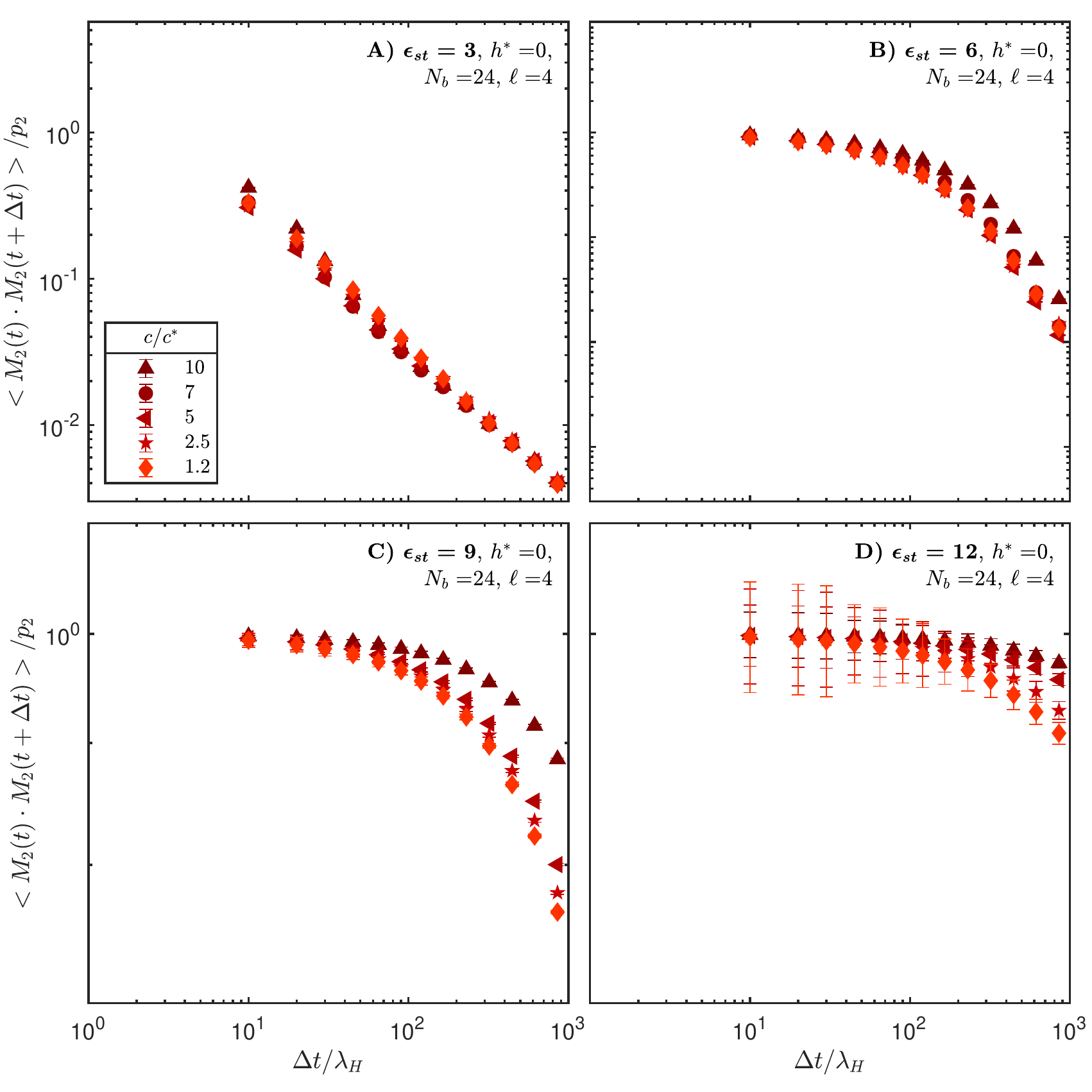}
    \caption{All $M_2$ ACF data for $\ell=0$, $h^*=0.2$.}
    \vspace{-20pt}
\end{figure*}
\begin{figure*}[t]
    \includegraphics[width=\textwidth]{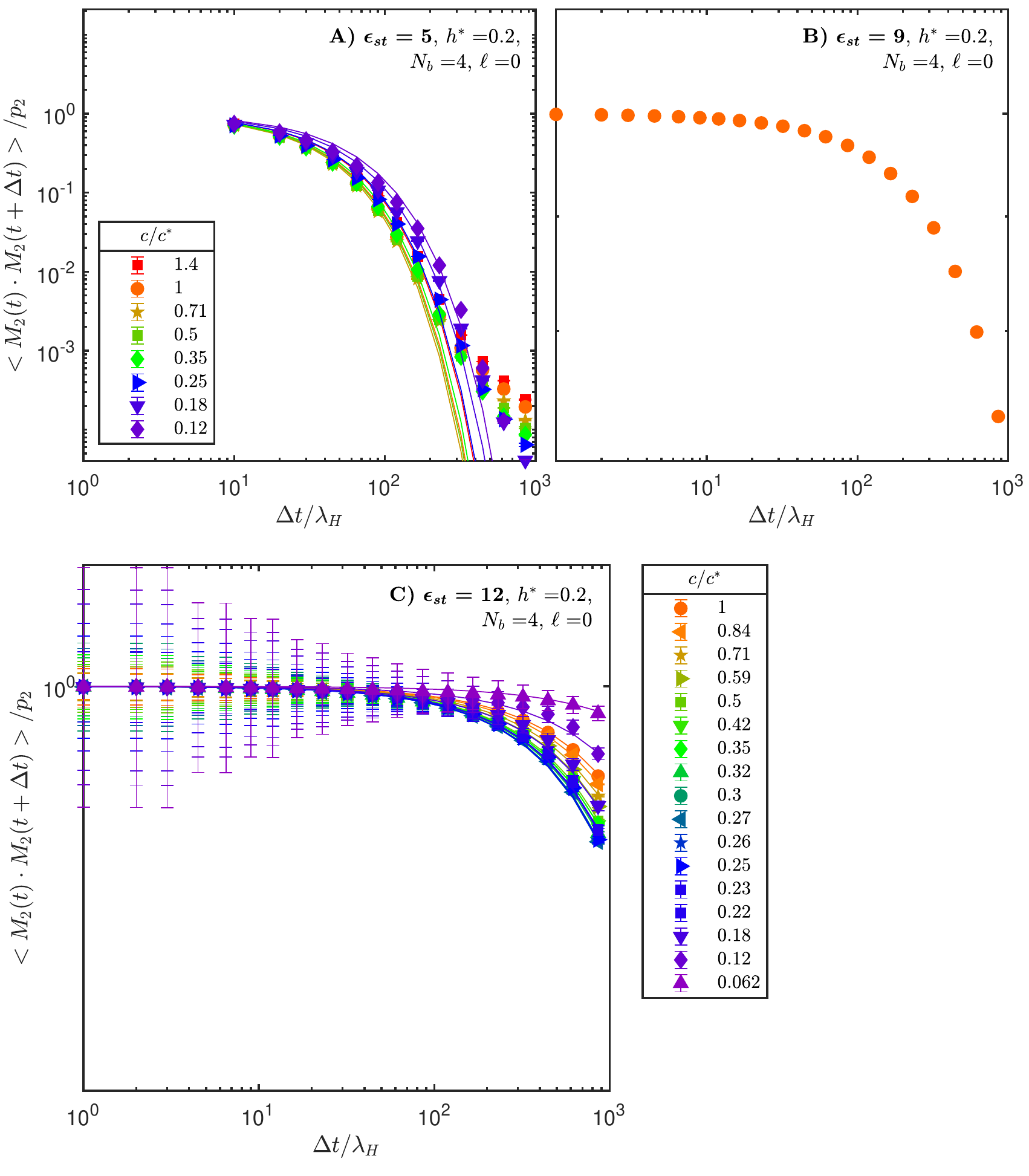}
    \caption{All $M_2$ ACF data using the HOOMD code for $\ell=4$, $h^*=0$.  Fit lines to the form $y=\exp{-t/\tau_M}$ are included to emphasize the single-exponential character of this ACF, and the ability of this fitting method to exclude the power-law tail.}
    \label{fig:M2acf_Nb24_h0}
    \vspace{-16pt}
\end{figure*}

\section{Associated Cluster Size Distribution}

\begin{figure*}[t]
\vspace{-0.5cm}
  \centerline{
 \resizebox{\textwidth}{!}{ \begin{tabular}{cc}
        \includegraphics[width=9cm,height=!]{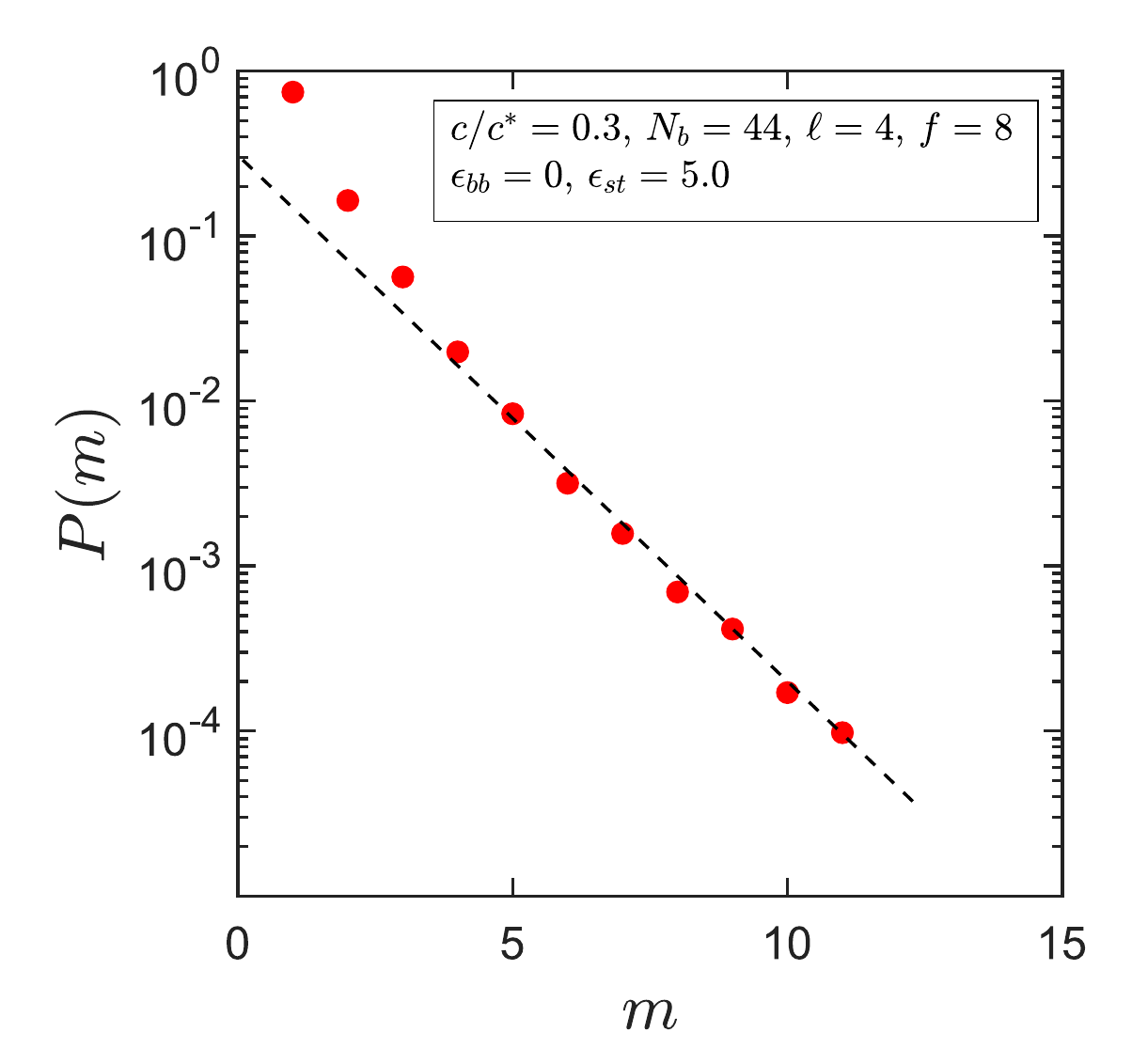} 
    & 
       \includegraphics[width=9cm,height=!]{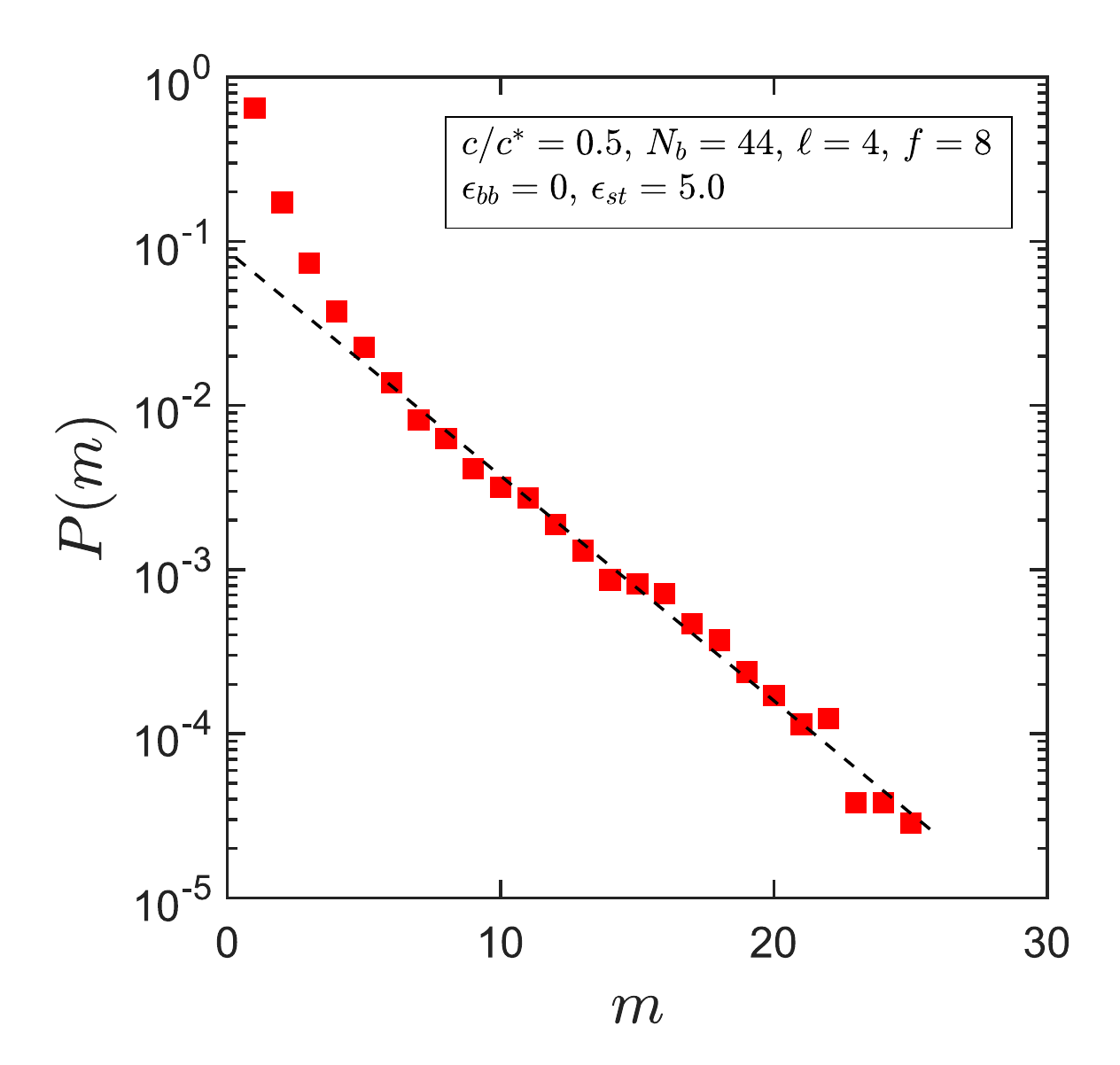} \\[5pt]
           (a)  & 
       (b) \\
       \multicolumn{2}{c}{\includegraphics[width=9cm,height=!]{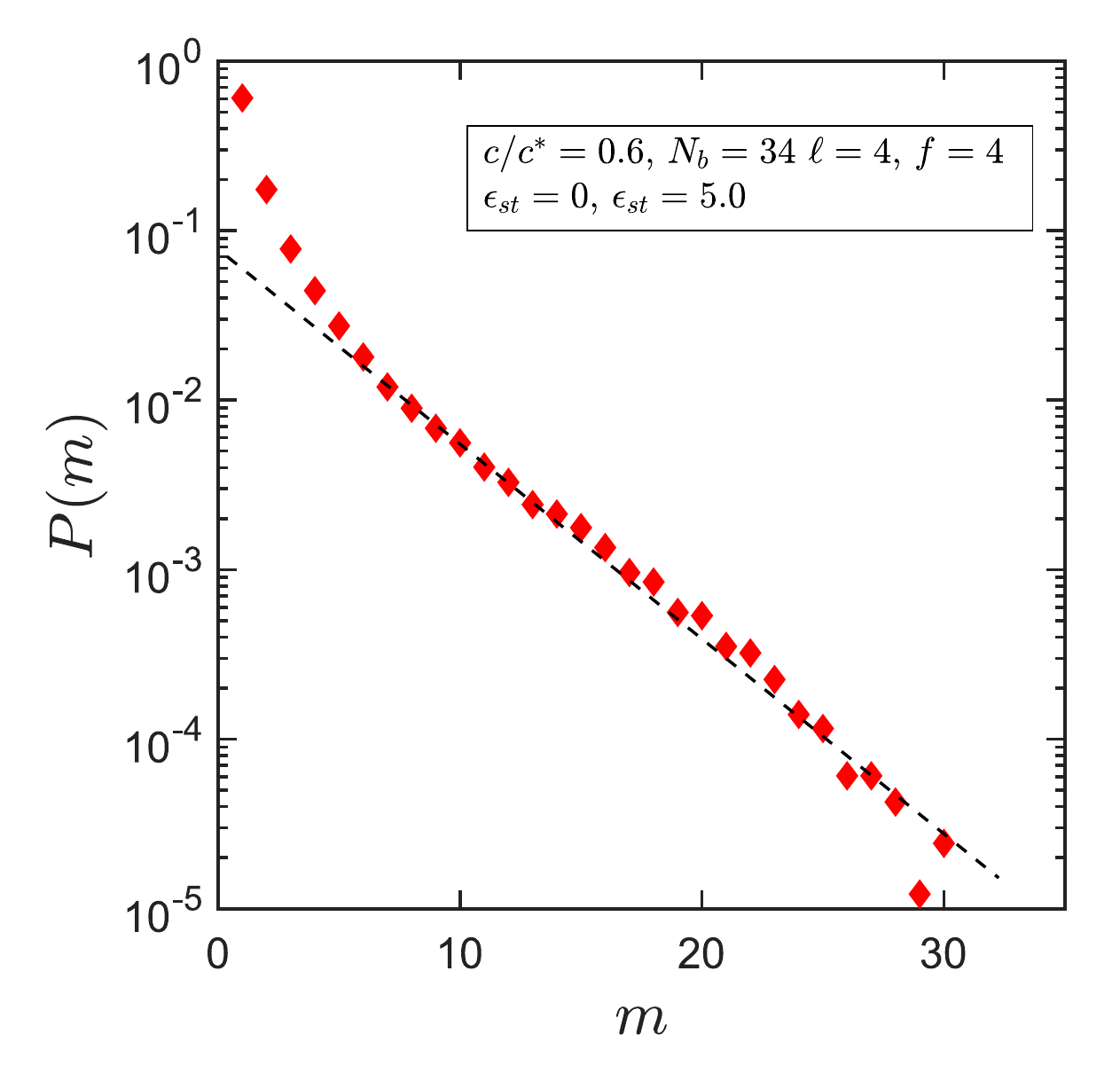} }\\[5pt]
       \multicolumn{2}{c}{(c)}     \\
  \end{tabular} }}
\caption{\small{Chain cluster size distribution in semi-log scale for systems of associative polymers with spacer length $\ell=4$, $\epsilon_{bb}=0$ and $\epsilon_{st}=5.0$ at (a) $c/c^*=0.3$, $N_b=44$, (b) $c/c^*=0.5$, $N_b=44$ and (c) $c/c^*=0.6$, $N_b=34$. The solid red symbols are simulation data and the exponential decay of the cluster size distribution is shown by the dashed black lines.} 
\label{fig:CustDist_exp}}
\vspace{-0.6cm}
\vspace{-20pt}
\end{figure*}

\begin{figure*}[t]
    \centerline{
    \begin{tabular}{c c}
        \includegraphics[width=73mm]{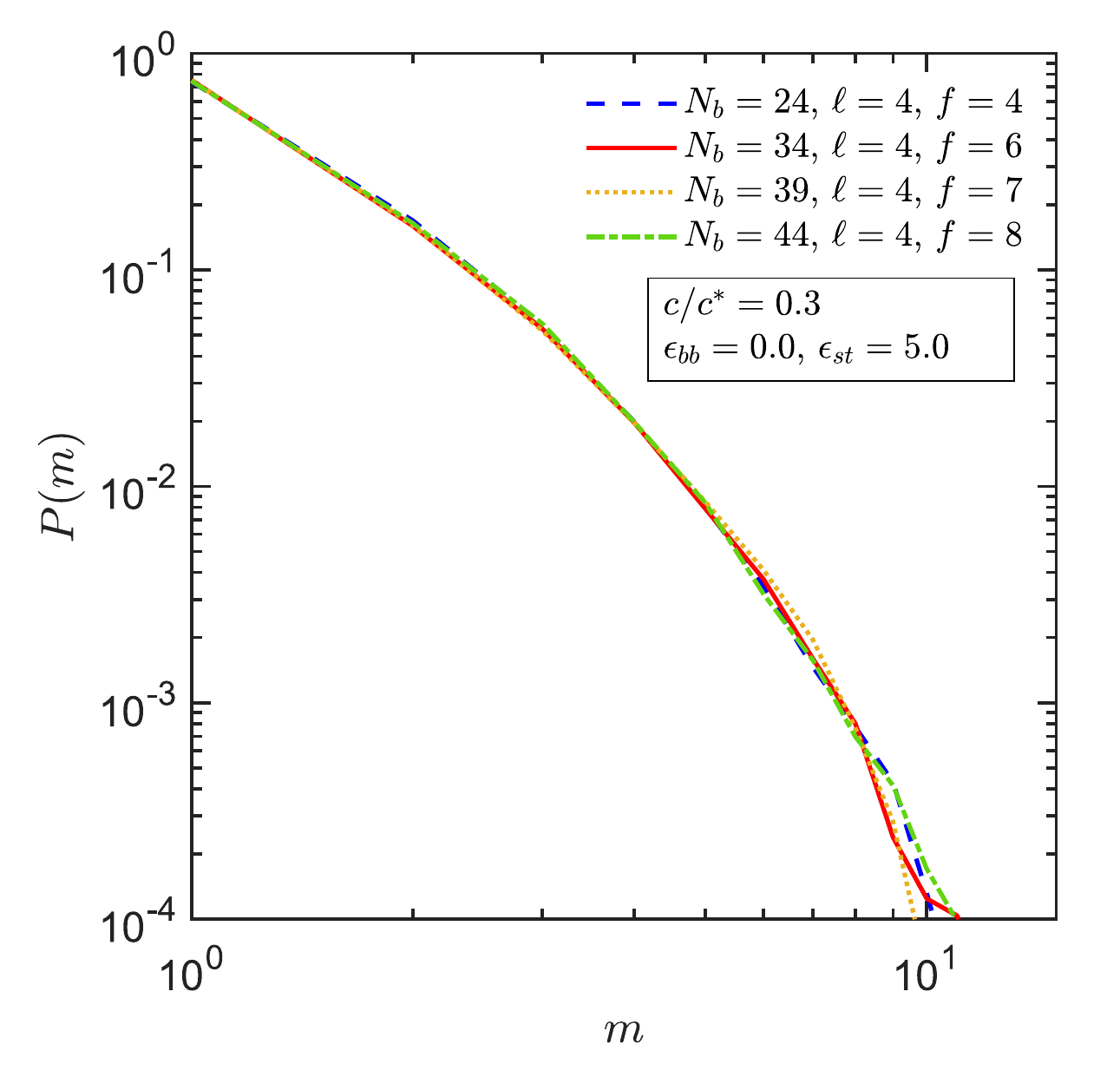} 
        & \includegraphics[width=74mm]{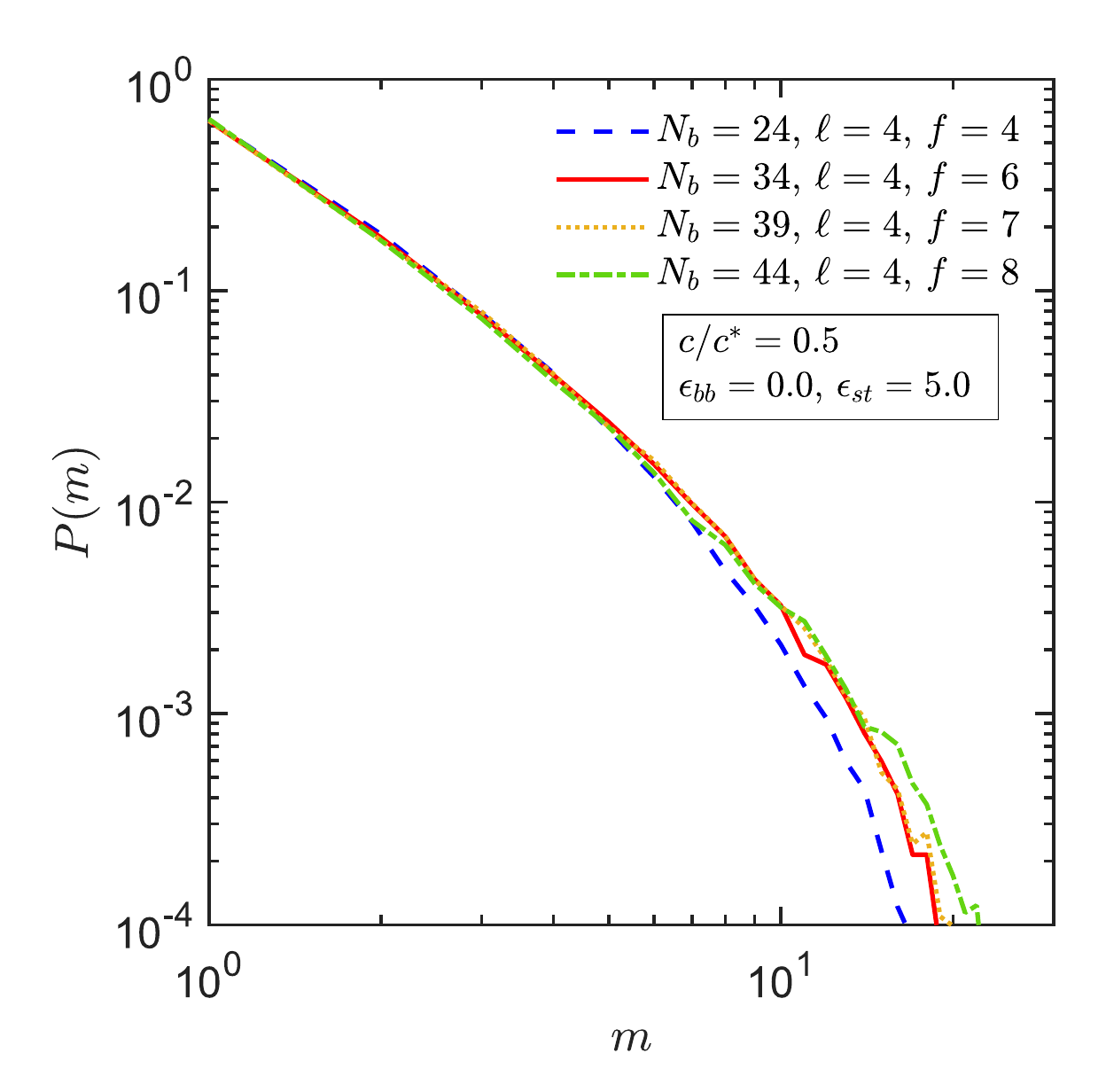} \\
        (a) & (b) \\
        \includegraphics[width=74mm]{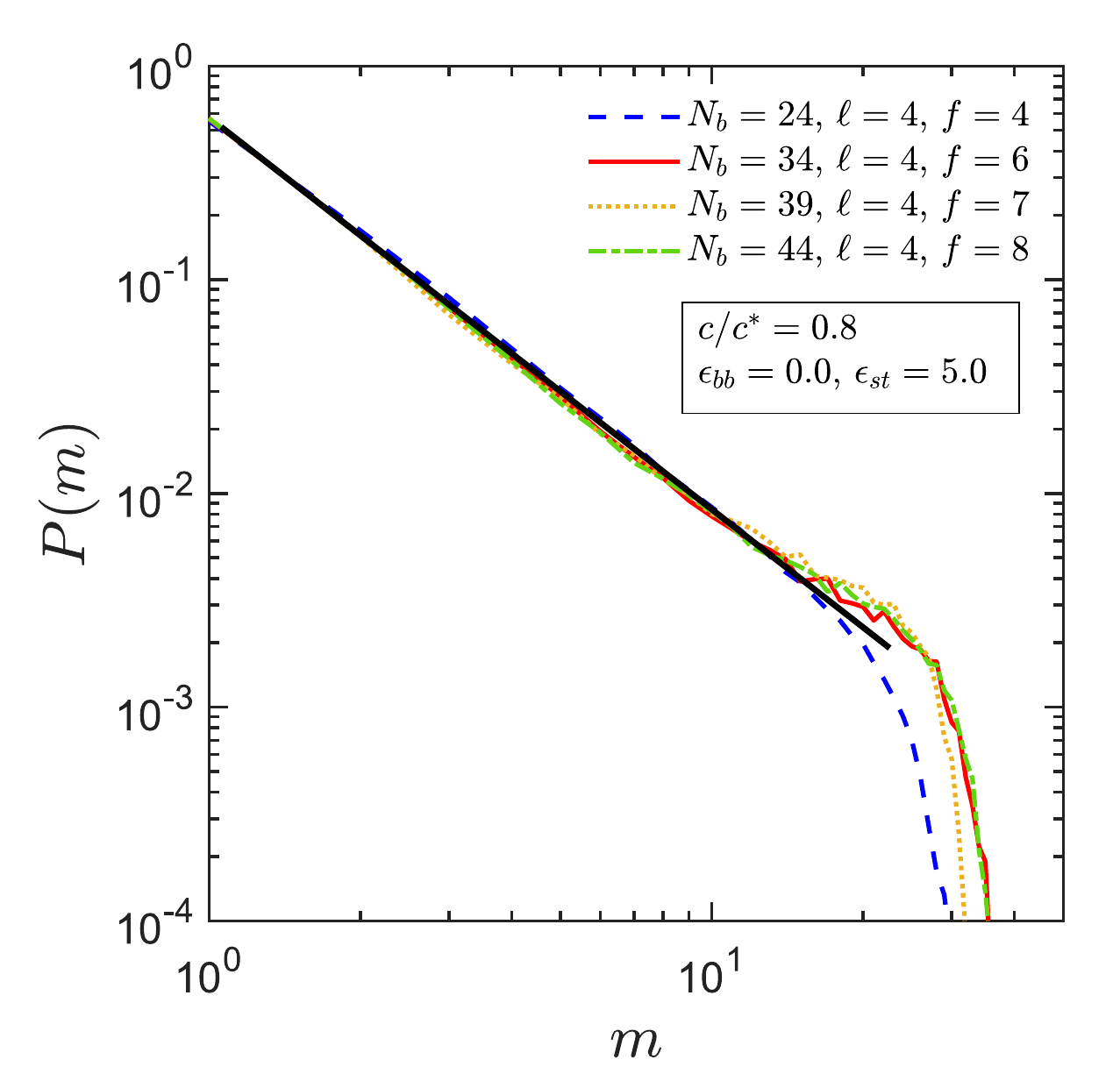} 
        & \includegraphics[width=71mm]{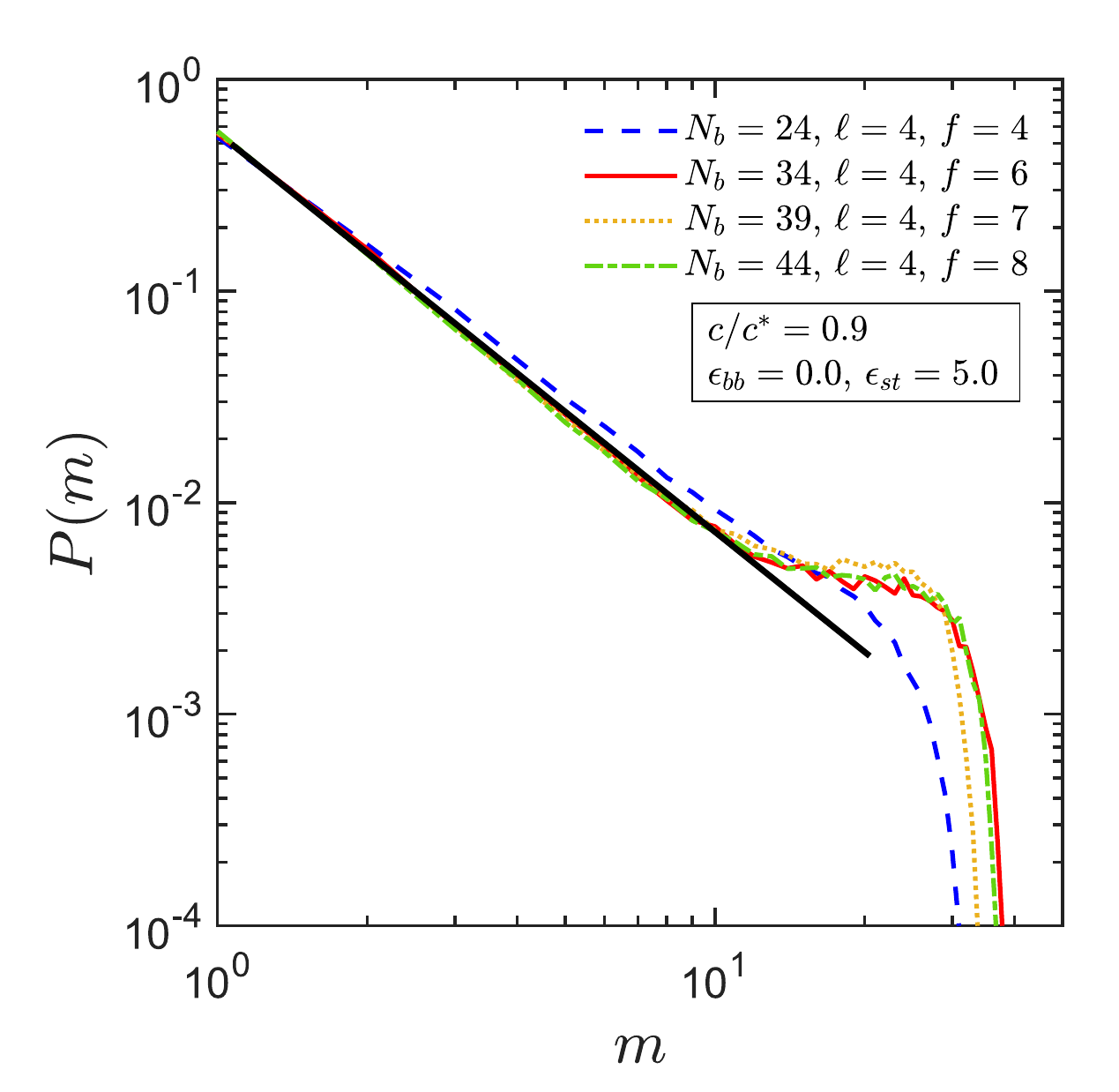} \\
        (c) & (d) \\
    \end{tabular}
    }
   \caption{\small{Chain cluster size distribution in log-log scale for systems of associative polymers with spacer length $\ell=4$, $\epsilon_{bb}=0$ and $\epsilon_{st}=5.0$ at different values of $c/c^* = \lbrace 0.3,\, 0.5,\, 0.8,\, 0.9\rbrace$ and different chain lengths, $N_b$. The power-law behaviour of the cluster size distribution is shown by the solid black lines in (c) and (d).} 
\label{fig:CustDist_PL}}
\vspace{-17pt}
\end{figure*}

\begin{figure*}[t]
    \includegraphics[width=\textwidth]{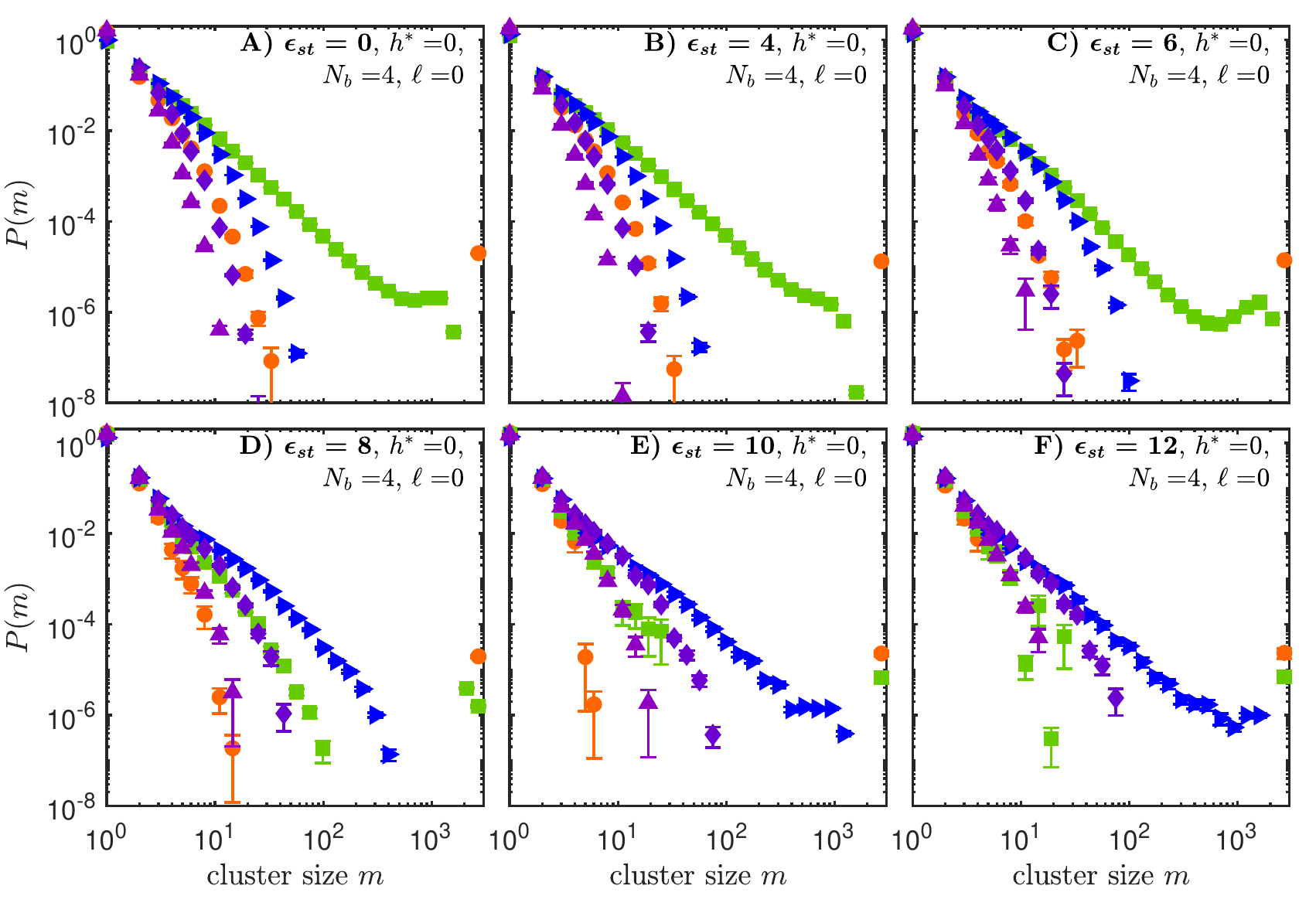}
    \caption{All cluster size distribution data for $\ell=0$, $h^*=0$. $c/c^*=$ 0.06 for purple triangles, 0.12 for purple diamonds, 0.25 for blue triangles, 0.5 for green squares, and 1 for red circles.}
    \label{fig:ClustDist_Nb4_h0}
    \vspace{-5pt}
\end{figure*}

In an associative system below the gel concentration, finite sized clusters of associated chains form. With increase in polymer concentration, the cluster size increases, which results in the formation of a system spanning network at the gel point. In finite-sized simulation systems, bimodality of the chain cluster size distribution is observed~\cite{AritraStatJoR}. Additionally, at the gel point the material structure becomes self-similar with no characteristic length scale and the cluster size distribution shows a power law scaling, $P(m) \sim m^{-{\tau_F}}$, where $P(m)$ is the probability of finding a cluster of size $m$ and the exponent $\tau_F$ is often called the Fisher exponent~\cite{Stauffer2007,Indei17}. Here, we have investigated the chain cluster size distribution, $P(m)$, of associative polymer solutions at different scaled concentrations, $c/c^*$, and different chain lengths, $N_b$, with spacer length, $\ell=4$, backbone interaction strength, $\epsilon_{bb}=0$ and sticker association strength, $\epsilon_{st}=5.0$. At low concentrations, such as $c/c^*=0.3,\,0.5,\, 0.6$, the probability of finding large cluster decreases exponentially as displayed in Figs.~\ref{fig:CustDist_exp}, whereas, at about $c/c^*=0.8$ and above the cluster size distributions, shown in Figs.~\ref{fig:CustDist_PL}~(c) and (d), indicate a power law scaling for a significantly wide range of values of cluster size, $m$. It is also noted that $P(m)$ becomes independent of chain length with increase in $N_b$. The shouldering of the chain cluster size distribution observed at higher value of $c/c^*(=0.9)$ is because of the onset of bimodality in the distribution. From this analysis we interpret that the onset of gelation transition, based on the power-law scaling of the cluster size distribution, happens at about $c/c^*\approx 0.8$, which lies in between the other two gel transition concentrations based on other static signatures identified in Ref.~\citenum{AritraStatJoR}. This is another static signature of gelation which along with other static signatures guide us to explore the range of concentration at which dynamic signatures are to be observed. Figs.~\ref{fig:ClustDist_Nb4_h0}-\ref{fig:ClustDist_Nb24_h0} show the clusters size distributions for various conditions studied. Note in Fig.~\ref{fig:ClustDist_Nb4_h0} that a power law and even bimodal distribution is possible even with $\epsilon_{st}=0$, at high concentration. This is due to the fact that stickers with $\epsilon_{st}=0$ have a 1:1 probability of being bound or unbound based simply on proximity, so an instantaneous cluster is still possible at high concentration, thought this cluster would have no rheological effect.

\begin{figure*}[t]
    \includegraphics[width=\textwidth]{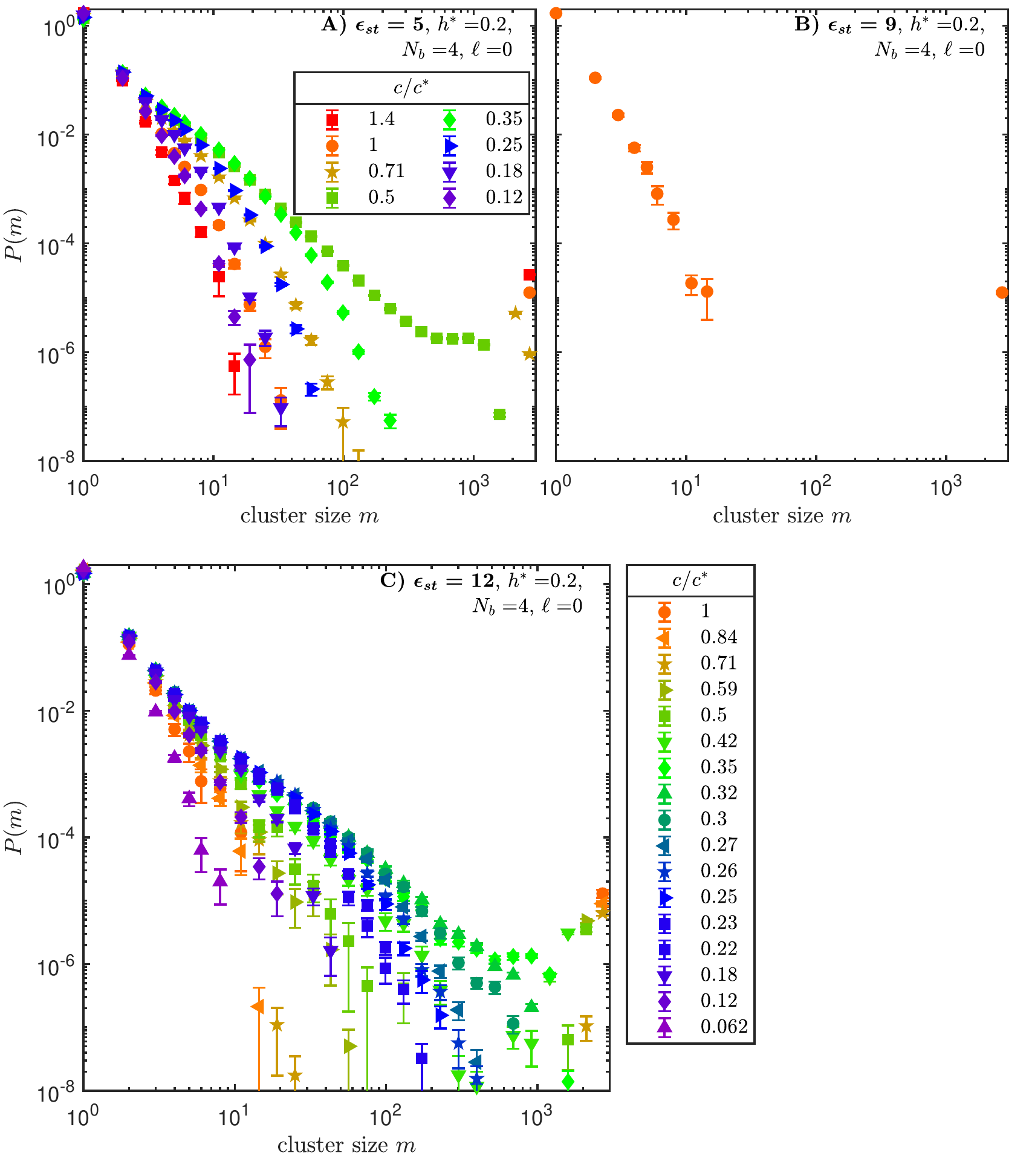}
    \caption{All cluster size distribution data for $\ell=0$, $h^*=0.2$.}
       \vspace{-20pt}
\end{figure*}

\begin{figure*}[t]
    \includegraphics[width=\textwidth]{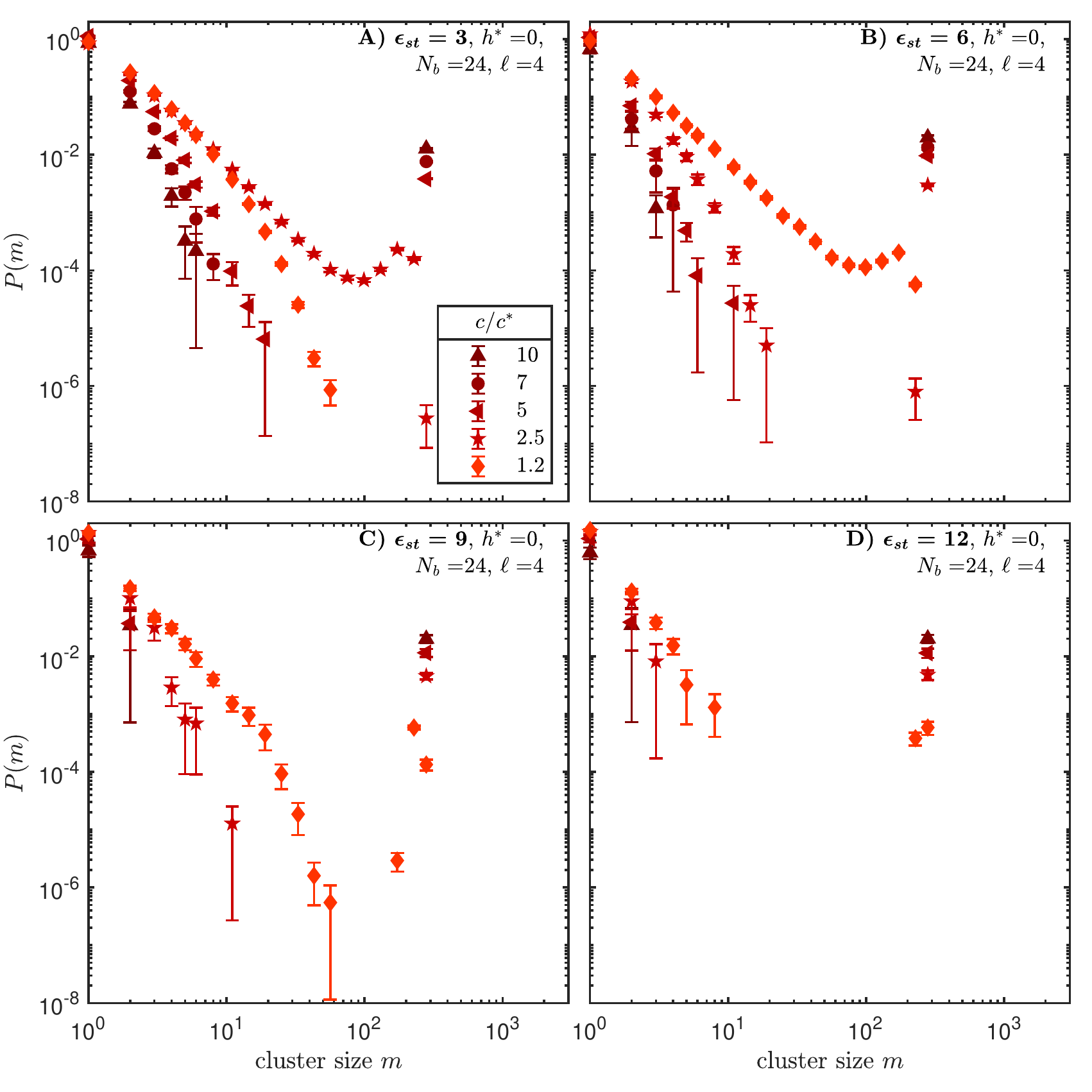}
    \caption{All cluster size distribution data using the HOOMD code for $\ell=4$, $h^*=0$.}
    \label{fig:ClustDist_Nb24_h0}
    \vspace{-10pt}
\end{figure*}

\section{Validation: Oscillatory Shear}

\begin{figure}[t]
 \begin{center}
   \resizebox{12cm}{!}{\includegraphics*[width=4cm]{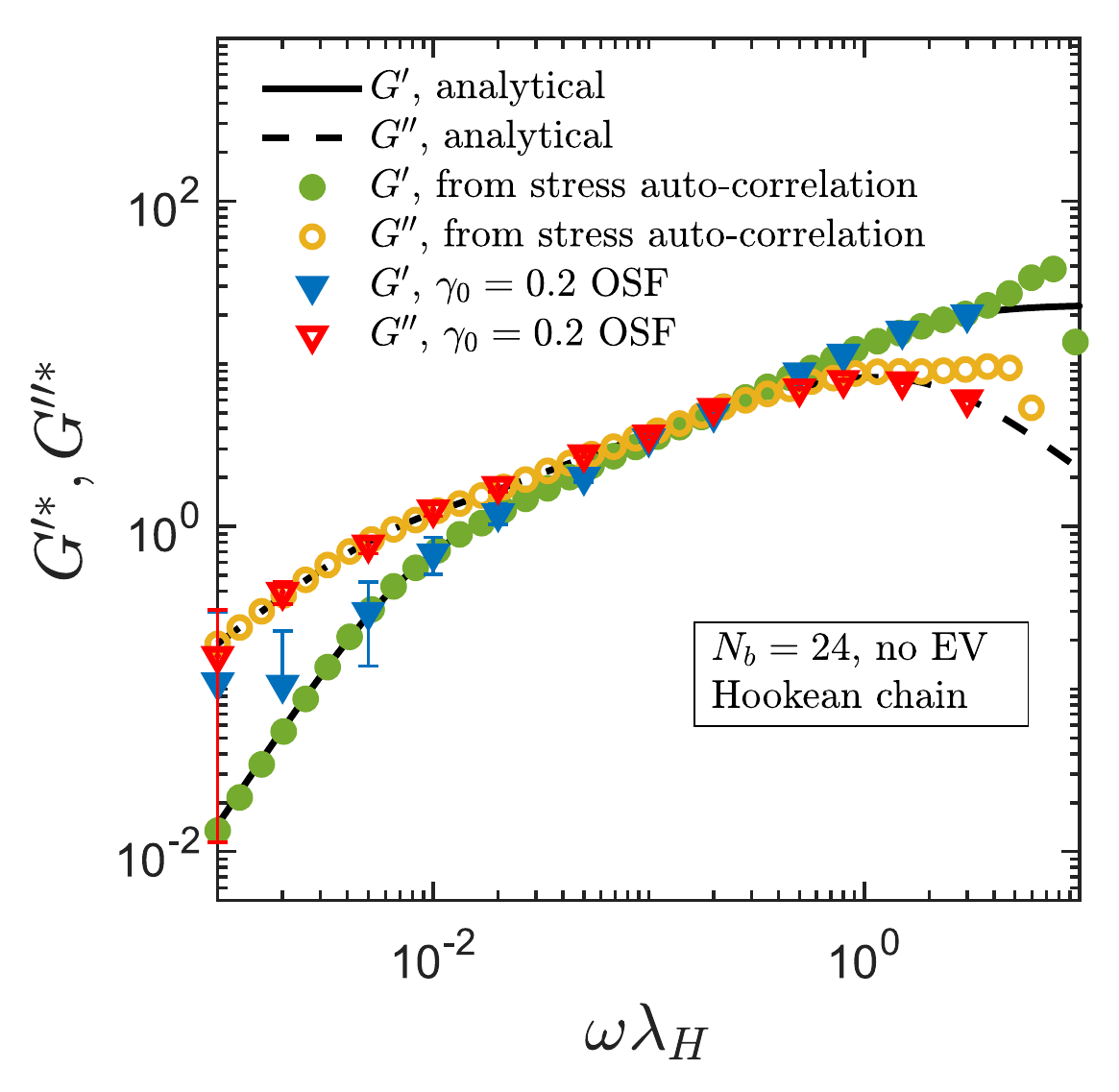}}
 \end{center}
 \vskip-15pt
 \caption{\small{Non-dimensionalized dynamic moduli ($G'^*$ \& $G''^*$) as a function of dimensionless frequency, $\omega\,\lambda_H$, for a dilute solution of Rouse chains with chain length, $N_b=24$. The filled (for $G'$) and open (for $G''$) symbols are from BD simulations at equilibrium and under oscillatory shear flow (OSF), where $\gamma_0$ is the amplitude of oscillation. The solid and broken lines are analytical solutions for $G'$ and $G''$, respectively, for Rouse chains in a dilute solution.}}
\label{fig:DyModRouse}
    \vspace{-20pt}
\end{figure}

\begin{figure*}[t]
    \centerline{
    \begin{tabular}{c c}
        \includegraphics[width=85mm]{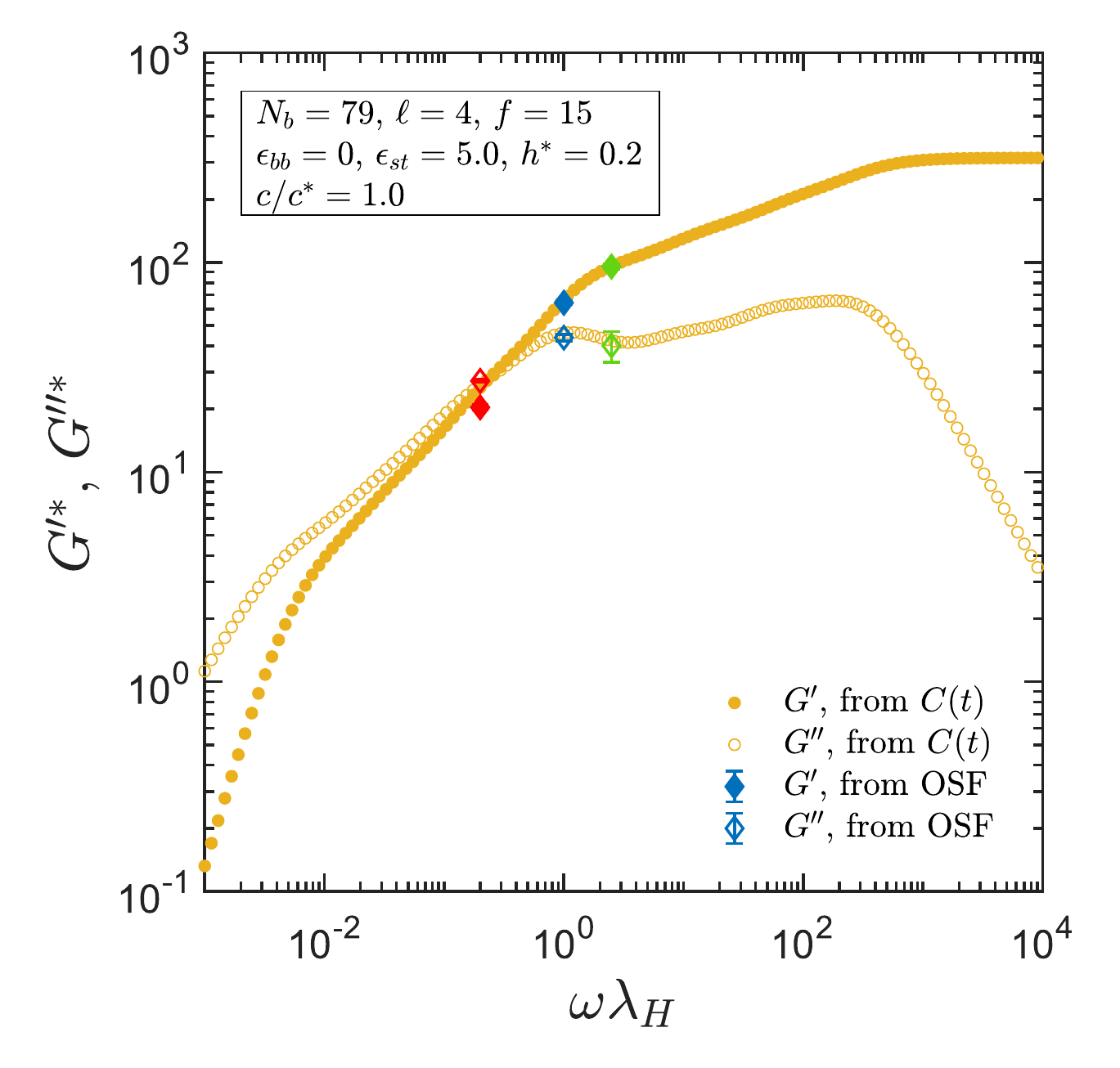} 
        & \includegraphics[width=85mm]{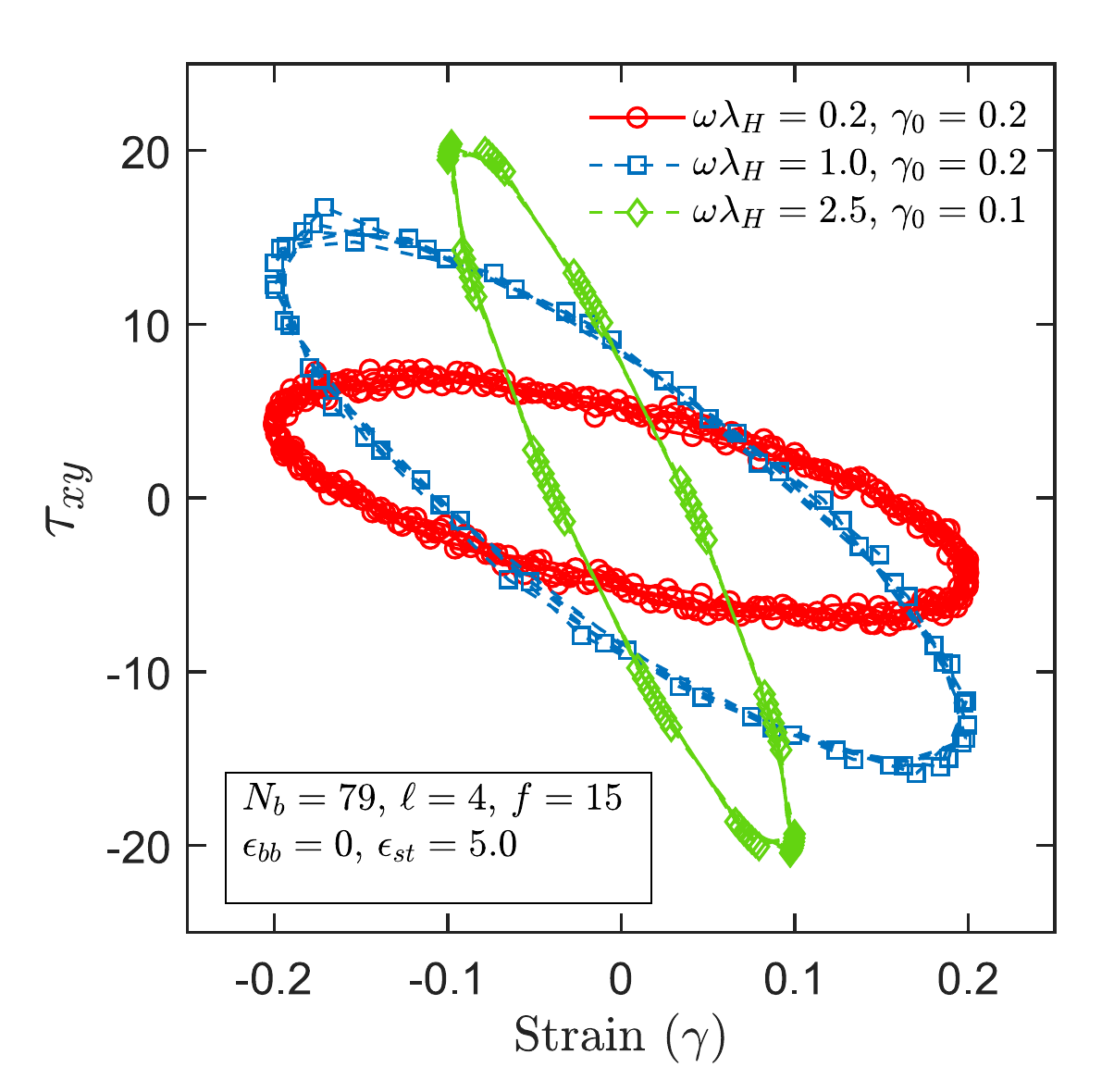} \\
        (a) & (b) \\
    \end{tabular}
    }
   \caption{(a) Non-dimensionalized dynamic moduli ($G'^*$ \& $G''^*$) as a function of dimensionless frequency, $\omega\,\lambda_H$, for associative polymer solutions with chain length, $N_b=79$, spacer length, $\ell=4$, $\epsilon_{bb}=0$ and $\epsilon_{st}=5.0$ at $c/c^*=1.0$. The filled (for $G'$) and open (for $G''$) symbols are from BD simulations at equilibrium and under oscillatory shear flow (OSF) at different oscillation frequencies, where $\gamma_0$ is the amplitude of oscillation. (b) Lissajous curve corresponding to the frequencies considered for the comparison of dynamic moduli calculated from equilibrium and OSF simulations.}
\label{fig:DyModN79}
    \vspace{-15pt}
\end{figure*}

It is possible that the equilibrium stress autocorrelation function $C(t)$ does not capture the plateau modulus of a gel, since $C(t)$ must go to zero as $t$ goes to infinty in equilibrium. That is, there could be a finite offset between $C(t)$ and $G(t)$ for a gel. To validate that the absence of an apparent gel transition at $\epsilon_{st}=5$ was not due to such a concern, oscillatory shear flow simulations were carried out at several frequencies to validate that there was no discrepancy between the modulii measured by transforming $C(t)$ vs those measured by preturbative simulations. Fig.~\ref{fig:DyModRouse} demonstrates the correctness of the the stress autocorrelation measurement and the oscillatory measurement by comparing with the exactly known Rouse spectrum of a 24 bead chain. Fig.~\ref{fig:DyModN79}a then demonstrates this validation for the non-trivial case of $N_b=79$ with several stickers. Lissajous curves for these oscillatory simulations are shown in Fig.~\ref{fig:DyModN79}b.

Here, we calculate dynamic moduli using these two methods for a system of Rouse chains in dilute limit and a solution of associative polymers with $N_b=79$, $\ell=4$, $f=15$, $\epsilon_{bb}=0$ and $\epsilon_{st}=5.0$ at $c/c^*=1.0$. In the Rouse case we compare the result with the analytical solutions as discussed in the following. For Rouse model of polymer solutions, consisting of bead-spring chains with $(N_b-1)$ Hookean springs per chain, the analytical expression for the constitutive equation is well known. For such a model there is a spectrum of relaxation times, $\lambda_j$, given by~\cite{Bird1987}
\begin{equation}
\lambda_j = \frac{\zeta/2H}{4\,\sin^2(j\pi/2N_b)}
\end{equation} 
where $\zeta$ is the friction coefficient and $H$ is spring constant. For a dilute solution of Rouse chains subjected to small amplitude oscillatory shear flow, the polymeric contribution of the complex modulus can be decomposed into the following real and imaginary parts,
\begin{eqnarray}
G'  & = n_pk_BT\sum\limits_{j=1}^{N_b-1}\dfrac{\lambda_j^2\,\omega^2}{1+(\lambda_j\,\omega)^2} \\
G'' & = n_pk_BT\sum\limits_{j=1}^{N_b-1}\dfrac{\lambda_j\,\omega}{1+(\lambda_j\,\omega)^2}
\end{eqnarray}
Here, $\omega$ is the frequency of oscillation and $n_p$ is the number density of chains. Fig.~\ref{fig:DyModRouse} shows very good agreement among different methods used to compute $G'$ and $G''$ for a dilute solution of Rouse chains. Both equilibrium and SAOS (with amplitude, $\gamma_0=0.2$) simulations produce the same results which is also in agreement with the analytical expressions.

Similarly, for the associative polymer solutions with chain length $N_b=79$ and spacer length $\ell=4$ at $c/c^*=1.0$, the dynamic moduli obtained from the equilibrium simulations are compared with that from SAOS. As shown in Fig.~\ref{fig:DyModN79}~(a), for the three values of oscillation frequencies considered here, $G'$ and $G''$ calculated from both the methods are in agreement with each other within errorbars. Note that even for a longer chain length ($N_b=79$) with short spacer ($\ell=4$), there exists a terminal relaxation behaviour with finite relaxation time as observed for the short chain simulations discussed in the main text. We also present the viscoelasticity of this system in terms of Lissajous curves, shown in Fig.~\ref{fig:DyModN79}~(b). Typically, with the increase in frequency of oscillations the Lissajous curve takes the shape of a more elongated ellipse (getting narrower along the minor axis and elongated along the major axis) until it becomes a straight line in the purely elastic limit. A similar trend is observed in case of the system of associative polymer solution considered here.

\section{Bayesian Information Criterion}

\begin{figure}[t]
 \begin{center}
   \resizebox{14cm}{!}{\includegraphics*[width=4cm]{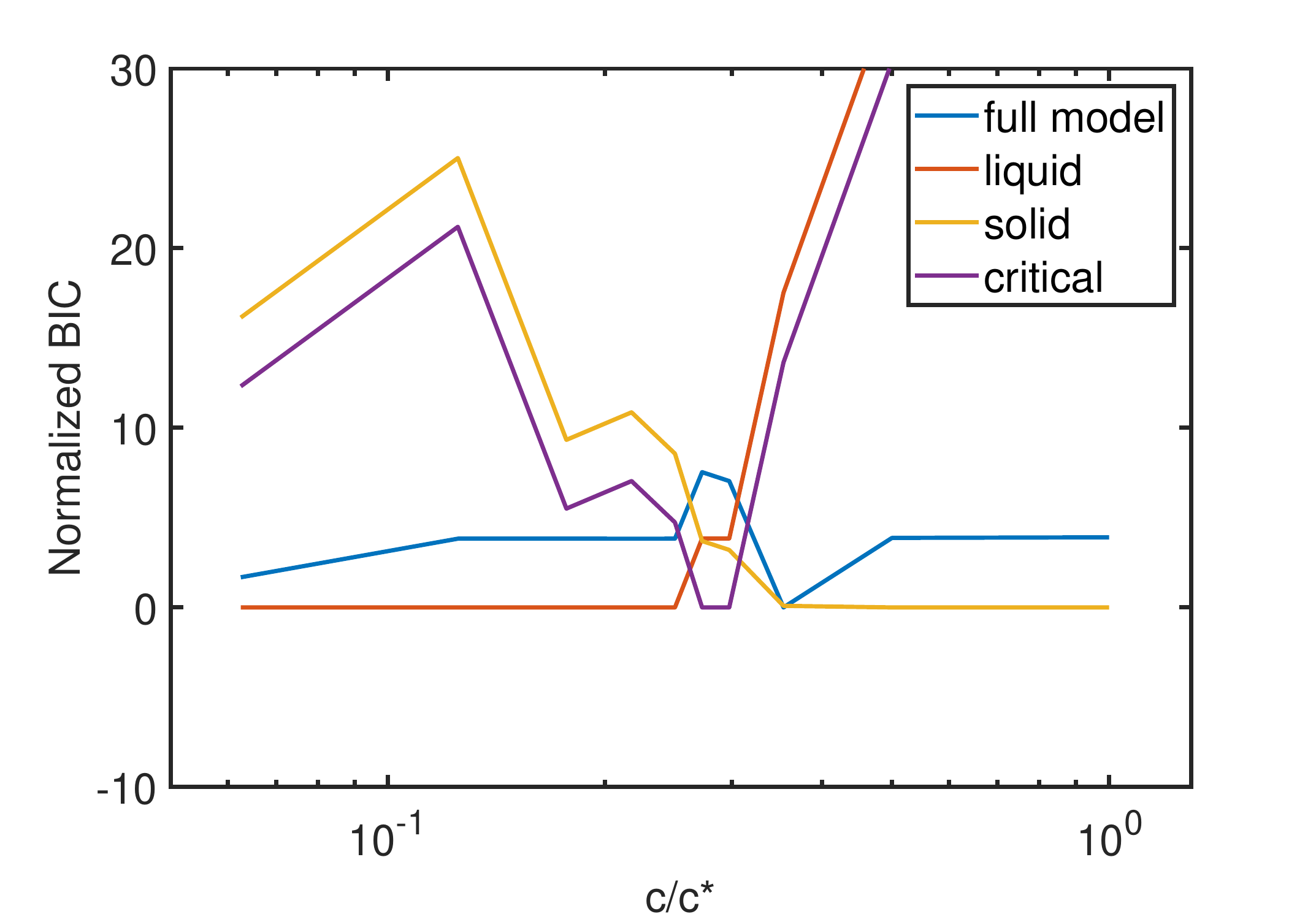}}
 \end{center}
 \vskip-15pt
 \caption{Bayesian information criterion for various generalizations of the Mittag-Leffler function as a model for the viscoelastic response of our simulated gel-forming system. These curves represent measurements for a system with $\ell=0,f=4,N_b=4,h^*=0.2,\epsilon_{st}=12$. Since the difference between BIC for different models is more meaningful than the absolute value, the minimum BIC for each concentration is subtracted from all models at that concentration. This emphasizes the optimal model at a particular concentration as the bottom line at that concentration. The height of each model above the 0 line suggests the degree to which that model is inappropriate due to too few or too many parameters.}
\label{fig:BIC}
    \vspace{-20pt}
\end{figure}

The model used in the main text to fit $G(t)$ data to extract relaxation times and moduli is
\begin{equation}
    G_{full}(t)=\left[\mathbb{G}E_\alpha\left(-\frac{\mathbb{G}}{\mathbb{V}}t^\alpha\right)+G_\epsilon\right]\exp\left(-\frac{t}{\tau_\epsilon}\right),
    \label{eqn:MLphys}
\end{equation}
In this equation, $G_\epsilon$ represents the height of the temporary elastic plateau due to the formation of a gel. If a gel has not formed, then a fitting routine will use this parameter to overfit the noise in $G(t)$, yielding a misleading value for $G_\epsilon$, which should be 0. Similarly, if the association lifetime is longer than the observation period, the terminal relaxation time $\tau_\epsilon$ is not apparent in the data, but the fitting routine will still attempt to determine the parameter. In order to evaluate which of these parameters can be meaningfully extracted from the data, We measured the Bayesian information criterion (BIC), presented in Fig.~\ref{fig:BIC}, for various subsets of the proposed model parameters. BIC is calculated as
\begin{equation}
    B=k\ln{n}-2\ln{\hat{L}},
    \label{eqn:BIC}
\end{equation}
where $n$ is the number of data points, $k$ is the number of model parameters, and $\hat{L}$ is the maximized likelihood of the model, here calculated as the total squared error of the fit. Each of the model variations tested in this way corresponds to a physically meaningful system state. For the case of a liquid with no discernible elastic response, we would have
\begin{equation}
    G_{liquid}(t)=\left[\mathbb{G}E_\alpha\left(-\frac{\mathbb{G}}{\mathbb{V}}t^\alpha\right)\right]\exp\left(-\frac{t}{\tau_\epsilon}\right).
    \label{eqn:MLliquid}
\end{equation}
For the case of the critical gel, with a power law tail of relaxations, so neither a terminal relaxation nor a plateau, we would have
\begin{equation}
    G_{critical}(t)=\left[\mathbb{G}E_\alpha\left(-\frac{\mathbb{G}}{\mathbb{V}}t^\alpha\right)\right].
    \label{eqn:MLcritical}
\end{equation}
In the case of a physical gel when the terminal flow due to dissociations is within the window of observation, the full model is appropriate. If the association lifetime is too long, then the elastic plateau appears permanent, and the system appears solid, leaving the expression
\begin{equation}
    G_{solid}(t)=\left[\mathbb{G}E_\alpha\left(-\frac{\mathbb{G}}{\mathbb{V}}t^\alpha\right)+G_\epsilon\right].
    \label{eqn:MLsolid}
\end{equation}

The comparisons in Fig.~\ref{fig:BIC} are consistent with the expected transition of the solution from liquid to critical to gel as concentration is increased. At low concentration, the liquid model is preferred. At concentration near 0.3$c^*$, the critical gel is favored. Just above the gel point, the network is not particularly robust, and there is ambiguity between the permanent or temporary model, the "solid" chemical gel model, or the "full" physical gel model. At high concentration with $\epsilon_{st}=12$, the gel appears to be permanent as far as data within our simulation window can show. Considering the possible distortion of parameter estimates due to over-fitting, calculations of viscosity in the sol state are carried out on a fit using the liquid model, where that model is preferred by the BIC. Analysis of $G_\epsilon$ is similarly limited to high concentrations. Note that in systems with low sticker strength, the BIC universally prefers the fluid model. Similar considerations were taken with the Akaike information criterion, which can prefer models with more parameters, but the results were the same.

\section{Scaling Exponents}

\begin{figure}[t]
 \begin{center}
   \resizebox{12cm}{!}{\includegraphics*[width=4cm]{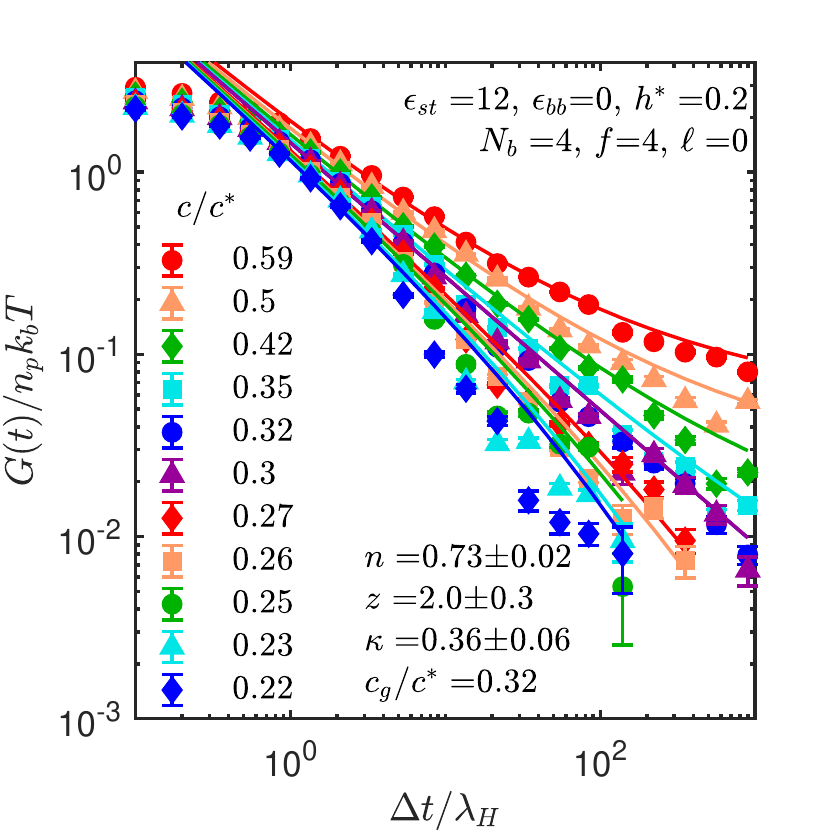}}
 \end{center}
 \vskip-15pt
 \caption{Simultaneous regression to computed $G(t)$ curves at several concentrations using the form (\ref{eqn:JoshiGoftForm}).
\label{fig:hyperscaling Goft}}
    \vspace{-20pt}
\end{figure}

Suman et al. (2021)\cite{10.1063/5.0038830} have arrived at the following expressions for $G(t)$ near the gel point,
\begin{align}
    G(t)=\left\{
    \begin{array}{l l l}
        &St^{-n}e^{-\alpha_St^{\kappa}};  & c<c_g \\
        &St^{-n};& c=c_g  \\
        & G(t) = St^{-n}(1+\alpha_G t^{\kappa}+...+\displaystyle\frac{\alpha_G^m t^{m\kappa}}{m!}) + G_e; &  c>c_g
    \end{array}\right.
    \label{eqn:JoshiGoftForm}
\end{align}
where $\alpha_S \sim (c_g -c)$, $\alpha_G \sim (c-c_g)$, $m=\mathrm{floor}(n/k)$, and $G_\epsilon\sim (c-c_g)^z$. A clear advantage of using these forms to extract scaling information is that the curves from many concentrations can be regressed simultaneously using consistent coefficients $c_g,S,n,\kappa$, and proportionality constants for $\alpha_S, \alpha_G$, and $G_\epsilon$, thus reducing the total number of free fitting parameters for the whole data set dramatically. The resulting fits and exponents are shown in Fig.~\ref{fig:hyperscaling Goft}. In the development of (\ref{eqn:JoshiGoftForm}) there are several time scales that are assumed to be well separated. In particular, the fastest Rouse relaxation and the dissociation time scale are assumed to be outside the observation window in the short and long time directions, respectively. However, in this work computational limitations and the study of dissociation dynamics required these time scales to be relatively close together. This produced $G(t)$ curves with several artifacts which are not accounted for in (\ref{eqn:JoshiGoftForm}). At short $\Delta t<\lambda_H$, we observe a plateau due to the relaxation time of a single spring toward the instantaneous shear modulus $G_0$, which is not accounted for in (\ref{eqn:JoshiGoftForm}). Accounting for this short-time plateau motivated our adoption of the Mitaag-Leffler form. Meanwhile, the longest Zimm relaxation time for a 4-bead chain at $h^*=0.2$ is $3.8\lambda_H$, and this time scale appears in our data as the floor, below which the terminal relaxation due to network modes cannot recede. So the observable network modes in the range 10-1000$\lambda_H$ are conflated with the single chain modes when $c$ isn't very close to $c_g$. Note in Fig.~12 of the main text that the terminal relaxation time $\tau_G$ grows from 10 to 1000 between $c/c^*=0.2-0.3$. The proximity of the single-chain relaxation time and the network relaxation time mean that the exponential cutoff $e^{-\alpha_St^{\kappa}}$ doesn't fit cleanly our sol phase data, as seen in the $c/c^*= 0.22$ and 0.23 curves in \ref{fig:hyperscaling Goft}. The result of these artifacts is that the apparent scaling exponents extracted using this fitting procedure do not comply with the hyperscaling relationship. Improving the confidence of the simulation scaling would therefore require many simulations in the range $c/c^*= 0.25$-0.3, and the duration of those simulations would need to be at least an order of magnitude longer to observe the termination of the network relaxation modes. Furthermore, the signal to noise ratio gets worse as $G(t)$ decays, so more simulation replicas would also be needed. Significant computational resources have already been consumed to locate \cg{} precisely enough to determine where this sort of focused data collection should be targeted. However, as discussed in the main text, applying a more phenomenological model and constraining the scaling fits for each set of parameters to satisfy the hyperscaling relationship reveals that our data are clearly consistent with this relationship.

\bibliography{bibfile_dyn}